\documentclass[12pt]{article} 
\usepackage{amsmath} 
\usepackage{amssymb,amsfonts,euscript} 
\usepackage{hyperref} 
\renewcommand{\baselinestretch}{1.2} 
\newcommand\nn{\nonumber} 
 
\newcommand{\A}{{\cal A}} 
\newcommand{\Ord}{{\cal{O}}} 
\newcommand{\C}{{\cal C}} 
 
\newcommand{\D}{{\cal D}}

\newcommand{\M}{{\cal M}}

\newcommand{\T}{{\cal T}} 
 
\newcommand{\F}{{\cal F}} 
\newcommand{\cL}{\cal{L}} 
\newcommand{\B}{{\cal B}} 
\newcommand{\lr}{{\cal{L}}_{\sgb(\s)}} 
 
\newcommand{\Zint}{\mathbb{Z}}

\newcommand{\bz}{\bar{z}} 
\newcommand{\bw}{\bar{w}}

\def\a{\alpha} 
\def\be{\beta} 
\def\l{\lambda} 
\def\m{\mu} 
\def\n{\nu} 
\def\r{\rho} 
\def\de{\delta} 
\def\k{\kappa}

\def\hgp{\hat{g}_+}

\def\part{\partial} 
\renewcommand{\sp}{,\hspace{.3in}} 

\newcommand{\sa}{\mathop{\vtop{\ialign{##\crcr 
$\hfil\displaystyle{\longrightarrow}\hfil$\crcr\noalign{\kern-1pt\nointerlineskip} 
\hspace{.12in}$^\sigma$\hskip6pt\crcr\noalign{\kern3pt}}}}} 
\newcommand{\slra}{\mathop{\vtop{\ialign{##\crcr 
$\hfil\displaystyle{\longleftrightarrow}\hfil$\crcr\noalign{\kern-1pt\nointerlineskip} 
\hspace{.12in}$^\sigma$\hskip6pt\crcr\noalign{\kern3pt}}}}} 
\newcommand{\sat}{\mathop{\vtop{\ialign{##\crcr 
$\hfil\displaystyle{\longrightarrow}\hfil$\crcr\noalign{\kern-1pt\nointerlineskip} 
\hspace{.12in}$^\sigma$\hskip6pt\crcr\noalign{\kern3pt}}}}} 
\newcommand{\pa}{\mathop{\vtop{\ialign{##\crcr 
$\hfil\displaystyle{\oplus}\hfil$\crcr\noalign{\kern+1pt\nointerlineskip} 
\hspace{.08in}$^{\alpha=0}$\hskip6pt\crcr\noalign{\kern3pt}}}}} 
\newcommand{\pan}{\mathop{\vtop{ialgin{##\crcr 
$\hfil\displaystyle{\oplus}\hfil$\crcr\noaligan{\kern+2pt\nointerlinkeskip} 
\hspace{.03in} $^{\alpha}$\hskip6pt\crcr\noalign{\kern3pit}}}}} 
\newcommand{\ka}{\mathop{\vtop{\ialign{##\crcr 
$\hfil\displaystyle{\longleftrightarrow}\hfil$\crcr\noalign{\kern-1pt\nointerlineskip} 
\hspace{.12in}$^K$\hskip6pt\crcr\noalign{\ker3pt}}}}} 
\newcommand{\bp}{\mathop{\vtop{ialign{##\crcr 
$\hfil\displaystyle{}\hfil$\crcr\noalign{\kern-13pt\nointerlineskip} 
\big{(}\hskip0pt\crcr\noalign{\kern3pt}}}}} 
\newcommand{\cbp}{\mathop{\vtop{ialign{##\crcr 
$\hfil\displaystyle{}\hfil$\crcr\noalign{\kern-13pt\nointerlineskip} 
\big{)}\hskip0pt\crcr\noalign{\kern3pt}}}}}

\newcommand{\s}{\sigma} 
\newcommand{\srange}{\sigma=0,...,N_c-1}

\renewcommand{\sp}{,\hspace{.3in}} 
 
\newcommand{\p}{^\prime} 
\newcommand{\ws}{\omega (h_\s)}

\newcommand{\hc}{$\hat{J}_{\gst}$} 
 
\newcommand{\tp}{{2\pi i}}

\newcommand{\dg}{\textup{dim}g}

\newcommand{\sgb}{{\mbox{\scriptsize{\gb}}}}

\newcommand{\gfraks}{{\mbox{\scriptsize{\mbox{${\mathfrak g}$}}}}} 
\def\gb            {\mbox{$\hat{\mathfrak g}$}}

\def\sm#1      {\mbox{\scriptsize $#1$}}

\def\z         {\mbox{$\mathbb  Z$}}

\def\srac#1#2{\smal{\frac{#1}{#2}}} 
\def\foot#1{\mbox{\footnotesize $#1$}}

\def\smal#1{\mbox{\small $#1$}} 
\def\big#1{\mbox{\large $#1$}}

\def\hjb{\hat{\bar{J}}}

\def\hjbb{ \hat{\bar{J}}^\sharp }

\def\dual{\underset{\s}{\longrightarrow}}

\def\sg{\smal{\EuScript{G}}} 
\def\sj{{\cal J}}

\def\sl{{\cal L}} 
 
\def\hc{^\dagger} 
\def\hcj{\dagger} 
\def\one{{\mathchoice {\rm 1\mskip-4mu l} {\rm 1\mskip-4mu} {\rm 1\mskip-4.5mu l} 
{\rm 1\mskip-5mu l}}} 
 
\def\e{\eta}

\def\nrm{{n(r)\m}}

\def\mnrn{{-n(r),\n}}

\def\mnrrs{{m+\srac{n(r)}{\r(\s)}}}

\def\sG{{\cal G}} 
\def\stG{{\tilde{\sG}}} 
\def\stF{{\tilde{\F}}} 
\def\gfrak{\mbox{$\mathfrak g$}} 
\def\goto{\longrightarrow} 
\def\hj{\hat{J}}

\def\schisig{{\foot{\chi(\s)}}} 

\def\hc{^\dagger} 
\def\st{{\cal T}} 
\def\0b{\ } 
\def\pl{\partial}

\def\srange{\s=0,\ldots,N_c-1} 
\def\sm{{\cal M}}

\def\Lg{{\frac{\e^{ab}}{2 k + Q_{\gfraks}}}} 
\def\Dg{{\Delta_{\gfraks}(T)}} 
\def\dg{{\Delta_{\gfraks}}} 
\def\jfj{{\frac{\hat{j}}{f_j(\s)}}} 
\def\fji{{\frac{1}{f_j(\s)}}} 
\def\modmj{{(m+ \jfj)}} 
\def\modnl{{(n+ \frac{\hat{l}}{f_l(\s)})}} 
\def\modmn{{(m + n +\frac{\hat{j}+\hat{l}}{f_j(\s)})}} 
\def\cg{{c_{\gfraks}}} 
\def\hg{{\tilde{h}_{\gfraks}}} 
\def\bigspc{{\quad \quad \quad \quad}} 
 
\def\gscfwt{{\hat{\Delta}_0 (\s)}} 
\def\gscfwtj{{\hat{\Delta}_{0j} (\s)}} 
\def\hlbb{{\hat{\bar{L}}^\sharp}} 
\def\hggz{{\hat{g} (\st,\bz,z,\s)}}

\def\hggp{{\hat{g}_+ (\st,z,\s)}} 
\def\lpl{{\overleftarrow{\pl}}} 
\def\lplw{{\overleftarrow{\pl_w}}} 
 
\def\lplm{{\overleftarrow{\pl_{\m}}}} 
\def\Lpone{{\hat{L}_{1,j} (\frac{1}{f_j (\s)})}} 
\def\Lmone{{\hat{L}_{f_j (\s)-1 ,j} (-1 +\frac{f_j (\s)-1}{f_j (\s)})}} 
\def\t{{\Theta}} 
\def\hgjz{{\hat{g}_j (\st,\bz,z,\s)}} 
\def\hgjm{{\hat{g}_j^- (\st,\bz,\s)}} 
\def\hgjp{{\hat{g}_j^+ (\st,z,\s)}} 
\def\pp{{^{\prime \prime}}} 
 
\def\hglm{{\hat{g}_l^- (\st,\bz,\s)}} 
\def\hglp{{\hat{g}_l^+ (\st,z,\s)}} 
\def\Aj0{{| A (j;\st) \rangle_\s}} 
\def\Ajop{{| A^+ (j;\st) \rangle_\s}} 
\def\Ajom{{| A^- (j;\st) \rangle_\s}}

\def\su{{ \mathfrak{su} }} 
\def\so{{ \mathfrak{so} }} 
\def\sl{{ \mathfrak{sl} }} 
\def\nrrs{{ \frac{n(r)}{\r(\s)}}}
\def\thickone{{\rm 1\mskip-4.5mu l}}
 
\makeatletter 
\renewcommand{\@makefnmark}{\mbox{$^{\ddagger\@thefnmark}$}} 
\renewcommand{\subsection}{\@startsection 
{subsection}{2}{0pt
}{-\baselineskip}{0.5\baselineskip} 
{\normalfont\normalsize\bf}} 
\renewcommand{\section}{\@startsection 
{section}{2}{0pt
}{-\baselineskip}{0.5\baselineskip} 
{\bf\large}} 

\makeatother 
\numberwithin{equation}{section} 
\numberwithin{table}{section}

\newcommand{\publititle}[8] 
{ 
  \vspace*{-3cm} 
  \begin{flushright} #1 \\ {\tt #2} \end{flushright} 
  \vfill 
  \begin{center}{\Large
    \bfseries #3}\end{center} 
  \vskip 8mm 
  \begin{center}{\large #4}\end{center} 
  \begin{center}{\normalsize #5}\end{center} 
  \vskip 8mm 
  \nopagebreak 
  \noindent #6 
  \vfill 
  \begin{flushleft} #7 
  \end{flushleft} 
  \hrule width 6.7cm \vskip.1mm 
  {\small #8} 
  \thispagestyle{empty} 
  \clearpage 
} 
 
\begin{document} 
 
\publititle{ ${}$ \\ UCB-PTH-02/30 \\ LBNL-51112 \\ hep-th/0208087}{}{Extended Operator Algebra and Reducibility\\ 
in the WZW Permutation Orbifolds} 
{M.B.Halpern$^{\S}$ and 
C. Helfgott$^{\dagger}$} 
{Department of Physics, 
University of California, 
Berkeley, California 94720, USA  \\ {\it and} 
Theoretical Physics Group,  Lawrence Berkeley National 
Laboratory \\ 
University of California, 
Berkeley, California 94720, USA \\
August 12, 2002
\\[2mm]} {Recently the operator algebra, including the twisted affine primary fields, and a set of twisted 
KZ equations were given for the WZW permutation orbifolds. In the first part of this paper we extend 
this operator algebra to include the so-called {\it orbifold Virasoro algebra} of each WZW permutation 
orbifold. These algebras generalize the orbifold Virasoro algebras (twisted Virasoro operators) found
some years ago in the cyclic permutation orbifolds. In the second part, we discuss the {\it 
reducibility} of the twisted affine primary fields of the WZW permutation orbifolds, obtaining a 
simpler set of {\it single-cycle twisted KZ equations}. Finally we combine the orbifold Virasoro algebra 
and the single-cycle twisted KZ equations to investigate the spectrum of each orbifold, identifying the 
analogues of the {\it principal primary states and fields} also seen earlier in cyclic permutation 
orbifolds. Some remarks about general WZW orbifolds are also included.
} {$^{\S}${\tt halpern@physics.berkeley.edu} \\ $^{\dagger}${\tt  helfgott@socrates.berkeley.edu} 
} 
 
\clearpage 
 
\renewcommand{\baselinestretch}{.4}\rm 
{\footnotesize 
\tableofcontents 
} 
\renewcommand{\baselinestretch}{1.2}\rm 
 
\section{Introduction} 

In the last few years there has been a quiet revolution in the local theory of {\it current-algebraic orbifolds}.
Building on the discovery of {\it orbifold affine algebras} \cite{Borisov:1997nc, Evslin:1999qb} in the cyclic 
permutation orbifolds, Refs.[3-5] gave the twisted currents and stress tensor in each twisted sector of any current-algebraic 
orbifold $A(H)/H$ - where $A(H)$ is any current-algebraic conformal field theory [6-11] with a finite symmetry group $H$. The 
construction treats all current-algebraic orbifolds at the same time, using the method of {\it eigenfields} and the 
{\it principle of local isomorphisms} to map OPEs in the symmetric theory to OPEs in the orbifold. The orbifold results are 
expressed in terms of a set of twisted tensors or {\it duality transformations}, which are discrete Fourier transforms constructed 
from the eigendata of the $H$-{\it eigenvalue problem}.

More recently, the special case of the {\it WZW orbifolds} $A_g (H)/H$ was worked out in further detail 
\cite{deBoer:2001nw,Halpern:2002ab}, while extending the operator algebra in this case to include the {\it twisted 
affine primary fields}, {\it twisted vertex operator equations} and {\it twisted Knizhnik-Zamolodchikov equations} 
for each sector of every WZW orbifold:

$\bullet$ The WZW permutation orbifolds \cite{deBoer:2001nw,Halpern:2002ab}

$\bullet$ Inner-automorphic WZW orbifolds \cite{deBoer:2001nw}

$\bullet$ The (outer-automorphic) charge conjugation orbifold on $\su (n \geq 3)$ \cite{Halpern:2002ab}

$\bullet$ Other outer-automorphic WZW orbifolds on simple $g$ \cite{Halpern:2002ab}

\noindent Ref.~\cite{Halpern:2002ab} also solved the twisted vertex operator equations in an abelian limit to obtain the 
{\it twisted vertex operators} for each sector of a very large class of abelian orbifolds. For discussion of WZW orbifolds 
at the level of characters, see Refs.~\cite{Verlinde:1989oa, Borisov:1997nc,Bantay:1998ek, Birke:1999ik}.

In addition to the operator formulation, the {\it general WZW orbifold action} was also given in Ref.~\cite{deBoer:2001nw}, 
with applications to special cases in Refs.~\cite{deBoer:2001nw,Halpern:2002ab}. The general WZW orbifold action provides 
the classical description of each sector of every WZW orbifold in terms of appropriate {\it group orbifold elements}, 
which are the classical limit of the twisted affine primary fields. Moreover, Ref.~\cite{Halpern:2002hw} gauged the 
general WZW orbifold action by general twisted gauge groups to obtain the {\it general coset orbifold action}.

In this paper, we consider the local theory of all {\it WZW permutation orbifolds} in further detail, with special 
attention to the generalization of certain features \cite{Borisov:1997nc}

$\bullet$ orbifold Virasoro algebra (twisted Virasoro operators)

$\bullet$ orbifold $\sl (2)$ Ward identities

$\bullet$ principal primary states and fields

\noindent observed earlier by different methods in $\Zint_\l$ cyclic permutation orbifolds with $\l=$ prime. The orbifold
Virasoro algebras for $\Zint_\l ,\,\,\l=$ prime were discovered independently in Refs. \cite{Borisov:1997nc, Dijk:1997dv}, 
and a recent application of these algebras is found in Ref.~\cite{Gomis:2002pp}. As we shall see, there is a {\it disjoint 
cycle} structure underlying all of these features, which is closely related to the {\it reducibility} of the twisted affine 
primary fields and the twisted KZ equations.

In further detail, Sec.~2 applies the principle of local isomorphisms to extend the operator algebra to include the 
{\it general orbifold Virasoro algebra} of the general WZW permutation orbifold. As in Ref.~\cite{Borisov:1997nc}, the 
general orbifold Virasoro algebra is associated to {\it orbifold} $\sl (2)${\it Ward identities} for the correlators of 
the twisted affine primary fields, and we find that these Ward identities are satisfied by any solution of the twisted 
KZ system.

After some discussion of the action formulation \cite{deBoer:2001nw,Halpern:2002hw} and an abelian limit 
\cite{Halpern:2002ab} of the WZW permutation orbifolds, we argue in Sec.~3 that the twisted affine primary fields of 
the WZW permutation orbifolds are generically {\it reducible}. This leads us to a new, simpler set of {\it single-cycle
twisted KZ equations} for the correlators of the blocks of the twisted affine primary fields (see Eq.~\eqref{FactKZ2}).

In Sec.~4, we combine the extended operator algebra and the new single-cycle twisted KZ system to study aspects of the 
{\it spectrum} of each WZW permutation orbifold. Following Ref.~\cite{Borisov:1997nc}, we find in particular that 
the {\it asymptotic formulae} for in- and out-states are modified relative to the conventional asymptotic formulae of 
untwisted CFT. Moreover, using these orbifold-modified asymptotic formulae, we find that the twisted affine primary 
fields create {\it twisted affine primary states}, which are also primary under the orbifold Virasoro algebra. The 
twisted affine primary states also serve as base states for certain sets of so-called {\it principal primary states} 
\cite{Borisov:1997nc} in the modules of the orbifold Virasoro algebra. We are also able to identify the matrix 
components of the twisted affine primary fields as the analogues of the old {\it principal primary fields} of 
Ref.~\cite{Borisov:1997nc}.

Finally, Sec.~5 considers asymptotic formulae and the reducibility of the twisted affine primary fields and states 
of general WZW orbifolds.

Because we are studying features specific to the WZW permutation orbifolds, this paper is in large part self-contained
-- using primarily the notation appropriate \cite{Halpern:2000vj,deBoer:2001nw,Halpern:2002ab} to the WZW permutation 
orbifolds. To understand this paper in the context of the general orbifold program, see Refs.~\cite{deBoer:1999na, 
Halpern:2000vj,deBoer:2001nw,Halpern:2002ab} and the general remarks in Sec.~5.
 
\section{Extended Operator Algebra in WZW Permutation Orbifolds} 
 
\subsection{Permutation-Invariant WZW Systems} 
 
In this subsection, we will extend the usual chiral algebra of permutation-invariant WZW systems to  
include what we will call the {\it partial stress tensors} of such systems. 
 
To begin, we recall the notation of Ref.~\cite{deBoer:2001nw} for permutation-invariant WZW systems: 
\begin{subequations} 
\begin{equation} 
g=\oplus_I \gfrak^I ,\quad \gfrak^I \simeq \gfrak ,\quad k^I =k   
\end{equation} 
\begin{equation} 
f_{aI,bJ}{}^{cL} = f_{ab}{}^c (\tau_I )_J{}^L ,\quad G_{aI,bJ} = k\e_{ab} \de_{IJ} 
\end{equation} 
\begin{equation} 
\label{TauDef1}
T = \oplus_I T^I ,\quad T_{aI} = T_a \tau_I ,\quad  
[T_a ,T_b ]= if_{ab}{}^c T_c ,\quad (\tau_I )_J{}^L \equiv \de_{IJ} \de_J{}^L ,\quad \tau_I \tau_J = \de_{IJ} \tau_I 
\end{equation} 
\begin{equation} 
I=0,\ldots,K-1 ,\quad a,b,c = 1,\ldots , \text{dim}\gfrak . 
\end{equation} 
\end{subequations} 
Here the index $I$ labels the copies $\gfrak^I$ of any simple $\gfrak$, whose structure constants and Killing metric  
are $f_{ab}{}^c$ and $\e_{ab}$ respectively, and $\{ T_a \}$ is any matrix irrep of $\gfrak$. Then Ref.~\cite{deBoer:2001nw} 
gives the left-mover OPEs of the system: 
\begin{subequations} 
\label{calg} 
\begin{equation} 
\label{curalg} 
J_{aI} (z) J_{bJ} (w) = \frac{\delta_{IJ} k \eta_{ab}}{(z-w)^2} + \frac{if_{ab}{}^c J_{cI}(w) }{z-w} + \Ord (z-w)^0 
\end{equation} 
\begin{equation} 
\label{utj} 
T(z) J_{aI} (w) = \left( \frac{1}{(z-w)^2} + \frac{\pl_w }{z-w} \right) J_{aI} (w) +\Ord (z-w)^0 
\end{equation} 
\begin{equation} 
\label{utt} 
T(z) T(w) = \frac{c_g/2}{(z-w)^4} + \left( \frac{2}{(z-w)^2} + \frac{\partial_w }{z-w} \right) T(w) 
+\Ord (z-w)^0 
\end{equation} 
\begin{equation} 
\label{Jgl} 
J_{aI} (z) g (T,\bw,w) = \frac{g(T,\bw,w)}{z-w}  T_a \tau_I + \Ord (z-w)^0 
\end{equation} 
\begin{equation} 
\label{Tgl} 
T (z) g (T,\bw,w) = \frac{g(T,\bw,w)\Dg}{(z-w)^2} + \frac{\part_w}{z-w} g(T,\bw,w) + \Ord (z-w)^0 
\end{equation} 
\begin{equation} 
\label{ConfWt} 
\Lg T_a T_b \equiv \Dg \one .
\end{equation} 
\end{subequations} 
Here $\{ J_{aI}(z)\}$ are the currents \cite{Kac:1967,Moody:1967gf,Bardakci:1971nb} of affine $g$, $g(T,\bz,z)$ is the 
affine primary field corresponding to matrix irrep $T$ of $\gfrak$, and the constant $\Dg$ is the conformal weight of 
irrep $T$ under the affine-Sugawara  construction $T_{\gfraks} (z)$ on $\gfrak$. The affine primary fields are 
matrix-valued, with index structure
\begin{equation} 
g(T,\bz,z)_{\a I}{}^{\be J} ,\quad \a,\be = 1,\ldots,\text{dim}\;T ,\quad I,J = 0,\ldots,K-1 . \label{Ind-Struct}
\end{equation} 
In fact, the affine primary fields are block-diagonal, with blocks labelled by the copies $I$, but we shall  
not use this fact explicitly until Sec. 3. The stress tensor $T(z)$ in \eqref{calg} is the affine-Sugawara 
construction [6, 7, 22-24] on $g$ 
\begin{subequations} 
\label{affSug} 
\begin{equation} 
T(z) \equiv T_g (z) = \Lg \sum_I : J_{aI}(z) J_{bI}(z) : 
\end{equation} 
\begin{equation} 
c_g \equiv K \cg ,\quad \cg \equiv \frac{x\;\text{dim}\gfrak}{x + \hg},\quad x = \frac{2 k} 
{\psi_{\gfraks}^2} ,\quad \hg = \frac{Q_{\gfraks}}{\psi_{\gfraks}^2} . 
\end{equation} 
\end{subequations} 
The right-mover analogue of the system \eqref{calg}, \eqref{affSug} is discussed in Ref.~\cite{deBoer:2001nw}. 
 
In this paper we will extend the system \eqref{calg}-\eqref{affSug} to include the set of {\it partial 
stress tensors} $T_I (z)$ 
\begin{subequations} 
\begin{equation} 
T_I (z) \equiv \Lg : J_{aI}(z) J_{bI} (z) : ,\quad T_I (z) \simeq T_{\gfraks} (z) 
\end{equation} 
\begin{equation} 
\label{sumT} 
T (z) = \sum_I T_I (z) 
\end{equation} 
\end{subequations} 
which sum to the full stress tensor $T(z)$. The partial stress tensors are the individual affine-Sugawara 
constructions on the copies $\gfrak^I$, and they satisfy the additional OPEs: 
\begin{subequations} 
\label{ExtOPE}
\begin{equation} 
T_I (z) J_{aJ}(w) = \de_{IJ} \left( \frac{1}{(z-w)^2} + \frac{\partial_w }{z-w} \right) J_{aI} (w) 
+\Ord (z-w)^0 
\end{equation} 
\begin{equation} 
T_I (z) T_J (w) = \de_{IJ} \left( \frac{\cg /2}{(z-w)^4} + \left( \frac{2}{(z-w)^2} + \frac{\pl_w}{z-w} 
\right) T_I (w) \right) +\Ord (z-w)^0 
\end{equation} 
\begin{equation} 
T(z) T_I (w) = \frac{\cg /2}{(z-w)^4} + \left( \frac{2}{(z-w)^2} + \frac{\pl_w }{z-w} \right) T_I (w)  
+\Ord (z-w)^0 
\end{equation} 
\begin{equation} 
T_I (z) g(T,\bw,w) = \left( \frac{g(T,\bw,w) \Dg}{(z-w)^2} + \frac{\pl_w g(T,\bw,w)} 
{(z-w)} \right) \tau_I +\Ord (z-w)^0 .
\end{equation} 
\end{subequations} 
The relations \eqref{utj}, \eqref{utt}, \eqref{Tgl} for the stress tensor $T(z)$ are obtainable 
from these OPE's by the identity \eqref{sumT}. 

We will also need the associated left-mover vertex operator equation (see e.g. Ref.~\cite{Halpern:1996et})
\begin{equation} 
\pl g(T,\bz,z) = \frac{2 \eta^{ab}}{2 k +Q_{\gfraks}} \sum_I : J_{aI} (z) g(T,\bz,z): T_b \tau_I 
\end{equation} 
which follows from the OPE's above.  

We turn next to the action of the symmetry group
\begin{equation} 
H({\rm permutation}) \subset{\rm Aut}(g) ,\quad H({\rm permutation}) \subset S_N,\,\,\, K\leq N
\end{equation} 
of the WZW model defined above.  Picking one representative $h_\s \in H$ of each conjugacy class,
the action is \cite{deBoer:2001nw} 
\begin{subequations} 
\label{Haction}
\begin{equation} 
J_{aI} (z)\p = \ws_I{}^J J_{aJ} (z) ,\quad \srange
\end{equation} 
\begin{equation}
\label{ExtHEig}
g (T,\bz,z)\p = W (T, h_\s) g (T,\bz,z) W^{\dagger} (T, h_\s) ,\quad W (T, h_\s)_{\a I}{}^{\be J} = 
\de_\a{}^\be \ws_I{}^J 
\end{equation} 
\begin{equation} 
T_I (z)\p = \ws_I{}^J T_J (z), \quad T (z)' = T (z) 
\end{equation} 
\end{subequations} 
where $\ws$ is the action in the adjoint representation of $h_\s \in H$, $W(T,\s)$ is the action of  
$h_\s$ in irrep $T$ and $N_c$ is the number of conjugacy classes of $H$.  The action \eqref{Haction} is 
an automorphism of the extended OPE's \eqref{calg}, \eqref{ExtOPE} if and only if three conditions hold 
\begin{subequations} 
\label{sumw} 
\begin{equation} 
\sum_I \ws_I{}^J = \sum_J \ws_I{}^J = 1 ,\quad 
\ws_I{}^L \ws_J{}^M \de_{LM} = \de_{IJ} 
\end{equation} 
\begin{equation}
\label{wCond2}
\ws_I{}^L \ws_J{}^M (\tau_L )_M{}^N \omega^{\dagger} (h_\s)_N{}^K = (\tau_I )_J{}^K .
\end{equation} 
\end{subequations} 
These conditions are satisfied by all permutation matrices, and only by permutation matrices. Together,
\eqref{wCond2} and the second part of \eqref{ExtHEig} provide the solution to the linkage relation \cite{deBoer:2001nw}
for permutation groups.

\subsection{Eigenfields and Duality Transformations \label{Eigf}} 

In current-algebraic orbifold theory, we consider next the $H$-eigenvalue problem \cite{deBoer:1999na,Halpern:2000vj}, 
which diagonalizes the action of $\ws$ in each sector. For permutation groups, the $H$-eigenvalue problem has the reduced 
form \cite{Halpern:2000vj}
\begin{subequations} 
\label{UForm1}
\begin{equation} 
\ws_I{}^J U^{\dagger} (\s)_J{}^{n(r)j} = U^{\dagger} (\s)_I{}^{n(r)j} E_{n(r)} (\s) ,\quad \srange 
\end{equation} 
\begin{equation} 
E_{n(r)} (\s) = e^{-\tp \frac{n(r)}{\r(\s)}}, \quad n(r) \equiv n(r(\s)) \in \Zint 
\end{equation} 
\end{subequations} 
where $\r(\s)$ is the order of $h_\s \in H$. All quantities are periodic $n(r) \goto n(r)\pm \r(\s)$ in
the spectral indices $n(r)$, and, correspondingly, the generic eigenvector matrix $U^\hcj (\s)$ is a discrete 
Fourier element. 

The solution to the $H$-eigenvalue problem allows us to define the standard eigencurrents \cite{deBoer:1999na,Halpern:2000vj}
and affine eigenprimary fields \cite{deBoer:2001nw} of sector $\s$
\begin{subequations} 
\begin{equation} 
\sj_{n(r)aj}(z,\s) \equiv \schisig_{n(r)j} U(\s)_{n(r)j}{}^IJ_{aI}(z) ,\quad \srange
\end{equation} 
\begin{equation} 
\sg(T,\bz,z,\s) \equiv U(T,\s) g(T,\bz,z) U^\hcj (T,\s) , \quad U(T,\s)_{n(r)\a j}{}^{\be I} \equiv 
\de_\a{}^\be U(\s)_{n(r)j}{}^I 
\end{equation} 
\begin{equation} 
\label{eigeng} 
\sg(T,\bz,z,\s)_{n(r)\a j}{}^{n(s)\be l} = U(\s)_{n(r)j}{}^I g(T,\bz,z)_{\a I}{}^{\be J} 
U^\hcj(\s)_J{}^{n(s)l} ,\quad \a ,\be =1,\ldots ,\text{dim}\;T 
\end{equation} 
\end{subequations} 
where following Ref.~\cite{Halpern:2002ab} we have chosen an $a$-independent normalization $\chi$. To these, we may 
add the set of {\it partial eigenstress tensors} $\t_{n(r)j}$,
\begin{subequations} 
\begin{equation} 
\t_{n(r)j} (z,\s) \equiv \schisig_{n(r)j} U(\s)_{n(r)j}{}^I T_I(z) 
\end{equation} 
\begin{equation} 
T (z) = \A^{n(r)j}(\s) \t_{n(r)j} (z,\s) = \sum_j \A^{0j}(\s) \t_{0j} (z,\s) 
\end{equation} 
\begin{equation} 
\label{Aselec}
\A^{n(r)j}(\s) \equiv \schisig_{n(r)j}^{-1} \sum_I U^{\dagger}(\s)_I{}^{n(r)j}  
= \de_{n(r), 0\,\text{mod }\r(\s)} \A^{0j}(\s) 
\end{equation} 
\end{subequations} 
where the $\A$-selection rule in \eqref{Aselec} is implied by Eqs. \eqref{sumw} and \eqref{UForm1}. 
The eigenfields $\sj ,\sg ,$ and $\t$ are constructed to have the following diagonal responses
\begin{subequations} 
\label{EigAuto}
\begin{equation} 
\sj_{n(r)aj}(z,\s)\p = E_{n(r)}(\s) \sj_{n(r)aj}(z,\s) = e^{-\tp \frac{n(r)}{\r(\s)}} \sj_{n(r)aj} (z,\s) 
\end{equation} 
\begin{equation} 
\sg(T,\bz,z,\s)\p = E(T,\s) \sg(T,\bz,z,\s) E(T,\s)^{\ast} 
\end{equation} 
\begin{equation} 
E(T,\s)_{n(r)\a j}{}^{n(s)\be l} \equiv \de_\a{}^\be \de_j^l \de_{n(r)-n(s) ,0\,\text{mod }\r(\s)} E_{n(r)}(\s) 
\end{equation} 
\begin{equation} 
\t_{n(r)j} (z,\s)\p = E_{n(r)}(\s) \t_{n(r)j} (z,\s) = e^{-\tp \frac{n(r)}{\r(\s)}} \t_{n(r)j} (z,\s) 
\end{equation} 
\begin{equation} 
T(z)\p = T(z) 
\end{equation} 
\end{subequations} 
to the automorphism group $H$.

The next step is to reexpress everything in terms of the eigenfields. For example, we may express the
stress tensor and partial eigenstress tensors in terms of the eigencurrents as follows:
\begin{subequations}
\begin{equation} 
T (z) = {\cL}_{\sgb (\s)}^{n(s)al ; n(t)bm}(\s) : \sj_{n(s)al}(z,\s) \sj_{n(t)bm}(z,\s) : 
\end{equation} 
\begin{eqnarray} 
{\cL}_{\sgb (\s)}^{n(s)al ; n(t)bm}(\s) & \equiv & \schisig_{n(s)al}^{-1} \schisig_{n(t)bm}^{-1} 
\frac{\e^{ab}}{2 k + Q_g} \sum_I U^{\dagger}(\s)_I{}^{n(s)l} U^{\dagger}(\s)_I{}^{n(t)m} \nn \\ 
& = & \de_{n(s)+n(t) ,0\,\text{mod }\r(\s)} {\cL}_{\sgb (\s)}^{n(s)al ; -n(s),bm}(\s) 
\end{eqnarray} 
\begin{equation} 
\t_{n(r)j} (z,\s) = {\cL}_{n(r)j}{}^{n(s)al; n(t)bm} (\s) : \sj_{n(s)al}(z,\s) \sj_{n(t)bm}(z,\s) : 
\end{equation} 
\begin{eqnarray} 
\label{PlTIIT}
{\cL}_{n(r)j}{}^{n(s)al ;n(t)bm} (\s) & \equiv & \frac{\schisig_{n(r)j}}{\schisig_{n(s)al} 
\schisig_{n(t)bm}} \Lg \sum_I U(\s)_{n(r)j}{}^I U^{\dagger}(\s)_I{}^{n(s)l} 
U^{\dagger}(\s)_I{}^{n(t)m} \nn \\ 
& = & \de_{n(r)-n(s)-n(t) ,0\,\text{mod }\r(\s)} {\cL}_{n(r)j}^{n(s)al ;n(r)-n(s),bm}(\s) 
\end{eqnarray} 
\begin{equation} 
{\cL}_{\sgb (\s)}^{n(s)al ; n(t)bm}(\s) = \A^{n(r)j}(\s) {\cL}_{n(r)j}^{n(s)al; n(t)bm} (\s) .
\end{equation} 
\end{subequations} 
Here the forms in (2.15a,b) are well-known \cite{deBoer:1999na,Halpern:2000vj} and ${\cL}_{\sgb (\s)} (\s)$ is 
called the twisted inverse inertia tensor. The twisted tensors ${\cL}_{\sgb (\s)} (\s)$ and ${\cL} (\s)$ in 
\eqref{PlTIIT} are examples of {\it duality transformations}, which are discrete Fourier transformations of 
tensors from the untwisted theory.

We also need to compute the OPE's of the eigenfields, 
\begin{subequations} 
\begin{eqnarray} 
\sj_{n(r)aj} (z,\s) \sj_{n(s)bl} (w,\s) = \frac{\sG_{n(r)aj;n(s)bl} (\s)}{(z-w)^2} + \bigspc \bigspc  
\bigspc \nn \\ 
\bigspc + \frac{ i\F_{n(r)aj;n(s)bl}{}^{n(t)cm} (\s) \sj_{n(t)cm} (w,\s)}{z-w} + \Ord (z-w)^0 
\end{eqnarray} 
\begin{eqnarray} 
\t_{n(r)j}(z,\s) \t_{n(s)l}(w,\s) = \stG_{n(r)j; n(s)l}(\s) \frac{\cg /2}{(z-w)^4} +  \bigspc \bigspc 
\bigspc \nn \\ 
\quad \quad +\stF_{n(r)j; n(s)l}{}^{n(t)m}(\s)  \left( \frac{2}{(z-w)^2} + \frac{\partial_w}{z-w} 
\right) \t_{n(t)m} (w,\s) + \Ord (z-w)^0 
\end{eqnarray} 
\begin{equation} 
T (z) \t_{n(r)j} (w,\s) = \B_{n(r)j} (\s) \frac{\cg /2}{(z-w)^4} + \left( \frac{2}{(z-w)^2} + 
\frac{\pl_w}{z-w} \right) \t_{n(r)j} (w,\s) + \Ord (z-w)^0 
\end{equation} 
\begin{equation} 
T(z) \sg(T,\bw,w,\s) = \sg(T,\bw,w,\s) \left( \frac{\D_{\sgb (\s)} (\st (T,\s))}{(z-w)^2} + \frac{\lplw}{z-w} 
\right) +\Ord (z-w)^0  
\end{equation} 
\end{subequations} 
where $\sG, \F,$ and $\D_{\sgb (\s)}$ are called respectively the twisted metric \cite{deBoer:1999na,Halpern:2000vj}, 
the twisted structure constants \cite{deBoer:1999na,Halpern:2000vj}, and the twisted conformal weight matrix \cite{deBoer:2001nw}. 
The twisted tensors $\stG, \stF,$ and $\B$ appear for the first time in this paper. The explicit forms of all these duality 
transformations are
\begin{subequations} 
\begin{equation} 
\sG_{n(r)aj;n(s)bl}(\s) = k \e_{ab} \stG_{n(r)j;n(s)l} (\s)  
\end{equation} 
\begin{equation} 
\stG_{n(r)j;n(s)l} (\s) \equiv \schisig_{n(r)j} \schisig_{n(s)l} \sum_I U(\s)_{n(r)j}{}^I U(\s)_{n(s)l}{}^I 
\end{equation} 
\begin{equation} 
\F_{n(r)aj;n(s)bl}{}^{n(t)cm}(\s) = f_{ab}{}^c \stF_{n(r)j;n(s)l}{}^{n(t)m} (\s) 
\end{equation} 
\begin{equation} 
\stF_{n(r)j;n(s)l}{}^{n(t)m}(\s) \equiv \frac{\schisig_{n(r)j} \schisig_{n(s)l}}{\schisig_{n(t)m}} 
U(\s)_{n(r)j}{}^I U(\s)_{n(s)l}{}^J (\tau_I )_J{}^L U^\hcj (\s)_L{}^{n(t)m} 
\end{equation} 
\begin{equation} 
\B_{n(r)j}(\s) \equiv \A^{n(s)l} (\s) \stG_{n(s)l;n(r)j} (\s) = \schisig_{n(r)j} \sum_I U(\s)_{n(r)j}{}^I 
,\quad \A^{n( r)j}(\s) \B_{n( r)j}(\s) = K 
\end{equation} 
\begin{equation} 
\label{TwCfWt} 
\D_{\sgb (\s)} (\st (T,\s)) \equiv U(T,\s) \Dg U^\hcj (T,\s) = {\cL}_{\sgb (\s)}^{n(r)aj;n(s)bl} \st_{n(r)aj} 
(T,\s) \st_{n(s)bl} (T,\s) 
\end{equation} 
\end{subequations} 
where $K$ is defined in \eqref{Ind-Struct}. In \eqref{TwCfWt} we also have the duality transformations called the 
twisted representation matrices $\st$, which in this case have the form
\begin{subequations} 
\begin{equation} 
\label{TwRepM1}
\st_{n(r)aj} (T,\s) \equiv \schisig_{n(r)j} U(\s)_{n(r)j}{}^I U(T,\s) T_a \tau_I U\hc (T,\s) = T_a t_{n(r)j} (\s) 
\end{equation} 
\begin{eqnarray} 
t_{n(r)j}(\s)_{n(s)l}{}^{n(t)m} &=& \schisig_{n(r)j} U(\s)_{n(r)j}{}^I U(\s)_{n(s)l}{}^J (\tau_I )_J{}^K 
U\hc (\s)_K{}^{n(t)m} \nn \\
&=& \schisig_{n(s)l}^{-1} \schisig_{n(t)m} \stF_{n(r )j;n(s)l}{}^{n(t)m}(\s) 
\end{eqnarray} 
\end{subequations} 
where the matrices $\{ \tau_I \}$ were defined in \eqref{TauDef1}.
 
To solve the $H$-eigenvalue problem and evaluate the duality transformations, it is convenient to use the 
(untwisted but $\s$-dependent) cycle basis defined in Ref.~\cite{Halpern:2002ab}. In this basis, each element $h_\s$ is 
expressed in terms of disjoint cycles of size $f_j(\s)$, where $j$ indexes the cycles and $\hat{j}$ 
indexes the position within the $j$th cycle.  Then we have \cite{Halpern:2002ab} 
\begin{subequations} 
\label{CycleB} 
\begin{equation} 
\label{TildeFs}
I \goto \hat{j}j, \quad J_{aI}(z) \goto \tilde{J}_{a\hat{j}j} (z,\s) ,\quad T_I (z) \goto \tilde{T}_{\hat{j}j} 
(z,\s) 
\end{equation} 
\begin{equation} 
n(r)j \goto \hat{j} j, \quad \sj_{n(r)aj}(z,\s) \goto \sj_{\hat{j}aj} (z,\s) ,\quad \t_{n(r)j} (z,\s) \goto 
\t_{\hat{j}j} (z,\s) 
\end{equation} 
\begin{equation} 
\frac{n(r)}{\r(\s)} = \frac{N(r)}{R(\s)} = \frac{\hat{j}}{f_j(\s)}, \quad \bar{\hat{j}} =0,\ldots, 
f_j(\s) -1 ,\quad \sum_j f_j (\s) = K 
\end{equation} 
\begin{equation} 
\label{Uform} 
U(\s)_{n(r)j}{}^I \goto U(\s)_{n(r)j}{}^{\hat{l}l} \goto U(\s)_{\hat{j}j}{}^{\hat{l}l} = 
\frac{\de_{jl}}{\sqrt{f_j(\s)}} \,\, e^{\frac{\tp \hat{j} \hat{l}}{f_j(\s)}} , \quad E_{\hat{j}}(\s) 
= e^{-\frac{\tp \hat{j}}{f_j(\s)}} 
\end{equation} 
\begin{equation} 
\schisig_{\hat{j}aj} = \schisig_{\hat{j}j} = \sqrt{f_j(\s)} 
\end{equation} 
\end{subequations} 
where $\bar{\hat{j}}$ is the pullback of the spectral index $\hat{j}$ to the fundamental domain. As a 
computational aid for the reader, we list some well-known examples 
\begin{subequations} 
\begin{align} 
& \Zint_\l : \! K = \lambda, \, f_j(\s)\! = \!\rho(\s), \, \bar{\hat{j}} = 0,\ldots, \rho(\s)\! -\!1, \, j = 0,\ldots, 
\frac{\lambda}{\rho(\s)}\! -\!1,\,\s = 0,\ldots, \rho(\s)\!-\!1 \label{62a} \\ 
& \Zint_\l, \,\,\l = \text{prime}: \quad \r(\s) = \l , \quad  \bar{\hat{j}}\! =\! 0,\dots, \l \!-\!1,  
\quad  j\!=\!0, \quad \s =1, \ldots, \l -1 \, \\  
& S_N :\,\, K=N, \quad \! f_j(\s) = \s_j, \quad \s_{j+1} \leq \s_j, \,\, j = 0, \dots, n(\vec{\s})-1, 
\quad\sum_{j=0}^{n(\vec{\s})-1} \s_j = N \label{62c} 
\end{align} 
\end{subequations} 
so that e.g. the sectors of the $S_N$ permutation orbifolds are labelled by the ordered partitions of $N$. 
 
All the eigenfields are independent of the untwisted basis, e.g. 
\begin{eqnarray} 
\sj_{n(r)aj} (z,\s) \goto \sj_{\hat{j}aj}(z,\s) &\!\!\!= \schisig_{\hat{j}aj}U(\s)_{\hat{j}j}{}^{I} 
J_{aI}(z,\s) & \!\!\! \nn \\
&\!\!\!= \schisig_{\hat{j}aj}U(\s)_{\hat{j}j}{}^{\hat{l}l} 
\tilde{J}_{a\hat{l}l}(z,\s) &\!\!\!=\! \sum_{\hat{j}^{\p}=0}^{f_j (\s)-1} e^{\frac{\tp \hat{j} \hat{j}^{\p}}{f_j 
(\s)}} \tilde{J}_{a\hat{j}^{\p}j} (z,\s) .
\end{eqnarray} 
Similarly, each duality transformation is independent of the choice of untwisted basis, and hence all of 
the OPEs of the eigenfields are likewise basis-independent. 

Using the explicit forms of $U(\s)$ and $\schisig$ in \eqref{CycleB}, we may evaluate all the duality 
transformations that appear in our development: 
\begin{subequations} 
\begin{equation} 
\A^{\hat{j}j}(\s) = \de_{\hat{j} ,0\,\text{mod } f_j(\s)}, \quad \B_{\hat{j}j}(\s) = f_j(\s) 
\de_{\hat{j} ,0\, \text{mod } f_j(\s)} 
\end{equation} 
\begin{equation} 
\sG_{\hat{j}aj;\hat{l}bl} (\s) = k \e_{ab} f_j(\s) \de_{jl} \de_{\hat{j} +\hat{l} ,0\,\text{mod } 
f_j(\s)}, \quad \stG_{\hat{j}j;\hat{l}l}(\s) = f_j(\s) \de_{jl} \de_{\hat{j} +\hat{l} ,0\,\text{mod } 
f_j(\s)} 
\end{equation} 
\begin{equation} 
\F_{\hat{j}aj;\hat{l}bl}{}^{\hat{m}cm}(\s) = f_{ab}{}^c \de_{jl} \de_l^m \de_{\hat{j} +\hat{l}-\hat{m} 
,0\,\text{mod } f_j(\s)}, \quad \stF_{\hat{j}j;\hat{l}l}{}^{\hat{m}m}(\s) = \de_{jl}\de_l^m \de_{\hat{j} 
+\hat{l}-\hat{m} ,0\,\text{mod } f_j(\s)} 
\end{equation} 
\begin{equation} 
\D_{\sgb (\s)} (\st (T,\s))_{\a \hat{j}j}{}^{\be \hat{l}l} = \de_{\a}{}^{\be} \de_j^l \de_{\hat{j} -\hat{l} 
,0\,\text{mod } f_j(\s)} \Dg 
\end{equation} 
\begin{equation} 
{\cL}_{\sgb (\s)}^{\hat{l}al; \hat{m}bm}(\s) = \Lg \frac{1}{f_j(\s)} \de^{lm} \de_{\hat{l}+\hat{m},0 
\, \text{mod } f_j(\s)} 
\end{equation} 
\begin{equation} 
{\cL}_{\hat{j}j}{}^{\hat{l}al;\hat{m}bm}(\s) = \Lg \frac{1}{f_j(\s)} \de_j^l \de^{lm} 
\de_{\hat{j}-\hat{l}-\hat{m},0\, \text{mod } f_j(\s)} 
\end{equation} 
\begin{equation} 
\label{PerEx}
\st_{\hat{j}aj}(T,\s) = T_a t_{\hat{j}j}(\s) ,\quad t_{\hat{j}j} (\s)_{\hat{l}l}{}^{\hat{m}m} \equiv 
\de_{jl} \de_l^m \de_{\hat{j} +\hat{l} - \hat{m} ,0\,\text{mod } f_j(\s)} ,\quad t_{\hat{j}\pm 
f_j(\s),j} (\s) = t_{\hat{j}j} (\s)
\end{equation} 
\begin{equation} 
[T_a ,T_b] = if_{ab}{}^c ,\quad t_{\hat{j}j} (\s) t_{\hat{l}l} (\s) = \de_{jl} t_{\hat{j}+\hat{l},j} (\s) 
,\quad [ t_{\hat{j}j} (\s), t_{\hat{l}l} (\s) ] = 0 ,\quad \forall \,j,l .
\end{equation} 
\end{subequations} 
Here $T$ is an irrep of $\gfrak$ and $\Dg$ is the conformal weight of representation $T$ under the 
affine-Sugawara construction on $\gfrak$.  All these quantities are periodic $\hat{j} \goto \hat{j} 
\pm f_j(\s)$ in any spectral index, as seen for example in \eqref{PerEx}. In what follows we often 
abbreviate $\st \equiv \st(T,\s)$.
 
\subsection{Local Isomorphisms and Twisted Operator Product Expansions} 
 
The next step is an application of the {\it principle of local isomorphisms} \cite{deBoer:1999na,Halpern:2000vj,
deBoer:2001nw} which is a map from the untwisted theory in the eigenfield basis to twisted sector $\s$ of the 
orbifold. In the present case, the local isomorphisms are written as:
\begin{subequations} 
\begin{equation} 
\sj_{\hat{j}aj} (z,\s) \dual \hj_{\hat{j}aj}(z,\s) ,\quad \sg(T,\bz,z,\s)_{\a \hat{j}j}{}^{\be 
\hat{l}l} \dual \hggz_{\a \hat{j}j}{}^{\be \hat{l}l} 
\end{equation} 
\begin{equation} 
\t_{\hat{j}j}(z,\s) \dual \hat{\t}_{\hat{j}j}(z,\s) , \quad T(z) \dual \hat{T}_\s (z)  
\end{equation} 
\begin{equation} 
a=1,\ldots ,\text{dim}\gfrak ,\,\,\, \a ,\be =1,\ldots ,\text{dim}\;T ,\,\,\, \bar{\hat{j}} =0,\ldots  
,f_j(\s)\!-\!1 ,\,\,\, \srange .
\end{equation} 
\end{subequations} 
Here the hatted fields $\hj ,\hat{g}$, and $\hat{\t}$ are called respectively the {\it twisted currents}, 
the {\it twisted affine primary fields}, and the {\it twisted partial stress tensors}. The principle of
local isomorphisms tells us first that the monodromies of the twisted fields 
\begin{subequations} 
\label{LMonos}
\begin{equation} 
\hj_{\hat{j}aj}(ze^\tp ,\s) = E_{\hat{j}}(\s) \hj_{\hat{j}aj}(z,\s) = e^{-\tp \frac{\hat{j}}{f_j(\s)}} 
\hj_{\hat{j}aj}(z,\s) 
\end{equation} 
\begin{equation} 
\hat{\t}_{\hat{j}j}(ze^\tp ,\s) = E_{\hat{j}}(\s) \hat{\t}_{\hat{j}j}(z,\s) = e^{-\tp \frac{\hat{j}} 
{f_j(\s)}} \hat{\t}_{\hat{j}j}(z,\s) 
\end{equation} 
\begin{equation} 
\hat{T}_\s (ze^\tp ) = \hat{T}_\s (z) 
\end{equation} 
\end{subequations} 
have the same form as the automorphic responses \eqref{EigAuto} of the eigenfields. The monodromies of 
the twisted affine primary fields $\hat{g}$ are determined by the twisted KZ equations. We remind the 
reader that all quantities are periodic $\hat{j} \goto \hat{j} \pm f_j(\s)$ in the spectral indices. 

The principle of local isomorphisms also tells us that the OPEs of the twisted fields are the same as the 
OPEs of the eigenfields. For example, the OPEs of the standard twisted fields
\begin{subequations} 
\label{TwCAlg}
\begin{equation} 
\label{JJtwist} 
\hj_{\hat{j}aj}(z,\s) \hj_{\hat{l}bl}(w,\s) = \de_{jl} \left( \frac{k f_j(\s) \e_{ab} \de_{\hat{j}+ 
\hat{l} ,0\,\text{mod } f_j(\s)}}{(z-w)^2} + \frac{if_{ab}{}^c \hj_{\hat{j}+\hat{l} ,cj} (w,\s)} 
{z-w} \right) + \Ord (z-w)^0 
\end{equation} 
\begin{equation} 
\hat{T}_\s (z) \hj_{\hat{j}aj}(w,\s) = \left( \frac{1}{(z-w)^2} + \frac{\partial_w}{z-w} \right) 
\hj_{\hat{j}aj}(w,\s) + \Ord (z-w)^0 
\end{equation} 
\begin{equation} 
\hat{T}_\s (z) \hat{T}_\s (w) = \frac{c_g /2}{(z-w)^4} + \left( \frac{2}{(z-w)^2} + \frac{\pl_w}{z-w} 
\right) \hat{T}_\s (w) + \Ord (z-w)^0 
\end{equation} 
\begin{equation} 
\hj_{\hat{j}aj} (z,\s) \hat{g}(\st,\bw,w,\s) = \frac{\hat{g}(\st,\bw,w,\s) T_a t_{\hat{j}j} (\s) } 
{z-w} +\Ord (z-w)^0 
\end{equation} 
\begin{equation} 
\hat{T}_\s (z) \hat{g}(\st,\bw,w,\s) = \frac{ \hat{g}(\st,\bw,w,\s) \Dg}{(z-w)^2} + \frac{\pl_w 
\hat{g}(\st,\bw,w,\s)}{z-w} + \Ord (z-w)^0 
\end{equation} 
\end{subequations} 
are given in Refs.~\cite{deBoer:2001nw,Halpern:2002ab}.

In our extension, we also obtain the OPEs involving the twisted partial stress tensors
\begin{subequations} 
\label{TOPEs}
\begin{equation} 
\label{TJtwist} 
\hat{\t}_{\hat{j} j} (z,\s) \hj_{\hat{l} al} (w,\s) = \de_{jl} \left( \frac{1}{(z-w)^2} + \frac{\pl_w} 
{z-w} \right) \hj_{\hat{j} + \hat{l}, aj} (w,\s) +\Ord (z-w)^0 
\end{equation} 
\begin{eqnarray} 
\label{TTtwist} 
\hat{\t}_{\hat{j} j} (z,\s) \hat{\t}_{\hat{l} l} (w,\s)= \de_{jl} {\big [} \frac{\de_{\hat{j} 
+\hat{l},0\, \text{mod } f_j(\s)} \cg f_j(\s) /2}{(z-w)^4} + \bigspc \bigspc \bigspc \nn \\ 
\bigspc \bigspc + \left( \frac{2}{(z-w)^2} + \frac{\pl_w}{z-w} \right) \hat{\t}_{\hat{j} +\hat{l}, j} 
(w,\s) {\big ]} + \Ord(z-w)^0 
\end{eqnarray} 
\begin{eqnarray} 
\hat{T}_\s (z) \hat{\t}_{\hat{j}j} (w,\s) =  \frac{\de_{\hat{j} ,0\, \text{mod } f_j(\s)} \cg 
f_j(\s) /2}{(z-w)^4} + \bigspc \bigspc \bigspc \quad \nn \\ 
\bigspc \bigspc + \left( \frac{2}{(z-w)^2} + \frac{\partial_w}{z-w} \right) \hat{\t}_{\hat{j}j} (w,\s) 
+\Ord (z-w)^0 
\end{eqnarray} 
\begin{eqnarray} 
\hat{\t}_{\hat{j}j}(z,\s) \hat{g}(\st,\bw,w,\s) =   \bigspc \bigspc \bigspc \bigspc \bigspc \nn \\ 
\quad \quad \left[ \frac{ \hat{g} (\st,\bw,w,\s) \Dg}{(z-w)^2} + \frac{\pl_w \hat{g} (\st,\bw,w,\s)} 
{z-w} \right] t_{\hat{j}j} (\s) + \Ord (z-w)^0 .
\end{eqnarray} 
\end{subequations} 
In terms of the twisted currents, the explicit forms of the stress tensor and the twisted partial 
stress tensors are
\begin{subequations} 
\begin{equation}
\label{TEqJJ} 
\hat{T}_\s (z) = \Lg \sum_j \frac{1}{f_j(\s)} \sum_{\hat{j}=0}^{f_j(\s)-1} : \hj_{\hat{j}aj} (z,\s) 
\hj_{-\hat{j},bj} (z,\s) : 
\end{equation} 
\begin{equation} 
\hat{\t}_{\hat{j}j} (z,\s) = \Lg \frac{1}{f_j (\s)} \sum_{\hat{l} =0}^{f_j(\s)-1} : \hj_{\hat{l} 
aj} (z,\s) \hj_{\hat{j}-\hat{l} ,bj} (z,\s) : 
\end{equation} 
\begin{equation} 
\label{TeqPlT} 
\hat{T}_\s (z) = \sum_j \hat{\t}_{0j} (z,\s) 
\end{equation} 
\end{subequations} 
where \eqref{TEqJJ} was first given in Ref.~\cite{Halpern:2002ab}. The normal ordering here is operator product normal 
ordering \cite{Evslin:1999qb,deBoer:1999na,Halpern:2000vj} of the twisted currents.

Note in particular the $\de_{jl}$ factor in \eqref{JJtwist}, \eqref{TJtwist}, and \eqref{TTtwist}. Taken 
together, these relations tell us that although each sector $\s$ may contain many cycles $j$, the dynamics 
of each cycle is independent.
 
As a final local result in the permutation orbifolds, we give the twisted left-mover vertex operator 
equation \cite{deBoer:2001nw,Halpern:2002ab} 
\begin{eqnarray} 
\label{LVOE} 
\pl \hggz = \frac{2 \eta^{ab}}{2k+ Q_{\gfraks}} \sum_j \fji \sum_{\hat{j}=0}^{f_j (\s)-1} : \hat{J}_{\hat{j}aj} 
(z) \hggz :_M T_b t_{-\hat{j},j} (\s) \quad \nn \\ 
\bigspc \bigspc -\frac{\Dg}{z} \sum_j (1-\fji) \hggz t_{0j} (\s) 
\end{eqnarray} 
where this mode normal ordering is defined in Ref.~\cite{deBoer:2001nw}.

\subsection{Extended Operator Algebra and the Orbifold Virasoro Subalgebra} 

Using the monodromies \eqref{LMonos}, we may now expand the twisted fields in terms of modes
\begin{subequations} 
\begin{equation} 
\hj_{\hat{j}aj} (z,\s) = \sum_{m \in \Zint} \hj_{\hat{j}aj} \modmj  z^{-\modmj -1}  
\end{equation} 
\begin{equation} 
\hj_{\hat{j}\pm f_j(\s),aj} (m +\frac{\hat{j}\pm f_j(\s)}{f_j(\s)}) = \hj_{\hat{j}aj} (m\pm 1 +\jfj ) 
\end{equation} 
\begin{equation} 
\label{TModes}
\hat{\t}_{\hat{j}j} (z,\s) = \sum_{m \in \Zint} \hat{L}_{\hat{j}j} \modmj z^{-\modmj -2}  
\end{equation} 
\begin{equation} 
\hat{L}_{\hat{j}\pm f_j(\s),j} (m+ \frac{\hat{j}\pm f_j(\s)}{f_j(\s)}) = \hat{L}_{\hat{j}j} (m\pm 1 +\jfj ) 
\end{equation} 
\begin{equation} 
\label{LsL0j} 
\hat{T}_\s (z) = \sum_{m \in \Zint} L_\s (m) z^{-m-2}, \quad L_\s (m) = \sum_j \hat{L}_{0j}(m) 
\end{equation} 
\end{subequations} 
and then the algebra of these twisted modes follows from the twisted OPE system \eqref{TwCAlg}, \eqref{TOPEs}.
The algebra of the twisted current modes and the Virasoro generators $L_\s (m)$ 
\begin{subequations} 
\begin{eqnarray} 
\label{JJAlg} 
\left[ \hj_{\hat{j}aj} \modmj , \hj_{\hat{l}bl} \modnl \right] = \bigspc \bigspc \bigspc \bigspc 
\bigspc \nn \\ 
= \de_{jl} \left( if_{ab}{}^c \hj_{\hat{j} +\hat{l}, cj} \modmn + \e_{ab}kf_j(\s) \modmj \de_{m+n 
+\frac{\hat{j}+\hat{l}}{f_j(\s)} ,0} \right) \quad 
\end{eqnarray} 
\begin{equation} 
\left[ L_\s (m), \hj_{\hat{j}aj} (n+ \frac{\hat{j}}{f_j(\s)}) \right] = - (n+ \frac{\hat{j}}{f_j(\s)}) 
\hj_{\hat{j}aj} (m+n+ \frac{\hat{j}}{f_j(\s)}) 
\end{equation} 
\begin{equation} 
\left[ L_\s (m), L_\s (n) \right] = (m-n) L_\s (m+n) + \frac{c_g}{12} m (m^2 -1) \de_{m+n ,0} 
\end{equation} 
\begin{equation} 
\label{JGAlg} 
\left[ \hj_{\hat{j}aj} \modmj, \hat{g} (\st (T,\s),\bz,z,\s) \right] = \hat{g} (\st (T,\s),\bz,z,\s) T_a 
t_{\hat{j}j} (\s) z^{m + \jfj} 
\end{equation} 
\begin{equation} 
\left[ L_\s (m), \hat{g} (\st (T,\s),\bz,z,\s) \right] = \hat{g} (\st (T,\s),\bz,z,\s) (\lpl_z z + (m+1) \Dg ) 
\end{equation} 
\end{subequations} 
was given in Refs.~\cite{deBoer:2001nw,Halpern:2002ab}. Here \eqref{JJAlg} is the {\it general orbifold affine algebra} 
[1-3, 5, 12, 13]. Note that the simple components $\hj_{\hat{j}aj}$ of the orbifold affine algebra act entirely within 
disjoint cycle $j$. The adjoint of these operators in a Cartesian basis for $\gfrak$
\begin{equation}
\label{JAdj}
\hj_{\hat{j}aj} (m+\jfj)^\hcj = \hj_{-\hat{j},aj} (-m -\jfj) ,\quad L_\s (m)^\hcj = L_\s (-m)
\end{equation}
was given in Refs.~\cite{Borisov:1997nc, Halpern:2000vj,deBoer:2001nw}.

In our extension, we also obtain the algebra of the twisted Virasoro modes $\hat{L}_{\hat{j}j}$,
\begin{subequations} 
\begin{equation} 
\left[ \hat{L}_{\hat{j}j} \modmj , \hj_{\hat{l}al} \modnl \right] = -\de_{jl} \modnl \hj_{\hat{j}+ 
\hat{l}, aj} \modmn 
\end{equation} 
\begin{eqnarray} 
\label{Lmodes} 
\left[ \hat{L}_{\hat{j}j} \modmj , \hat{L}_{\hat{l}l} \modnl \right]  = \de_{jl} \{ (m-n + 
\frac{\hat{j}-\hat{l}}{f_j(\s)}) \hat{L}_{\hat{j}+\hat{l},j} \modmn + \quad \nn \\ 
+ \frac{\cg f_j(\s)}{12} \modmj ((m+\jfj )^2 -1) \de_{m+n+ \frac{\hat{j}+\hat{l}}{f_j (\s)} ,0} \} 
\quad \quad 
\end{eqnarray} 
\begin{eqnarray} 
\left[ L_\s (m), \hat{L}_{\hat{j}j} (n + \frac{\hat{j}}{f_j(\s)}) \right] & = & (m-n - \jfj) 
\hat{L}_{\hat{j}j} (m +n +\frac{\hat{j}}{f_j(\s)}) + \nn \\ 
& + & f_j(\s) \frac{\cg}{12} m (m^2-1) \de_{m+n+\jfj ,0} 
\end{eqnarray} 
\begin{align} 
& \left[ \hat{L}_{\hat{j}j} \modmj , \hat{g} (\st (T,\s),\bz,z,\s) \right] = \nn \\ 
& \quad \quad = \hat{g} (\st (T,\s),\bz,z,\s) \left( \lpl_z z + (\modmj +1) \Dg \right) t_{\hat{j}j}(\s) z^{m+ \jfj} 
\end{align} 
\begin{equation}
a=1,\ldots \text{dim} \gfrak ,\quad \bar{\hat{j}}=0,\ldots ,f_j(\s)-1 ,\quad \srange
\end{equation}
\end{subequations} 
where \eqref{Lmodes} is the {\it general orbifold Virasoro algebra}. The orbifold Virasoro algebra 
for the special case $H= \Zint_\l ,\l =$ prime
\begin{subequations}
\label{ChrLL}
\begin{equation}
\hat{L}^{(r)} (m+\frac{r}{\l}) \equiv \hat{L}_{\hat{j},j=0} (m+\frac{\hat{j}}{\l}) |_{\hat{j}=r}
,\quad \bar{r}=0,\ldots ,\l -1,\,\,\, \s=1,\ldots \l-1
\end{equation}
\begin{eqnarray}
[\hat{L}^{(r)} (m+\frac{r}{\l}) ,\hat{L}^{(s)} (n+\frac{s}{\l}) ]&=&(m-n +\frac{r-s}{\l}) \hat{L}^{(r+s)}
(m+n+\frac{r+s}{\l})+ \nn \\
&+& \frac{\l c}{12} (m+\frac{r}{l}) ((m+\frac{r}{\l})^2 -1) \de_{m+n+\frac{r+s}{\l} ,0}
\end{eqnarray}
\end{subequations}
was given earlier in Ref.~\cite{Borisov:1997nc} and independently in Ref.~\cite{Dijk:1997dv}. Of course, we obtain in 
this paper only the special case of \eqref{ChrLL} with the affine-Sugawara central charge $c=\cg$. By the same token, 
the orbifold Virasoro algebra \eqref{Lmodes} with $\cg \rightarrow c$ will also occur in more general copy permutation 
orbifolds, where $c$ is the central charge of each copy of a general affine-Virasoro construction [9-11] on affine $\gfrak$.

We comment briefly on subalgebras of the orbifold Virasoro algebra. We note first the {\it 
semisimple integral Virasoro subalgebra}
\begin{equation} 
\label{ssIVsA}
\hat{j} = 0: \quad [\hat{L}_{0j} (m) , \hat{L}_{0l} (n)] = \de_{jl} \{ (m-n) \hat{L}_{0j} (m+n) + 
\frac{f_j(\s) \cg}{12} m (m^2-1) \de_{m+n,0} \} 
\end{equation} 
whose generators are in fact the Virasoro operators of each cycle separately (see \eqref{LsL0j}).
The orbifold Virasoro algebra also contains a closed subalgebra whose generators are 
\begin{equation} 
\label{SL2gen} 
\hat{L}_{0,j} (0),\,\,\hat{L}_{1,j} (\frac{1}{f_j (\s)}) ,\,\,\hat{L}_{f_j (\s)-1, j} (-1 + \frac{f_j 
(\s)-1}{f_j (\s)}) = \hat{L}_{-1,j} (0 + \frac{-1}{f_j (\s)}) ,\quad \forall j 
\end{equation} 
and one finds that this subalgebra is a set of mutually-commuting {\it centrally-extended} $\sl (2)$
algebras:
\begin{subequations} 
\label{ceSL2} 
\begin{equation} 
\left[ \hat{L}_{1,j} (\frac{1}{f_j (\s)}) ,\hat{L}_{f_l (\s)-1, l} (-1 + \frac{f_l (\s)-1}{f_l (\s)}) 
\right] = \de_{jl} \left( \frac{2}{f_j (\s)} \hat{L}_{0,j} (0) + \frac{\cg}{12} (\frac{1}{f_j (\s)^2} 
-1) \right) 
\end{equation} 
\begin{equation} 
\left[ \hat{L}_{0,j} (0) ,\hat{L}_{1,l} (\frac{1}{f_l (\s)}) \right] = - \de_{jl} \frac{1}{f_j (\s)} 
\hat{L}_{1,j} (\frac{1}{f_j (\s)}) 
\end{equation} 
\begin{equation} 
\left[ \hat{L}_{0,j} (0) ,\hat{L}_{f_l (\s)-1, l} (-1 + \frac{f_l (\s)-1}{f_l (\s)}) \right] = 
\de_{jl} \frac{1}{f_j (\s)} \hat{L}_{f_j (\s)-1, j} (-1 + \frac{f_j (\s)-1}{f_j (\s)}) .
\end{equation} 
\end{subequations} 
Relative to Ref.~\cite{Borisov:1997nc}, we see that each of these $\sl (2)$'s is associated to a fixed cycle $j$. As
we shall see in the following subsections, there are $\sl (2)$ Ward identities \cite{Borisov:1997nc} associated to
this subalgebra -- but not to the standard $\sl (2)$ subalgebra generated by $\{ L_\s (|m| \leq 1) \}$.

\subsection{The Scalar Twist-Field States} 
 
For the orbifold affine algebras, it is known [1-3], that the {\it scalar twist-field state} $\tau_\s (0) |0\rangle 
= |0\rangle_\s$ of sector $\s$ satisfies 
\begin{subequations}
\begin{equation} 
\label{GSCond} 
\hj_{\hat{j}aj} (m+\jfj \geq 0) |0 \rangle_\s = {}_\s \langle 0| \hj_{\hat{j}aj} (m+\jfj \leq 0) =0
\end{equation} 
\begin{equation}
a= 1,\ldots ,\text{dim}\gfrak ,\quad \bar{\hat{j}} =0,\ldots ,f_j (\s)-1 ,\quad \srange
\end{equation}
\end{subequations}
and the scalar twist-field state is in fact the {\it ground state} of sector $\s$ under the Virasoro generator $L_\s (0)$. 
Then a more useful form of the Virasoro generators is the mode-ordered form \cite{Halpern:2000vj,deBoer:2001nw,Halpern:2002ab} 
\begin{subequations} 
\begin{align} 
\label{jmodes} 
& : \hj_{\hat{j}aj} \modmj \hj_{\hat{l}bl} \modnl :_M \equiv \theta (m +\jfj \geq 0) \hj_{\hat{l}bl} 
\modnl \hj_{\hat{j}aj} \modmj + \nn \\ 
& \bigspc \bigspc + \theta (m +\jfj < 0) \hj_{\hat{j}aj} \modmj \hj_{\hat{l}bl} \modnl  
\end{align} 
\begin{align} 
\label{LJJ}
& L_\s (m) = \left( \Lg \sum_j \frac{1}{f_j(\s)} \sum_{\hat{j}=0}^{f_j(\s)-1} \sum_{p \in \Zint} : 
\hj_{\hat{j}aj} (p + \jfj) \hj_{-\hat{j},bj} (m-p -\jfj) :_M \right) +  \nn \\ 
&  \bigspc \bigspc+ \gscfwt \de_{m,0} 
\end{align} 
\begin{equation} 
L_\s (m \geq 0) | 0 \rangle_\s = \de_{m,0} \gscfwt |0 \rangle_\s ,\quad {}_\s \langle 0| 
L_\s (m \leq 0) = {}_\s \langle 0| \gscfwt \de_{m,0} 
\end{equation} 
\begin{equation} 
\label{GSCnf} 
\gscfwt \equiv \frac{\cg}{24} \sum_j (f_j(\s) - \frac{1}{f_j(\s)}) = \frac{\cg}{24} (K - \sum_j 
\frac{1}{f_j (\s)}) 
\end{equation} 
\end{subequations} 
where $\gscfwt$ is the conformal weight of the scalar twist-field state of sector $\s$. Similarly, we
find for the generators of the orbifold Virasoro algebra
\begin{subequations} 
\begin{align} 
\label{oVirAlg} 
& \hat{L}_{\hat{j}j} \modmj =\!\left( \Lg \frac{1}{f_j (\s)} \sum_{\hat{l}=0}^{f_j (\s)-1} \sum_{p \in \Zint} : 
\hj_{\hat{l} aj} (p +\frac{\hat{l}}{f_j (\s)})  \hj_{\hat{j}-\hat{l} ,bj} (m-p + \frac{\hat{j}- 
\hat{l}} {f_j (\s)}) :_M \right) \nn \\ 
& \bigspc \bigspc \bigspc + \gscfwtj \de_{m+\jfj ,0}
\end{align} 
\begin{eqnarray} 
\label{GSEqn} 
\hat{L}_{\hat{j}j} (m +\jfj \geq -\frac{1}{f_j (\s)}) |0 \rangle_\s = \de_{m +\jfj ,0} \gscfwtj |0 
\rangle_\s \\ 
{}_\s \langle 0| \hat{L}_{\hat{j}j} (m +\jfj \leq \frac{1}{f_j (\s)}) = {}_\s \langle 0| \gscfwtj 
\de_{m+ \jfj ,0} 
\end{eqnarray} 
\begin{equation} 
\label{GSCnfJ} 
\gscfwtj \equiv \frac{\cg}{24} (f_j (\s) -\frac{1}{f_j (\s)}) ,\quad \gscfwt = \sum_j \gscfwtj 
\end{equation} 
\end{subequations} 
where $\gscfwtj$ are the {\it partial conformal weights} of sector $\s$. In further detail, \eqref{GSCond} and 
\eqref{oVirAlg} were used to establish ($2.39$b,c), which tell us that the scalar twist-field state is primary 
under the orbifold Virasoro algebra. The adjoint \cite{Borisov:1997nc, Halpern:2000vj} of the twisted Virasoro operators 
\begin{equation}
\hat{L}_{\hat{j}j} (m+\jfj)^\hcj = \hat{L}_{-\hat{j},j} (-m-\jfj)
\end{equation}
follows from Eqs.\eqref{JAdj} and \eqref{oVirAlg}.
 
\subsection{The Orbifold $\sl (2)$ Ward Identities and the Twisted KZ System} 
 
For the generators \eqref{SL2gen} of the centrally-extended $\sl (2)$, the relations \eqref{GSEqn} reduce to 
\begin{subequations} 
\begin{equation} 
\hat{L}_{1,j} (\frac{1}{f_j (\s)}) |0 \rangle_\s = {}_\s \langle 0| \hat{L}_{1,j} (\frac{1}{f_j (\s)}) = 0 
\end{equation} 
\begin{equation} 
\hat{L}_{f_j(\s) -1,j} (-1+ \frac{f_j (\s)-1}{f_j (\s)}) |0 \rangle_\s = {}_\s \langle 0| 
\hat{L}_{f_j(\s) -1,j} (-1+ \frac{f_j (\s)-1}{f_j (\s)}) = 0 
\end{equation} 
\begin{equation} 
\hat{L}_{0,j} (0) |0 \rangle_\s = \gscfwtj |0 \rangle_\s ,\quad {}_\s \langle 0| \hat{L}_{0,j} (0) = 
{}_\s \langle 0| \gscfwtj 
\end{equation} 
\end{subequations} 
where the partial conformal weight $\gscfwtj$ is given in \eqref{GSCnfJ}. Although the scalar twist-field state 
is {\it not invariant} under this $\sl (2)$, nevertheless \cite{Borisov:1997nc} we find the following identities 
\begin{eqnarray} 
\label{Acorr} 
\hat{A} (\st,\bz,z,\s) &\equiv & {}_\s \langle 0| \hat{g} (\st^{(1)},\bz_1 ,z_1 ,\s) 
\ldots \hat{g} (\st^{(N)} ,\bz_N ,z_N ,\s) |0 \rangle_\s \nn \\ 
&\equiv & \langle \hat{g} (\st^{(1)},\bz_1 ,z_1 ,\s) \ldots \hat{g} (\st^{(N)} ,\bz_N ,z_N ,\s) \rangle_\s 
\end{eqnarray} 
\begin{subequations} 
\begin{equation} 
\langle [\hat{L}_{0,j} (0), \hat{g}(\st^{(1)},\bz_1 ,z_1 ,\s) \ldots \hat{g}(\st^{(N)} ,\bz_N ,z_N ,\s)] 
\rangle_\s = 0 
\end{equation} 
\begin{equation} 
\langle [\Lpone , \hat{g} (\st^{(1)},\bz_1 ,z_1 ,\s) \ldots \hat{g} (\st^{(N)} ,\bz_N ,z_N ,\s)] \rangle_\s = 0 
\end{equation} 
\begin{equation} 
\langle [\Lmone , \hat{g} (\st^{(1)},\bz_1 ,z_1 ,\s) \ldots \hat{g} (\st^{(N)} ,\bz_N 
,z_N ,\s)] \rangle_\s = 0 
\end{equation} 
\end{subequations} 
for the correlators $\hat{A}$ in the scalar twist-field state of sector $\s$. These give the {\it orbifold} $\sl(2)$
{\it Ward identities}
\begin{subequations} 
\label{NewSLWI} 
\begin{equation} 
\label{NewWI0} 
\hat{A} (\st,\bz,z,\s) \sum_{\m =1}^{N} \left( \lplm z_\m + \Delta_{\gfraks} (T^{(\m )}) \right) 
t_{0j}^{(\m )} (\s) = 0 ,\quad \forall j 
\end{equation} 
\begin{equation} 
\label{NewWIP} 
\hat{A} (\st,\bz,z,\s) \sum_{\m =1}^{N} \left[ \left( \lplm z_\m + (1 +\frac{1}{f_j (\s)}) \Delta_{\gfraks} 
(T^{(\m )}) \right) t_{1j}^{(\m )} (\s) z_{\m}^{\frac{1}{f_j (\s)}} \right] =0, \,\,\, \forall j 
\end{equation} 
\begin{equation} 
\label{NewWIM} 
\hat{A} (\st,\bz,z,\s) \sum_{\m =1}^{N} \left[ \left( \lplm z_\m + (1 -\frac{1}{f_j (\s)}) \Delta_{\gfraks} 
(T^{(\m )}) \right) t_{f_j (\s)-1,j}^{(\m )}(\s) z_{\m}^{-\frac{1}{f_j (\s)}} \right] =0, \,\,\, \forall j 
\end{equation} 
\end{subequations} 
which are the analogues of the orbifold $\sl (2)$ Ward identities given for $H=\Zint_\l , \l=$ prime in 
Ref.~\cite{Borisov:1997nc}. The Ward identity in \eqref{NewWI0} implies the known $L_\s (0)$ Ward identity \cite{deBoer:2001nw} 
\begin{subequations} 
\label{KnownWI} 
\begin{equation} 
\langle [L_\s (0), \hat{g} (\st^{(1)},\bz_1 ,z_1 ,\s) \ldots \hat{g}(\st^{(N)},\bz_N ,z_N ,\s)] 
\rangle_\s = 0 
\end{equation} 
\begin{equation} 
\hat{A} (\st,\bz,z,\s) \sum_{\mu =1}^{N} \left( \lplm z_\m + \Delta_{\gfraks} (T^{(\mu )}) \right) =0 
\end{equation} 
\end{subequations} 
as a consequence of Eq. \eqref{LsL0j} and the relation $\sum_j t_{0j} (\s) = \thickone$ . 
 
It is known \cite{deBoer:2001nw} that the $L_\s (0)$ Ward identity \eqref{KnownWI} follows from the twisted left-mover 
KZ system \cite{deBoer:2001nw,Halpern:2002ab} 
\begin{subequations} 
\label{twistKZ} 
\begin{equation} 
\pl_{\m} \hat{A} (\st,\bz,z,\s) = \hat{A} (\st,\bz,z,\s) \hat{W}_{\m} (\st,z,\s) ,\quad \hat{W}_\m (\st,z,\s) = 
\sum_j \hat{W}^j_\m (\st,z,\s)  
\end{equation} 
\begin{eqnarray} 
\label{twKZconn} 
\hat{W}^j_\m (\st,z,\s) \equiv \frac{2}{2k +Q_{\gfraks}} \sum_{\n \neq \m} \frac{\eta^{ab} T_a^{(\n)} 
T_b^{(\m)}}{f_j (\s) z_{\m \n}} \sum_{\hat{j}=0}^{f_j(\s) -1} \left( \frac{z_\n}{z_\m} \right)^{\frac{\hat{j}}
{f_j(\s)}} t_{\hat{j}j}^{(\n)}(\s) t_{-\hat{j} ,j}^{(\m)}(\s) \quad \quad \nn \\ 
- \frac{\dg (T^{(\m)})}{z_\m} (1- \frac{1}{f_j(\s)}) t_{0j}^{(\m)}(\s) \bigspc \bigspc 
\end{eqnarray} 
\begin{equation} 
\label{GWI1} 
\hat{A} (\st,\bz,z,\s) \left( \sum_{\m =1}^N T_a^{(\m)} t_{0j}^{(\m)} (\s) \right) =0 ,\quad \forall j ,\,\, 
a=1,\ldots, \text{dim}\gfrak ,\,\, \srange .
\end{equation} 
\end{subequations} 
In this notation, all matrix products $M^{(\n)} N^{(\m)} \equiv M^{(\n)} \otimes N^{(\m)} ,\,\, \n \neq \m$ are tensor 
products. The form of the twisted KZ connection in ($2.46$a,b) is easily obtained from the original form given in 
Refs.~\cite{deBoer:2001nw,Halpern:2002ab} by the relation $\sum_j t_{0j} (\s) = \thickone$. The twisted partial connections 
$\hat{W}^j_\m$ defined above are abelian flat: 
\begin{subequations} 
\begin{equation} 
\pl_\m \hat{W}^j_\n - \pl_\n \hat{W}^j_\m = 0 
\end{equation} 
\begin{equation} 
\left[ \hat{W}^j_\m (\s), \hat{W}^{j\p}_\n (\s) \right] = 0 ,\quad \text{when } j \neq j\p 
\end{equation} 
\begin{equation}  
\label{248c} 
\left[ \hat{W}^j_\m (\s),\hat{W}^j_\n (\s) \right] =0 ,\quad \forall \,j, \m ,\n . 
\end{equation} 
\end{subequations} 
Here Eqs.~(2.47a,b) are relatively easy to check, and the more difficult proof of \eqref{248c} 
is essentially identical to the proof given for $H= \Zint_{\l}$ in Ref.~\cite{deBoer:2001nw}. We remind the reader
that the $z_\m^{-1}$ term in \eqref{twKZconn} can be traced back to the fact that the twist-field states 
$|0\rangle_\s = \tau_\s (0) |0\rangle ,\,\, L_\s (-1) |0\rangle_\s \neq 0$ are not translation invariant.
 
Moreover, one finds after some algebra (see App.~A) that the entire system \eqref{NewSLWI} 
of orbifold $\sl (2)$ Ward identities also follows from the twisted KZ system \eqref{twistKZ}. This
phenomenon is familiar from the untwisted KZ system \cite{KZ:1984kz}, but we emphasize that it is an unsolved 
problem to find the general solution of the orbifold $\sl (2)$ Ward identities.
 
\subsection{Right-Movers and Rectification} 
 
As current-algebraic orbifold theory [1-5, 12, 13, 17] is presently formulated, the action of the 
automorphism group on the right-mover fields 
\begin{subequations} 
\begin{equation} 
\bar{J}_{aI} (\bz)\p = \ws_{I}{}^{J} \bar{J}_{aJ} (\bz) ,\quad 
\bar{T}_{I} (\bz)\p = \ws_{I}{}^{J} \bar{T}_{J} (\bz) 
\end{equation} 
\begin{equation} 
\bar{T} (\bz)\p = \bar{T} (\bz)  
\end{equation} 
\end{subequations} 
is taken to be the same as the action on the left-movers. This defines what can be called the class of 
current-algebraic {\it symmetric} orbifolds, as contrasted with the class of current-algebraic asymmetric 
orbifolds which will not be studied here. 
 
Because the orbifold is symmetric, the right-mover eigenfields have the same forms as the left-movers, e.g. 
\begin{equation} 
\bar{\t}_{n(r)j} (z,\s) = \schisig_{n(r)j} U(\s)_{n(r)j}{}^I \bar{T}_I (z) 
\end{equation} 
and the principle of local isomorphisms tells us that the twisted right-mover OPE system is a copy of the 
twisted left-mover system above. Moreover, the principle of local isomorphisms gives the monodromies of the 
twisted fields: 
\begin{subequations} 
\label{RMonos} 
\begin{equation} 
\hjb_{\hat{j}aj} (\bz e^{-\tp} ,\s) = E_{\hat{j}} (\s) \hjb_{\hat{j}aj} (\bz ,\s) = e^{-\frac{\tp 
\hat{j}}{f_j (\s)}} \hjb_{\hat{j}aj} (\bz ,\s) 
\end{equation} 
\begin{equation} 
\hat{\bar{\t}}_{\hat{j}j} (\bz e^{-\tp},\s) = e^{-\frac{\tp \hat{j}}{f_j (\s)}} \hat{\bar{\t}}_{\hat{j} 
j} (\bz ,\s) 
\end{equation} 
\begin{equation} 
\hat{\bar{T}} (\bz e^{-\tp},\s) = \hat{\bar{T}} (\bz ,\s) .
\end{equation} 
\end{subequations} 
The rule here \cite{deBoer:2001nw} is that the monodromies of the right-movers are the same as those of the left-movers when 
the same path is followed. The monodromies \eqref{RMonos} give the mode expansions 
\begin{subequations} 
\begin{equation} 
\hjb_{\hat{j}aj} (\bz,\s) = \sum_{m \in \Zint} \hjb_{\hat{j}aj} \modmj \bz^{\modmj -1} ,\quad 
\hat{\bar{T}}_\s (\bz) = \sum_{m \in \Zint} \bar{L}_\s (m) \bz^{-m-2} 
\end{equation} 
\begin{equation} 
\hat{\bar{\t}}_{\hat{j}j} (\bz,\s) = \sum_{m \in \Zint} \hat{\bar{L}}_{\hat{j}j} \modmj \bz^{\modmj-2} 
\end{equation} 
\end{subequations} 
and, together with the twisted right-mover OPE system, one obtains the right-mover chiral algebra 
of each sector of the orbifold. 

As discussed on the sphere and the torus in Refs.~\cite{deBoer:2001nw,Halpern:2002ab}, the twisted right-mover current 
algebra shows a {\it sign reversal of the central terms}, and in fact we find the same phenomenon in the right-mover 
orbifold Virasoro algebra. 
 
In the rectification problem \cite{deBoer:2001nw,Halpern:2002ab} one asks whether the twisted right-mover current algebra
of sector $\s$ is equivalent to a copy of the twisted left-mover current algebra of that sector. For the WZW permutation
orbifolds in particular, this rectification has been demonstrated \cite{deBoer:2001nw} and we have checked that a similar 
rectification is possible for the right-mover orbifold Virasoro algebra. The rectified right-mover operators $\hjbb ,\hlbb$ 
are 
\begin{subequations} 
\begin{equation} 
\label{jjrect} 
\hjbb_{\hat{j}aj} \modmj \equiv \hjb_{-\hat{j},aj} (-m -\frac{\hat{j}}{f_j (\s)}) ,\quad 
\hlbb_{\hat{j}j} \modmj \equiv \hat{\bar{L}}_{-\hat{j},j} (-m -\frac{\hat{j}}{f_j (\s)}) 
\end{equation} 
\begin{equation} 
\label{LsbL0j} 
\bar{L}_\s (m) = \sum_j \hat{\bar{L}}_{0,j} (-m) = \sum_j \hlbb_{0,j} (m) 
\end{equation} 
\end{subequations} 
\begin{subequations} 
\begin{eqnarray} 
\label{jjcomm} 
\left[ \hjbb_{\hat{j}aj} \modmj , \hjbb_{\hat{l}bl} \modnl \right] = \bigspc \bigspc \bigspc \bigspc 
\bigspc \nn \\ 
= \de_{jl} \left( if_{ab}{}^c \hjbb_{\hat{j} +\hat{l}, cj} \modmn + \e_{ab}kf_j(\s) \modmj \de_{m+n 
+\frac{\hat{j}+\hat{l}}{f_j(\s)} ,0} \right) \quad 
\end{eqnarray} 
\begin{equation} 
\left[ \hlbb_{\hat{j}j} \modmj , \hjbb_{\hat{l}al} \modnl \right] = -\de_{jl} \modnl \hjbb_{\hat{j}+ 
\hat{l}, aj} \modmn 
\end{equation} 
\begin{eqnarray} 
\left[ \hlbb_{\hat{j}j} \modmj , \hlbb_{\hat{l}l} \modnl \right]  = \de_{jl} \{ (m-n + 
\frac{\hat{j}-\hat{l}}{f_j(\s)}) \hlbb_{\hat{j}+\hat{l},j} \modmn + \quad \nn \\ 
+ \frac{\cg f_j(\s)}{12} \modmj ((m+\jfj )^2 -1) \de_{m+n+ \frac{\hat{j}+\hat{l}}{f_j (\s)} ,0} \} 
\quad \quad 
\end{eqnarray} 
\end{subequations} 
and the right-mover algebra commutes with the left-mover algebra. For completeness, we also give the form 
of the rectified orbifold Virasoro generators in terms of the rectified right-mover modes $\hjbb$ 
\begin{subequations} 
\begin{align} 
& \bar{L}_\s (m) = \left( \Lg \sum_j \frac{1}{f_j(\s)} \sum_{\hat{j}=0}^{f_j(\s)-1} \sum_{p \in \Zint} : 
\hjbb_{\hat{j}aj} (p + \jfj) \hjbb_{-\hat{j},bj} (m-p -\jfj) :_M \right) \nn \\ 
& \bigspc \bigspc + \gscfwt \de_{m,0} 
\end{align} 
\begin{align} 
& \hlbb_{\hat{j}j} \modmj = \left( \Lg \frac{1}{f_j (\s)} \sum_{\hat{l}=0}^{f_j(\s)-1} \sum_{p \in \Zint} : 
\hjbb_{\hat{l} aj} (p +\frac{\hat{l}}{f_j (\s)})  \hjbb_{\hat{j}-\hat{l} ,bj} (m-p + 
\frac{\hat{j}- \hat{l}} {f_j (\s)}) :_M \right) \nn \\ 
& \bigspc \bigspc \bigspc + \gscfwtj \de_{m+\jfj ,0} 
\end{align} 
\end{subequations} 
where the $M$ normal ordering used here for the modes of $\hjbb$ is the same as that given in 
\eqref{jmodes} for the modes of $\hj$. These forms are nothing but right-mover copies of the corresponding
left-mover results in Eqs.~\eqref{LJJ} and \eqref{oVirAlg}.

On the other hand, the algebra of the rectified right-mover modes is not exactly the same as that of the 
left-mover modes. For example, the relations 
\begin{subequations} 
\begin{equation} 
\label{RJGAlg} 
\left[ \hjbb_{\hat{j}aj} \modmj ,\hat{g} (\st(T,\s) ,\bz,z,\s) \right] = -\bz^\modmj T_a t_{-\hat{j},j} (\s)\hat{g} 
(\st(T,\s) ,\bz,z,\s)
\end{equation} 
\begin{align} 
&\left[ \hlbb_{\hat{j}j} \modmj ,\hat{g} (\st(T,\s) ,\bz,z,\s) \right] = \nn \\
&\bigspc \quad \quad \bz^\modmj (\bz \bar{\pl} + (\modmj+1) \Dg ) t_{-\hat{j},j} (\s)\hat{g} (\st(T,\s) ,\bz,z,\s)
\end{align} 
\begin{equation}
[ \bar{L}_\s (m) ,\hat{g} (\st(T,\s) ,\bz,z,\s)] = \bz^m (\bz \bar{\pl} +(m+1)\Dg ) \hat{g} (\st (T,\s),\bz,z,\s)
\end{equation}
\end{subequations} 
are found for the commutators with the twisted affine primary fields. 
 
In terms of the rectified right-mover current modes, the right-mover ground state conditions are copies of 
those for the left-movers:
\begin{subequations} 
\label{RGSCond} 
\begin{equation} 
\hjbb_{\hat{j}aj} (m + \jfj \geq 0) |0 \rangle_\s = {}_\s \langle 0| \hjbb_{\hat{j}aj} (m +\jfj \leq 0) =0
\end{equation} 
\begin{eqnarray} 
\hlbb_{\hat{j}j} (m +\jfj \geq -\fji) |0 \rangle_\s = \de_{m+\jfj ,0} \gscfwtj |0\rangle_\s \nn \\ 
{}_\s \langle 0| \hlbb_{\hat{j}j} (m+\jfj \leq \fji) = {}_\s \langle 0| \gscfwtj \de_{m+\jfj ,0} . 
\end{eqnarray} 
\end{subequations} 
This leads immediately to the right-mover orbifold $\sl (2)$ Ward identities
\begin{subequations} 
\label{rWIs} 
\begin{equation} 
\label{rWI0}
\sum_{\m =1}^N \left( \bz_\m \bar{\pl}_\m + \Delta_{\gfraks} (T^{(\mu)}) \right) t_{0j}^{(\m )} (\s) \hat{A} 
(\st,\bz,z,\s) =0 ,\quad \forall j  
\end{equation} 
\begin{equation} 
\sum_{\m =1}^N \left[ \bz_{\m}^\fji \left( \bz_\m \bar{\pl}_\m + (1 +\fji) \Delta_{\gfraks} (T^{(\mu)}) \right)
t_{-1,j}^{(\m )} (\s) \right] \hat{A} (\st,\bz,z,\s) =0 ,\quad \forall j 
\end{equation} 
\begin{equation} 
\sum_{\m =1}^N \left[ \bz_{\m}^{-\fji} \left( \bz_\m \bar{\pl}_\m + (1 -\fji) \Delta_{\gfraks} (T^{(\mu)}) 
\right) t_{1j}^{(\m )} (\s) \right] \hat{A} (\st,\bz,z,\s) =0 ,\quad \forall j 
\end{equation} 
\begin{equation} 
\label{rWIold} 
\sum_{\m =1}^N \left( \bz_\m \bar{\pl}_\m + \Delta_{\gfraks} (T^{(\mu)}) \right) \hat{A} (\st,\bz,z,\s) =0  
\end{equation} 
\end{subequations} 
where Eq.~\eqref{rWIold}, which follows from \eqref{rWI0}, was given earlier in Ref.~\cite{deBoer:2001nw}. 
 
The twisted right-mover vertex operator equation is \cite{deBoer:2001nw,Halpern:2002ab}
\begin{eqnarray} 
\label{rVOE1} 
\bar{\pl} \hggz = -\frac{2 \eta^{ab}}{2k +Q_{\gfraks}} \sum_j \fji \sum_{\hat{j}=0}^{f_j(\s)-1} 
T_b t_{-\hat{j},j}(\s) :\hjb_{\hat{j}aj} (\bz) \hggz :_{\bar{M}} \nn \\ 
\bigspc \bigspc \bigspc -\frac{\Dg}{\bz} \sum_j (1-\fji ) t_{0j}(\s) \hggz 
\end{eqnarray} 
where $\bar{M}$ normal ordering is defined in Ref.~\cite{deBoer:2001nw}, and the twisted right-mover KZ system is 
\cite{deBoer:2001nw,Halpern:2002ab}:
\begin{subequations} 
\label{RtwistKZ}
\begin{equation} 
\bar{\pl}_\m \hat{A} (\st,\bz,z,\s) = \hat{\bar{W}}_\m (\st,\bz,\s) \hat{A} (\st,\bz,z,\s) ,\quad 
\hat{\bar{W}}_\m (\st,\bz,\s) = \sum_j \hat{\bar{W}}_{\m}^j (\st,\bz,\s) 
\end{equation} 
\begin{eqnarray} 
\hat{\bar{W}}_{\m}^j (\st,\bz,\s) \equiv \frac{2}{2k +Q_{\gfraks}} \sum_{\n \neq \m} \frac{\eta^{ab} 
T_a^{(\n)} T_b^{(\m)}}{f_j (\s) \bz_{\m \n}} \sum_{\hat{j}=0}^{f_j(\s) -1} \left( \frac{\bz_\n}
{\bz_\m} \right)^{\frac{\hat{j}}{f_j(\s)}} t_{\hat{j}j}^{(\m)}(\s) t_{-\hat{j} ,j}^{(\n)}(\s) \nn \\ 
 - \frac{\dg (T^{(\m)})}{\bz_\m} (1- \frac{1}{f_j(\s)}) t_{0j}^{(\m)}(\s) \bigspc \bigspc 
\end{eqnarray} 
\begin{equation} 
\left( \sum_{\m =1}^N T_a^{(\m)} t_{0j}^{(\m)} (\s) \right) \hat{A} (\st,\bz,z,\s) = 0 ,\quad \forall j
,\,\, a=1,\ldots ,\text{dim}\gfrak ,\,\, \srange .
\end{equation} 
\end{subequations} 
Finally, we have checked that the right-mover orbifold $\sl (2)$ Ward identities \eqref{rWIs} are satisfied as a 
consequence of this system. 

This completes our discussion of the orbifold $\sl (2)$ Ward identities associated to the orbifold Virasoro 
algebra of each WZW permutation orbifold. As discussed in Sec.~4, the orbifold Virasoro algebras are also 
important tools in the study of the spectra of the WZW permutation orbifolds. 
 
\section{Reducibility in WZW Permutation Orbifolds} 
 
\subsection{Reducibility of the Twisted Affine Primary Fields} 
 
We will argue in this subsection that the WZW permutation orbifolds admit a `natural' solution 
in which the twisted affine primary fields are {\it reducible} according to the disjoint cycles $j$ of 
$h_\s \in H$, with the following {\it block-diagonal structure}  
\begin{subequations} 
\begin{equation} 
\label{Reduce} 
\hat{g} (\st (T,\s),\bz,z,\s)_{\a \hat{j}j}{}^{\be \hat{l}l} = \de_j^l \,\hat{g}_j (\st (T,\s), 
\bz,z,\s)_{\a \hat{j}}{}^{\be \hat{l}} ,\quad \forall T \text{ for each } \srange 
\end{equation} 
\begin{equation} 
\a ,\be = 1,\ldots ,\text{dim}\;T ,\quad \bar{\hat{j}} ,\bar{\hat{l}} =0,\ldots ,f_j (\s)-1 
\end{equation} 
\end{subequations} 
where $\hat{g}_j$ are the blocks.
We will also argue (as in the untwisted case \cite{Halpern:1996et}) that the twisted affine primary fields can be  
factorized $\hat{g}_j =\hat{g}_j^- \cdot \hat{g}_j^+$ into twisted left- and right-mover fields. 
Finally we will see that all these fields are not functions of $\hat{j},\hat{l}$ independently, but 
only of the {\it difference variable} $\hat{l}-\hat{j}$. 

In fact, evidence has been accumulating for the reducibility of the twisted affine primary fields of 
the permutation orbifolds - in particular at the classical \cite{deBoer:2001nw,Halpern:2002hw} and at 
the abelian \cite{Halpern:2002ab} levels. We review this evidence first by way of motivation. 
  
\noindent $\bullet$ {\textbf{Motivation: Functional Integral Formulation}} 
 
For all WZW orbifolds, the {\it WZW orbifold action} \cite{deBoer:2001nw,Halpern:2002hw} is a functional of the 
classical {\it group orbifold elements} which are the classical (high-level) limit of the twisted affine primary 
fields. In the case of the WZW permutation orbifolds it has been known for some time that the classical group orbifold 
elements are {\it reducible} according to the disjoint cycles, exactly as shown in \eqref{Reduce}. For this discussion, 
it is conventional to work on the cylinder $(\xi ,t)$, where one has the following relations \cite{deBoer:2001nw,Halpern:2002hw}:
\begin{subequations} 
\begin{equation} 
\label{TwRepM} 
\st_{\hat{j}aj} (T,\s) = T_a t_{\hat{j}j} (\s)  
\end{equation} 
\begin{equation} 
\hat{g} (\st (T,\s), \xi, t,\s) = e^{ i\sum_{j\hat{j}}\hat{\be}^{\hat{j}aj} (\xi,t,\s) \st_{\hat{j}aj} (T,\s)} 
,\quad \forall T \text{ for each } \srange 
\end{equation} 
\begin{equation} 
\label{Mono} 
\hat{g} (\st,\xi +2\pi ,t,\s)_{\hat{j}j}{}^{\hat{l}l} = e^{-\tp \frac{\hat{j}-\hat{l}}{f_j(\s)}} \hat{g} 
(\st,\xi,t,\s)_{\hat{j}j}{}^{\hat{l}l} 
\end{equation} 
\begin{equation} 
\hat{g} (\st, \xi, t,\s)_{\hat{j}j}{}^{\hat{l}l} = \de_j^l \,\hat{g}_j (T\tau (j,\s),\xi,
t,\s)_{\hat{j}}{}^{\hat{l}} ,\quad \hat{j} ,\hat{l} = 0,\ldots ,f_j (\s)-1 
\end{equation} 
\begin{equation} 
\hat{g}_j (\st,\xi,t,\s) \equiv \hat{g}_j (T\tau (j,\s),\xi,t,\s) = e^{i \sum_{\hat{j}}  
\hat{\be}^{\hat{j}aj} (\xi,t,\s) T_a \tau_{\hat{j}} (j,\s)} 
\end{equation} 
\begin{equation} 
\label{TauDef} 
t_{\hat{j}j} (\s)_{\hat{l}l}{}^{\hat{m}m} = \de_{jl} \de_l^m \tau_{\hat{j}} (j,\s)_{\hat{l}}{}^{\hat{m}}
,\quad \tau_{\hat{j}} (j,\s)_{\hat{l}}{}^{\hat{m}} \equiv \de_{\hat{j}+\hat{l}-\hat{m} ,0\,\text{mod } f_j(\s)}  
\end{equation} 
\begin{equation} 
\label{Monog} 
\hat{g}_j (\st,\xi +2\pi ,t,\s)_{\hat{j}}{}^{\hat{l}} = e^{-\tp \frac{\hat{j}-\hat{l}}{f_j (\s)}} 
\hat{g}_j (\st,\xi,t,\s)_{\hat{j}}{}^{\hat{l}} .
\end{equation} 
\end{subequations} 
Here $\hat{g} (\st,\s)$ and $\hat{g}_j (\st,\s)$ are the group orbifold elements and their blocks respectively. The 
twisted representation matrices $\st (T,\s)$ in \eqref{TwRepM} are the same as those defined in \eqref{TwRepM1}, and 
we have suppressed the Lie indices $\a ,\,\be$ throughout. 
 
Reducibility of the group orbifold elements implies that the WZW orbifold action for the permutation orbifolds
is {\it separable} in $j$ \cite{deBoer:2001nw,Halpern:2002hw}: 
\begin{subequations} 
\label{Action} 
\begin{equation} 
\hat{S}_{\hat{g}(\s)} [\M ,\hat{g}] = \sum_j \hat{S}_{\s ,j} [\hat{g}_j] ,\quad \hat{g}_j \equiv \hat{g}_j 
(\st,\xi,t,\s) ,\quad \srange 
\end{equation} 
\begin{equation} 
\hat{S}_{\s ,j} [\hat{g}_j] \equiv -\frac{k}{y(T)} \sum_{\hat{j}=0}^{f_j (\s)-1} \sum_{\a =1}^{\text{dim } T} 
[\frac{1}{8\pi} \int d^2 \xi \hat{g}_j^{-1} \pl_+ \hat{g}_j \hat{g}_j^{-1} \pl_- \hat{g}_j + \frac{1}{12\pi} 
\int_{\Gamma} (\hat{g}_j^{-1} d\hat{g}_j )^3 ]_{\hat{j}\a}{}^{\hat{j}\a} 
\end{equation} 
\begin{equation}
d^2 \xi \equiv dt d\xi ,\quad \pl_{\pm} \equiv \pl_t \pm \pl_{\xi}
\end{equation} 
\begin{equation} 
\hat{S}_{\s,j} [\hat{g}_j (\st,\xi +2\pi ,t,\s)] = \hat{S}_{\s,j} [\hat{g}_j (\st,\xi,t,\s)] ,\,\,\forall j . 
\end{equation} 
\end{subequations} 
Moreover, separability of the action tells us that, at least in the classical limit, the dynamics of the block  
$\hat{g}_j$ is independent of the dynamics of the block $\hat{g}_l$ when $j \neq l$. 
 
The quantum theory admits a `natural' solution of the same type because the Virasoro operators $L_\s (m)$,  
$\bar{L}_\s (m)$ in \eqref{LsL0j} and \eqref{LsbL0j} are sums of commuting terms $\hat{L}_{0j} (m)$.  In a  
functional integral formulation with the action \eqref{Action}, this can be realized by choosing the naive  
measure which is factorized in $j$, 
\begin{equation} 
\label{jMeas} 
\D \hat{g} \equiv \prod_j \prod_{\xi ,t} d\hat{g}_j (\xi,t) . 
\end{equation} 
Then one has factorization of the correlators according to distinct values of $j$, e.g. when $j \neq l$: 
\begin{eqnarray} 
\label{Factor4} 
\lefteqn{ \langle \hat{g}_j (\st^{(1)},\xi_1,t_1 ,\s) \hat{g}_l (\st^{(2)},\xi_2,t_2 ,\s) \hat{g}_j (\st^{(3)},\xi_3,t_3 ,\s) 
\hat{g}_l (\st^{(4)},\xi_4,t_4 ,\s) \rangle_\s = } \nn \\ 
& & \langle \hat{g}_j (\st^{(1)},\xi_1,t_1 ,\s) \hat{g}_j (\st^{(3)},\xi_3,t_3 ,\s)  
\rangle_\s \times \langle \hat{g}_l (\st^{(2)},\xi_2,t_2 ,\s) \hat{g}_l (\st^{(4)},\xi_4,t_4 ,\s) \rangle_\s . 
\end{eqnarray} 
Similarly, all correlators in the functional integral formulation can be expressed in terms of the  
independent ``single-cycle'' correlators 
\begin{equation} 
\langle \hat{g}_j (\st^{(1)} ,\xi_1 ,t_1,\s) \ldots \hat{g}_j (\st^{(N)},\xi_N ,t_N,\s) \rangle_\s 
\end{equation} 
for each fixed value of $j$. 
 
Finally we notice that the blocks of the classical group orbifold elements are not functions of $\hat{j} ,\hat{l}$ 
independently, but rather only of the {\it difference variable} $\hat{l}-\hat{j}$:
\begin{equation} 
\label{DiffV} 
\hat{g}_j (\st,\xi,t,\s)_{\hat{j}+\hat{m}}{}^{\hat{l}+\hat{m}} = \hat{g}_j (\st,\xi,t,\s)_{\hat{j}} 
{}^{\hat{l}} ,\quad \forall \,\hat{j},\hat{l},\hat{m} .
\end{equation} 
This shift-invariance condition follows from the difference variable structure of the matrices $\tau_{\hat{j}} (j,\s)$ in  
\eqref{TauDef}. As we shall see, the shift invariance will also generalize to the twisted affine  
primary fields in the operator formulation. 
 
\noindent $\bullet$ {\textbf{Motivation: Abelian Permutation Orbifolds}} 
 
Ref.~\cite{deBoer:2001nw} gave the twisted vertex operator equations for all WZW orbifolds, and Ref.~\cite{Halpern:2002ab} 
solved those equations in an abelian limit to obtain the {\it twisted vertex operators} of a large class of abelian 
orbifolds 
\begin{equation} 
\frac{A_{{\rm Cartan}\,g}(H)}{H} \sp H \subset {\rm Aut}({\rm Cartan}\,g) \sp {\rm Cartan}\, g \subset g 
\end{equation} 
where the ambient algebra $g$ supplies the representation space for the twisted sectors of each orbifold. 
We are interested here in the case of the {\it abelian permutation orbifolds}, where Cartan $g$ has the form
\begin{equation}
{\rm Cartan} g =\oplus_I {\rm Cartan} \gfrak^I, \quad{\rm Cartan} \gfrak^I \simeq {\rm Cartan} \gfrak 
\end{equation}
and $H$(permutation) acts among the copies Cartan $\gfrak^I$ of Cartan $\gfrak$.

Our second reducibility argument is based on the explicit form of the twisted left-mover vertex operators 
given in Ref.~\cite{Halpern:2002ab} for the case of the abelian permutation orbifolds: 
\begin{subequations} 
\label{TwVerOp} 
\begin{equation} 
\label{TVerOp} 
\hat{g}_+ (\st,z,\s) = \hat{\Gamma} (\st ,\hj_0 (0),\s) \prod_j \hat{V}_j (\st,z,\s)  ,\quad \st (T,\s) = T t(\s)
\end{equation} 
\begin{eqnarray} 
\label{Vj} 
\lefteqn{ \hat{V}_j (\st,z,\s) \equiv z^{-\Delta (T) (1-\fji) t_{0j}(\s)} e^{iT_a t_{0j} (\s) \hat{q}^{aj} (\s)} z^{\eta^{ab} T_b  
t_{0j} (\s) \hj_{0aj} (0)/k f_j(\s)} \times } \nn \\ 
& & \quad \times \exp  \left( -\frac{\eta^{ab} T_b}{k} \fji \sum_{\hat{j}=0}^{f_j (\s)-1} \sum_{m \leq -1} \hj_{\hat{j}aj}  
\modmj \frac{z^{-\modmj}}{m+\jfj} t_{-\hat{j},j} (\s) \right) \nn \\ 
& &\quad \times \exp  \left( -\frac{\eta^{ab} T_b}{k} \fji \sum_{m \geq 1} \hj_{0aj} (m) \frac{z^{-m}}{m} t_{0j} (\s) 
\right) \bigspc \bigspc \bigspc \bigspc \nn \\ 
& & \quad \times \exp \left( -\frac{\eta^{ab} T_b}{k} \fji \sum_{\hat{j}=1}^{f_j (\s)-1} \sum_{m \geq 0} \hj_{\hat{j}aj}  
\modmj \frac{z^{-\modmj}}{m+\jfj} t_{-\hat{j},j} (\s) \right)  
\end{eqnarray} 
\end{subequations} 
\begin{subequations} 
\begin{equation} 
[ \hat{q}^{aj} (\s), \hj_{\hat{l}bl} (m+\frac{\hat{l}}{f_l (\s)}) ] = i\de_l^j \de_b^a \de_{m+\frac{\hat{l}} 
{f_l (\s)} ,0} ,\quad [\hat{q}^{aj} (\s), 
\hat{q}^{bl} (\s)]=0 
\end{equation} 
\begin{equation} 
\label{AbJJ} 
[ \hj_{\hat{j}aj} \modmj ,\hj_{\hat{l}bl} \modnl ] = \de_{jl} kf_j (\s) \eta_{ab} \modmj \de_{m+n+\frac{\hat{j}+\hat{l}}{f_j(\s)} ,0} 
\end{equation} 
\begin{equation} 
\hj_{\hat{j}aj} (m+\jfj \geq 0) |0 \rangle_\s = {}_\s \langle 0| \hj_{\hat{j}aj} (m+\jfj \leq 0) =0 
\end{equation} 
\begin{equation} 
[T_a ,T_b ]=0,\quad \frac{(\eta^{ab} T_a T_b )_{\a} {}^{\be}}{2k} = \Delta (T) \de_{\a}^{\be} ,\quad \forall T \text{ for each } \srange  
\end{equation}  
\begin{equation} 
 a,b = 1,\ldots , \text{dim(Cartan } \gfrak )  ,\quad \bar{\hat{j}} =0,\ldots ,f_j(\s)-1 . 
\end{equation} 
\end{subequations} 
Here \eqref{AbJJ} is the abelian analogue of the general orbifold affine algebra \eqref{JJAlg}, and 
$\hat{\Gamma} (\st,\hj_0 (0),\s)$ in \eqref{TVerOp} is the Klein transformation - which is a so-far undetermined 
function in this development. 
 
As far as block structure of $\hat{g}_+$ is concerned, the only subtlety here is in the Klein transformation $\hat{\Gamma}$. 
We know that the Klein transformation has a diagonal block structure, $\hat{\Gamma} =\oplus_j \hat{\Gamma}_j $, due to its 
dependence on the matrices $\{ t_{\hat{j}j} (\s)\}$ in $\st(T,\s)$. We will however also assume that $\hat{\Gamma}$ is a 
function only of $\{ \st_{0aj} \}$ and moreover that $\hat{\Gamma}_j$ is a function only of $\hj_{0aj} (0)$, that is 
\begin{equation} 
\label{GammaJ} 
\hat{\Gamma} (\st,\hj_0 (0),\s) \goto \hat{\Gamma} (\st_0,\hj_0 (0),\s)_{\hat{j}j}{}^{\hat{l}l} =  
\de_j^l \,\hat{\Gamma}_j (T\tau_0 (j,\s) ,\hj_{0j} (0),\s)_{\hat{j}}{}^{\hat{l}} 
\end{equation} 
where the matrices $\tau_{\hat{j}}(j,\s)$ are defined in \eqref{TauDef}. These assumptions are natural because the 
$\hat{q}$-dependent factor of the twisted vertex operator excites only the zero modes $\hj_{0aj} (0)$ of the twisted currents 
\begin{equation} 
\left( \hj_{\hat{j}aj} (m+\frac{\bar{\hat{j}}}{f_j(\s)} \geq 0) - \de_{jl} \,\de_{m,0} \,\de_{\bar{\hat{j}},0} \,\st_{0aj} (T,\s) \right) 
\{ e^{i\sum_b \hat{q}^{bl} \st_{0bl} (T,\s)} |0\rangle_\s \} =0  
\end{equation} 
by the amount $\st_{0j} (T,\s) $. 
 
Then we find with \eqref{GammaJ} that the twisted left-mover vertex operators \eqref{TwVerOp}  
exhibit the same block structure and shift invariance 
\begin{subequations} 
\begin{equation} 
\hat{g}_+ (\st,z,\s)_{\hat{j}j}{}^{\hat{l}l} = \de_j^l \,\hat{g}_j^+ (\st,z,\s)_{\hat{j}}{}^{\hat{l}} 
\end{equation} 
\begin{equation} 
\label{GJGamma} 
\hat{g}_j^+ (\st,z,\s) = \hat{\Gamma}_j (T\tau_0 (j,\s),\hj_{0j} (0),\s) \,\hat{V}_j (T\tau (j,\s),z,\s) 
\end{equation} 
\begin{equation}
\hat{V}_j (T\tau (j,\s) ,z,\s) \equiv \hat{V}_j (Tt(\s) ,z,\s) |_{t_{\hat{j}j} (\s) \rightarrow \tau_{\hat{j}}(j,\s)}
\end{equation}
\begin{equation} 
\hat{g}_j^+ (\st,z,\s)_{\hat{j}+\hat{m}}{}^{\hat{l}+\hat{m}} = \hat{g}_j^+ (\st,z,\s)_{\hat{j}}{}^{\hat{l}} 
,\quad \forall \,\hat{j},\hat{l},\hat{m}  
\end{equation} 
\end{subequations} 
which were found for the group orbifold elements above. The same block structure and shift invariance 
are found for the twisted right-mover vertex operators $\hat{g}_- (\st,\bz,\s)$ given in Ref.~\cite{Halpern:2002ab},  
as well as for the twisted non-chiral vertex operator $\hat{g} = \hat{g}_- \hat{g}_+$.  
 
The new feature here is the {\it chiral} block-diagonal structure, which will also generalize to the twisted 
affine primary fields of the full WZW permutation orbifolds. (Chiral block-diagonal structure 
can also be argued \'a la Witten \cite{Witten:1984ew} at the action level by considering the classical equations 
of motion of the group orbifold elements.) 
 
Having completed our presentation of the classical and abelian evidence, we now give a {\it derivation} 
of reducibility in the general WZW permutation orbifold, using the principle of local isomorphisms. 
 
\noindent $\bullet$ {\textbf{Derivation by Local Isomorphisms}} 
 
For this discussion we return (see Subsec.~$2.1$) to the untwisted permutation-invariant theory, where 
the untwisted affine primary fields are block-diagonal \cite{deBoer:2001nw}
\begin{subequations}
\begin{equation} 
\label{Blockg} 
g(T,\bz,z)_{\a I}{}^{\be J} = \de_I{}^J g_I (T,\bz,z)_\a{}^\be 
\end{equation} 
\begin{equation}
J_{aI} (z) g_J (T,\bw,w) = \de_{IJ} \frac{g_J (T,\bw,w)}{z-w} T_a + \Ord (z-w)^0
\end{equation}
\begin{equation}
T_I (z) g_J (T,\bw,w) =\de_{IJ} \left( \frac{g_J (T,\bw,w) \Dg}{(z-w)^2} +\frac{\pl_w g_J (T,\bw,w)}{z-w}
\right) + \Ord (z-w)^0 
\end{equation}
\end{subequations}
because the representation $T$ is reducible (see Eq.~\eqref{TauDef1}).
The blocks $g_I$ are the affine primary fields of each simple affine $\gfrak^I$. In parallel with the
earlier relabellings in \eqref{TildeFs}, the blocks $g_I$ are relabelled as $\{ \tilde{g}_{\hat{j}j} \}$ in
the $\s$-dependent cycle basis \cite{Halpern:2002ab}. Furthermore, it is known \cite{Halpern:1996et} that the 
untwisted affine primary fields of each copy of simple $\gfrak$ can be factorized into left- and right-moving 
chiral components: 
\begin{subequations}
\begin{equation}
g_I (T,\bz,z)_\a {}^{\be} \goto \tilde{g}_{\hat{j}j} (T,\bz,z,\s)_\a {}^{\be} 
\end{equation}
\begin{equation} 
\label{qgpspc} 
\tilde{g}_{\hat{j}j} (T,\bz,z,\s)_{\a} {}^{\be} = g_{\hat{j}j}^- (T,\bz,\s)_\a \cdot g_{\hat{j}j}^+ (T,z,\s)^{\be} 
,\quad \a ,\be =1,\ldots ,\text{dim}\;T . 
\end{equation}
\end{subequations} 
Here the dot connotes multiplication in the quantum group space, whose indices are suppressed; in what  
follows, we will often suppress the Lie indices $\a ,\be$ as well. 
 
We are now prepared to look for extra structure in the affine eigenprimary fields $\sg$, defined earlier in Eq.  
\eqref{eigeng}. Using \eqref{Blockg} in \eqref{eigeng} gives  
\begin{eqnarray} 
\sg (T,\bz,z,\s)_{\hat{j}j}{}^{\hat{l}l} &=& U(\s)_{\hat{j}j}{}^I \de_I{}^J g_I (T,\bz,z) U\hc (\s)_J 
{}^{\hat{l}l} \nn \\ 
&=& \sum_m \sum_{\hat{m}=0}^{f_m (\s)-1} U(\s)_{\hat{j}j}{}^{\hat{m}m} \tilde{g}_{\hat{m}m} (T,\bz,z,\s) 
U\hc (\s)_{\hat{m}m}{}^{\hat{l}l} .
\end{eqnarray} 
Then using the explicit form of $U(\s)$ in \eqref{Uform} we find after some algebra that the affine eigenprimary 
fields can be factorized into left- and right-mover chiral blocks: 
\begin{subequations} 
\label{EigenPF} 
\begin{equation} 
\sg (\st,\bz,z,\s)_{\hat{j}j}{}^{\hat{l}l} = \de_j^l \,\sg_j (\st,\bz,z,\s)_{\hat{j}}{}^{\hat{l}} 
\end{equation} 
\begin{eqnarray} 
\lefteqn{ \sg_j (\st,\bz,z,\s)_{\hat{j}}{}^{\hat{l}} = \fji \sum_{\hat{j}\p =0}^{f_j(\s) -1} e^{\frac{\tp  
(\hat{j}-\hat{l}) \hat{j}\p }{f_j(\s)}} \tilde{g}_{\hat{j}\p j} (T,\bz,z,\s) } \nn \\ 
& &= \fji \sum_{\hat{j}\p ,\hat{j}\pp =0}^{f_j(\s)-1} (e^{\tp \hat{j} \hat{j}\p / f_j(\s)} g_{\hat{j}\p j}^- (T,\bz,\s)) 
\cdot (e^{-\tp \hat{l} \hat{j}\pp /f_j(\s)} g_{\hat{j}\pp j}^+ (T,z,\s)) \de_{\hat{j}\p -\hat{j}\pp ,0\,\text{mod } f_j(\s)} \nn \\ 
& &= \sum_{\hat{m}=0}^{f_j(\s)-1} \sg_j^- (\st,\bz,\s)_{\hat{j}}{}^{\hat{m}} \cdot \sg_j^+ (\st,z,\s)_{\hat{m}}{}^{\hat{l}} 
\end{eqnarray} 
\begin{align} 
& \sg_j^+ (\st,z,\s)_{\hat{m}}{}^{\hat{l}} \equiv \fji \sum_{\hat{j}\p =0}^{f_j(\s)-1} e^{\tp (\hat{m}-\hat{l}) \hat{j}\p /f_j(\s)} 
g_{\hat{j}\p j}^+ (T,z,\s)  = \sg_j^+ (\st,z,\s)_{\hat{m} +\hat{j}}{}^{\hat{l}+\hat{j}} \nn \\ 
& \sg_j^- (\st,\bz,\s)_{\hat{j}}{}^{\hat{m}} \equiv \fji \sum_{\hat{j}\p =0}^{f_j(\s)-1} e^{\tp (\hat{j}-\hat{m}) \hat{j}\p /f_j(\s)}  
g_{\hat{j}\p j}^- (T,\bz,\s) =\sg_j^- (\st,\bz,\s)_{\hat{j}+\hat{l}}{}^{\hat{m}+\hat{l}} .  
\end{align} 
\end{subequations} 
Note that each of these blocks, $\sg_j$ as well as $\sg_j^{\pm}$, is a function only of the indicated  
difference variable.   
 
Next, we apply the principle of local isomorphisms  
\begin{subequations} 
\begin{equation} 
\sg_j (\st,\bz,z,\s) \dual \hgjz 
\end{equation} 
\begin{equation} 
\sg_j^+ (\st,z,\s) \dual \hgjp ,\quad \sg_j^- (\st,\bz,\s) \dual \hgjm 
\end{equation} 
\end{subequations} 
which promotes the properties \eqref{EigenPF} of the affine eigenprimary fields to isomorphic properties of the  
twisted affine primary fields $\hat{g} ,\hat{g}_j$ and $\hat{g}_j^{\pm}$: 
\begin{subequations} 
\label{GBlock} 
\begin{eqnarray} 
\label{hugz} 
\hggz_{\a \hat{j}j}{}^{\be \hat{l}l} &=& \de_j^l \,\hgjz_{\a \hat{j}}{}^{\be \hat{l}} ,\quad \srange \\  
\label{gjmp} 
\hgjz_{\a \hat{j}}{}^{\be \hat{l}}& =& \sum_{\hat{m}=0}^{f_j(\s)-1} \hgjm_{\a \hat{j}}{}^{\hat{m}} \cdot \hgjp_{\hat{m}}{}^{\be \hat{l}} 
\end{eqnarray} 
\begin{eqnarray} 
\label{GDiffV} 
\hgjp_{\hat{m}+\hat{j}}{}^{\be ,\hat{l}+\hat{j}} =\hgjp_{\hat{m}}{}^{\be \hat{l}} ,\quad \hgjm_{\a ,\hat{j}+ 
\hat{l}}{}^{\hat{m}+\hat{l}} = \hgjm_{\a \hat{j}}{}^{\hat{m}}, \nn \\ 
\hgjz_{\a ,\hat{j}+\hat{m}}{}^{\be ,\hat{l}+\hat{m}} = \hgjz_{\a \hat{j}}{}^{\be \hat{l}} ,\quad 
\bar{\hat{j}},\bar{\hat{l}},\bar{\hat{m}} = 0 ,\ldots ,f_j(\s)-1 . 
\end{eqnarray} 
\end{subequations} 
We emphasize that the left-right orbifold factorization \eqref{gjmp} followed directly from left-right
factorization in the symmetric theory and the principle of local isomorphisms, whereas only the consistency
of such a factorization was checked in Ref.~\cite{deBoer:2001nw}. Note also that the quantum group space, denoted by the 
dot, is the same quantum group space we started with in the untwisted factorization \eqref{qgpspc}. We 
remind the reader that, in addition to the shift invariance \eqref{GDiffV}, all quantities are also 
periodic $\hat{j} \goto \hat{j} \pm f_j(\s)$ in any spectral index. 
 
Finally, we remark that, although the twisted affine primary fields of the WZW permutation orbifolds 
are generically reducible as above, there are special cases which are irreducible. This happens in sector 
$\s$ of the orbifold whenever $h_\s \in H$(permutation) is composed of a single cycle ($j=0$ and $f_0 (\s)=K$);
for example, this phenomenon occurs in all twisted sectors of the $\z_{\lambda}$ cyclic permutation orbifolds
when $\lambda =$ prime. 
 
\subsection{Reduction of the Extended Operator Algebra} 
 
The factorized and block-diagonal form \eqref{GBlock} is consistent with the operator development in Sec.2,
and will lead us to a simpler reduced dynamics.

To begin, we will substitute the block structure \eqref{GBlock} into the OPEs and mode algebras of the WZW 
permutation orbifolds, and we will do this in two stages. First, we can use \eqref{hugz} to express all of our  
equations in terms of the unfactorized blocks $\hgjz$ of the twisted affine primary fields. For example, the 
reduced results 
\begin{subequations} 
\begin{equation} 
[ \hj_{\hat{j}aj} \modmj ,\hat{g}_l (\st,\bz,z,\s) ] = \de_{jl} \,\hat{g}_j (\st,\bz,z,\s) T_a \tau_{\hat{j}} 
(j,\s) z^{m+\jfj} 
\end{equation} 
\begin{equation} 
[ \hjbb_{\hat{j}aj} \modmj ,\hat{g}_l (\st,\bz,z,\s)] = -\de_{jl} \,\bz^{m+\jfj} T_a \tau_{-\hat{j}} (j,\s) 
 \hat{g}_j (\st,\bz,z,\s)  
\end{equation} 
\begin{align} 
& \pl \hgjz = \frac{2\eta^{ab}}{2k +Q_{\gfraks}} \fji \sum_{\hat{j}=0}^{f_j(\s)-1} :\hj_{\hat{j}aj} (z)  
\hgjz :_M T_b \tau_{-\hat{j}} (j,\s) \nn \\ 
& \bigspc \bigspc -\frac{\Dg}{z} (1-\fji) \hgjz  
\end{align} 
\end{subequations} 
follow from their unreduced counterparts in \eqref{JGAlg}, \eqref{RJGAlg} and \eqref{LVOE}. We note that 
all the reduced relations of this type can be easily obtained by the mnemonic 
\begin{equation} 
\label{Mnem} 
\hat{g} \goto \hat{g}_j ,\quad \hat{g}_{\pm} \goto \hat{g}_j^{\pm} ,\quad 
t_{\hat{j}j}(\s) \goto \tau_{\hat{j}} (j,\s) ,\quad t_{0j} (\s) \goto \tau_0 (j,\s) = \one  
\end{equation} 
from their unreduced counterparts. The matrices $\{ \tau_{\hat{j}} (j,\s) \}$ are defined in \eqref{TauDef}.
 
Next, we can use \eqref{gjmp} to factorize all the reduced OPEs and mode algebras into relations 
on the left- and right-mover blocks $\hat{g}_j^{\pm}$ of the twisted affine primary fields. We give here only 
the reduced and factorized form of the extended operator algebra: 
\begin{subequations} 
\label{Algg} 
\begin{equation} 
\label{CurAlgg} 
[ \hj_{\hat{j}aj} \modmj ,\hglp ] = \de_{jl} \,\hgjp T_a \tau_{\hat{j}} (j,\s) z^{m + \jfj}  
\end{equation} 
\begin{eqnarray} 
\label{VirAlgg} 
[ \hat{L}_{\hat{j}j} \modmj ,\hglp ] = \bigspc \bigspc \bigspc \bigspc \bigspc \nn \\ 
=\de_{jl} \,\hgjp \left( \lpl z + (m + \jfj +1) \Dg \right) \tau_{\hat{j}} (j,\s) z^{m+\jfj} 
\end{eqnarray} 
\begin{equation} 
[ L_\s (m) ,\hgjp] = \hgjp (\lpl z +(m+1) \Dg ) z^m 
\end{equation} 
\begin{equation} 
[ \hjbb_{\hat{j}aj} \modmj ,\hglm ] = -\de_{jl} \,\bz^{m + \jfj} T_a \tau_{-\hat{j}} (j,\s) \hgjm  
\end{equation} 
\begin{eqnarray} 
[ \hlbb_{\hat{j}j} \modmj ,\hglm ] = \bigspc \bigspc \bigspc \bigspc \bigspc \nn \\ 
=\de_{jl} \,\bz^{m+\jfj} \left( \bz \bar{\pl} + (m+ \jfj +1) \Dg \right) \tau_{-\hat{j}} (j,\s) \hgjm 
\end{eqnarray} 
\begin{equation} 
[ \bar{L}_\s (m) ,\hgjm] = \bz^m (\bz \bar{\pl} + (m+1) \Dg ) \hgjm 
\end{equation} 
\begin{equation} 
[ \hat{L} ,\hat{g}^- ] = [ \hj ,\hat{g}^- ] =0 ,\quad [\hlbb ,\hat{g}^+ ] =[\hjbb ,\hat{g}^+ ] =0 . 
\end{equation} 
\end{subequations} 
Similarly, we obtain the reduced twisted vertex operator equations 
\begin{subequations} 
\begin{eqnarray} 
\label{RedVOE} 
\pl \hgjp = \frac{2 \eta^{ab}}{2k + Q_{\gfraks}} \fji \sum_{\hat{j}=0}^{f_j(\s)-1} : \hj_{\hat{j}aj} (z) 
\hgjp :_M T_b \tau_{-\hat{j}} (j,\s) \quad \quad \nn \\ 
- \frac{\Dg}{z} (1-\fji )\hgjp 
\end{eqnarray}  
\begin{eqnarray} 
\bar{\pl} \hgjm = -\frac{2 \eta^{ab}}{2k + Q_{\gfraks}} \fji \sum_{\hat{j}=0}^{f_j(\s)-1} T_b \tau_{-\hat{j}} (j,\s) 
:\hjb_{\hat{j}aj} (\bz) \hgjm :_{\bar{M}} \quad \quad \nn \\ 
- \frac{\Dg}{z} (1-\fji )\hgjm 
\end{eqnarray} 
\end{subequations} 
for the left- and right-mover blocks of the twisted affine primary fields.   
 
\subsection{The Single-Cycle Form of the Twisted KZ System} 
 
Using Eq.~\eqref{GBlock} and the twisted KZ system \eqref{twistKZ}, \eqref{RtwistKZ}, the same procedure leads after some 
algebra to the reduced and factorized form of the twisted KZ system: 
\begin{subequations} 
\label{FactKZ1} 
\begin{eqnarray} 
\pl_\m \hat{A}_+ (\vec{j}; \st,z,\s) = \hat{A}_+ (\vec{j}; \st,z,\s) \hat{W}_\m (\vec{j}; \st,z,\s) \\ 
\bar{\pl}_\m \hat{A}_- (\vec{j}; \st,\bz,\s) = \hat{\bar{W}}_\m (\vec{j}; \st,\bz,\s) \hat{A}_- (\vec{j}; \st,\bz,\s) 
\end{eqnarray} 
\begin{eqnarray} 
\hat{W}_\m (\vec{j};\st,z,\s) \equiv \frac{2}{2k +Q_{\gfraks}} \sum_{\n \neq \m} \frac{\eta^{ab} T_a^{(\m)} 
T_b^{(\n)}}{f_{j_\m}(\s) z_{\m \n}} \sum_{\hat{j}=0}^{f_{j_\m} (\s)-1} \left( \frac{z_\n}{z_\m} \right)^{\frac{\hat{j}}{f_{j_\m} (\s)}} 
\tau_{\hat{j}}^{(\n)} (j_\n ,\s) \tau_{-\hat{j}}^{(\m)} (j_\m ,\s) \de_{j_\m ,j_\n} \nn \\ 
-\frac{\Delta_{\gfraks} (T^{(\m)})}{\bz_\m} (1-\frac{1}{f_{j_\m} (\s)}) \bigspc \bigspc 
\end{eqnarray} 
\begin{eqnarray} 
\hat{\bar{W}}_\m (\vec{j};\st,\bz,\s) \equiv \frac{2}{2k +Q_{\gfraks}} \sum_{\n \neq \m} \frac{\eta^{ab} T_a^{(\n)} 
T_b^{(\m)}}{f_{j_\m}(\s) \bz_{\m \n}} \sum_{\hat{j}=0}^{f_{j_\m} (\s)-1} \left( \frac{\bz_\n}{\bz_\m} \right)^{\frac{\hat{j}}{f_{j_\m}(\s)}} 
\tau_{\hat{j}}^{(\m)} (j_\m ,\s) \tau_{-\hat{j}}^{(\n)} (j_\n ,\s) \de_{j_\m ,j_\n} \nn \\ 
-\frac{\Delta_{\gfraks} (T^{(\m)})}{\bz_\m} (1-\frac{1}{f_{j_\m} (\s)})  \bigspc \bigspc 
\end{eqnarray} 
\begin{equation} 
\label{RedGWIs} 
\hat{A}_+ (\vec{j};\st,z,\s) \left( \sum_{\m =1}^N T_a^{(\m)} \de_{j,j_\m} \right)= \left( 
\sum_{\m=0}^N T_a^{(\m)} \de_{j,j_\m} \right) \hat{A}_- (\vec{j};\st,\bz,\s) =0 ,\quad \forall j . 
\end{equation} 
\end{subequations} 
Here the reduced left- and right-mover correlators $\hat{A}_{\pm}$ are defined as 
\begin{subequations} 
\begin{eqnarray} 
\hat{A}_+ (\vec{j} ;\st ,z ,\s) \equiv {}_\s \langle 0| \hat{g}_{j_1}^+ (\st^{(1)} ,z_1 ,\s), 
\ldots \hat{g}_{j_N}^+ (\st^{(N)} ,z_N ,\s) |0 \rangle_\s \bigspc \nn \\ 
\hat{A}_- (\vec{j} ;\st ,\bz ,\s) \equiv {}_\s \langle 0| \hat{g}_{j_1}^- (\st^{(1)} ,\bz_1 ,\s), 
\ldots \hat{g}_{j_N}^- (\st^{(N)} ,\bz_N ,\s) |0 \rangle_\s \bigspc 
\end{eqnarray} 
\begin{eqnarray} 
\hat{A} (\vec{j};\st,\bz,z,\s) &\equiv & \langle \hat{g}_{j_1} (\st^{(1)},\bz_1,z_1,\s) \ldots 
\hat{g}_{j_N} (\st^{(N)},\bz_N,z_N,\s) \rangle_\s \bigspc \nn \\ 
&=& \hat{A}_- (\vec{j};\st,\bz,\s) \cdot \hat{A}_+ (\vec{j};\st,z,\s) \bigspc \bigspc 
\end{eqnarray} 
\begin{equation} 
\label{RedCorr}
\hat{A} (\st,\bz,z,\s)_{\a_1 \hat{j}_1 j_1 ;\ldots ;\a_N \hat{j}_N j_N}^{\be_1 \hat{l}_1 l_1 ;\ldots ;\be_N 
\hat{l}_N l_N} =\de_{j_1}^{l_1} \ldots \de_{j_N}^{l_N} \hat{A} (\vec{j};\st,\bz,z,\s)_{\a_1 \hat{j}_1; \ldots ;\a_N 
\hat{j}_N}^{\be_1 \hat{l}_1; \ldots ;\be_N \hat{l}_N} 
\end{equation} 
\end{subequations} 
where $\hat{A} (\st,\bz,z,\s)$ in \eqref{RedCorr} is the full, non-chiral unreduced correlator defined earlier 
in \eqref{Acorr}. 
 
The Kronecker factors $\de_{j_\m ,j_\n}$ in the reduced connections $\hat{W}_\m (\vec{j}) ,\hat{\bar{W}}_\m (\vec{j})$  
and the reduced global Ward identity \eqref{RedGWIs} tell us that the reduced affine primary fields 
$\hat{g}_j$ as well as $\hat{g}_j^{\pm}$ decouple for distinct values of $j$. This allows us to choose the 
`natural' solution in which the correlators factorize according to distinct cycles $j$, as in the functional 
integral formulation above. As examples, we note that the factorized forms 
\begin{subequations} 
\label{FactEx} 
\begin{equation} 
\label{FactEx1}
\langle \hat{g}_j^+ (\st^{(1)} ,z_1,\s) \hat{g}_l^+ (\st^{(2)} ,z_2,\s) \rangle_\s = \langle \hat{g}_j^+ 
(\st^{(1)} ,z_1,\s) \rangle_\s \, \langle \hat{g}_l^+ (\st^{(2)} ,z_2,\s) \rangle_\s ,\,\, j \neq l 
\end{equation} 
\begin{align} 
& \langle \hat{g}_j^+ (\st^{(1)},z_1 ,\s) \hat{g}_l^+ (\st^{(2)},z_2 ,\s) \hat{g}_j^+ (\st^{(3)},z_3 ,\s) 
\hat{g}_l^+ (\st^{(4)},z_4 ,\s) \rangle_\s = \nn \\ 
&\quad =\langle \hat{g}_j^+ (\st^{(1)},z_1 ,\s) \hat{g}_j^+ (\st^{(3)},z_3 ,\s) \rangle_\s  
\times \langle \hat{g}_l^+ (\st^{(2)},z_2 ,\s) \hat{g}_l^+ (\st^{(4)},z_4 ,\s) \rangle_\s ,\,\,j \neq l  
\end{align} 
\end{subequations} 
are solutions of the reduced twisted KZ system - in parallel with \eqref{Factor4}. Such relations hold as well for 
$\hat{g}_j^+ \rightarrow \hat{g}_j^-$ and $\hat{g}_j^+ \rightarrow \hat{g}_j$. Moreover, it is known \cite{deBoer:2001nw} 
that all the one-point correlators vanish for non-trivial $\st$ so that, e.g. 
\begin{subequations} 
\begin{equation} 
\label{OnePt0}
\langle \hat{g} (\st,\bz,z,\s) \rangle_\s = \langle \hat{g}_j (\st,\bz,z,\s) \rangle_\s =0 ,\quad \st \neq 0
\end{equation} 
\begin{equation} 
\label{FacTwoPt} 
\langle \hat{g}_j (\st^{(1)} ,\bz_1,z_1,\s) \hat{g}_l (\st^{(2)} ,\bz_2,z_2,\s) \rangle_\s =\de_{jl} 
\langle \hat{g}_j (\st^{(1)} ,\bz_1,z_1,\s) \hat{g}_j (\st^{(2)} ,\bz_2,z_2,\s) \rangle_\s 
\end{equation} 
\end{subequations} 
where \eqref{FacTwoPt} follows from \eqref{FactEx1} and \eqref{OnePt0}. Similarly, one finds that 
$\langle \hat{g}_j \hat{g}_k \hat{g}_l \rangle_\s$ vanishes unless $j=k=l$. 
 
In the natural solution we therefore need only consider the {\it single-cycle correlators} at fixed values of $j$ 
\begin{subequations} 
\begin{eqnarray} 
\hat{A}_+ (j;\st (T,\s),z,\s) \equiv \langle \hat{g}_j^+ (\st^{(1)} (T^{(1)},\s) ,z_1 ,\s) \ldots \hat{g}_j^+  
(\st^{(N)} (T^{(N)},\s) ,z_N,\s) \rangle_\s \nn \\ 
\hat{A}_- (j;\st (T,\s),\bz,\s) \equiv \langle \hat{g}_j^- (\st^{(1)} (T^{(1)},\s),\bz_1 ,\s) \ldots \hat{g}_j^-  
(\st^{(N)} (T^{(N)},\s),\bz_N,\s) \rangle_\s 
\end{eqnarray} 
\begin{equation} 
\forall j,\,\, \forall T \text{ for each } \srange 
\end{equation} 
\end{subequations} 
from which all other correlators can be constructed, as in \eqref{FactEx}. Moreover, we find from \eqref{FactKZ1}  
that the single-cycle correlators satisfy the {\it single-cycle twisted KZ system} 
\begin{subequations} 
\label{FactKZ2} 
\begin{eqnarray} 
\pl_\m \hat{A}_+ (j;\st (T,\s),z,\s) = \hat{A}_+ (j;\st (T,\s),z,\s) \hat{W}_\m (j;\st (T,\s),z,\s) \nn \\ 
\bar{\pl}_\m \hat{A}_- (j;\st (T,\s),\bz,\s) = \hat{\bar{W}}_\m (j;\st (T,\s),\bz,\s) \hat{A}_- (j;\st (T,\s),\bz,\s) 
\end{eqnarray} 
\begin{eqnarray} 
\hat{W}_\m (j;\st (T,\s),z,\s) = \frac{2}{2k +Q_{\gfraks}} \sum_{\n \neq \m} \frac{\eta^{ab} T_a^{(\m)} T_b^{(\n)}} 
{f_j (\s) z_{\m \n}} \sum_{\hat{j}=0}^{f_j(\s)-1} \left( \frac{z_\n}{z_\m} \right)^{\jfj} \tau^{(\n)}_{\hat{j}} (j,\s) 
\tau^{(\m)}_{-\hat{j}} (j,\s) \nn \\ 
- \frac{\Delta_{\gfraks} (T^{(\m)})}{z_\m} (1-\fji) \quad \quad 
\end{eqnarray} 
\begin{eqnarray} 
\hat{\bar{W}}_\m (j;\st (T,\s),\bz,\s) = \frac{2}{2k +Q_{\gfraks}} \sum_{\n \neq \m} \frac{\eta^{ab} T_a^{(\n)} T_b^{(\m)}} 
{f_j (\s) \bz_{\m \n}} \sum_{\hat{j}=0}^{f_j(\s)-1} \left( \frac{\bz_\n}{\bz_\m} \right)^{\jfj} \tau^{(\m)}_{\hat{j}} (j,\s) 
\tau^{(\n)}_{-\hat{j}} (j,\s) \nn \\ 
- \frac{\Delta_{\gfraks} (T^{(\m)})}{\bz_\m} (1-\fji) \quad \quad 
\end{eqnarray} 
\begin{equation} 
\label{SCGWI} 
\hat{A}_+ (j;\st (T,\s),z,\s) \left( \sum_{\m =1}^N T_a^{(\m)} \right) = \left( \sum_{\m =1}^N T_a^{(\m)} \right) 
\hat{A}_- (j;\st (T,\s),\bz,\s) =0 ,\,\, \forall j 
\end{equation} 
\end{subequations} 
which operates entirely within each disjoint cycle. Note in particular that the single-cycle  
global Ward identity in \eqref{SCGWI} has the same form as the usual global Ward identity \cite{KZ:1984kz} found in  
untwisted KZ systems. 
 
The single-cycle twisted KZ system \eqref{FactKZ2}, \eqref{GDiffV} is one of the central results of this paper. 

As a first application of \eqref{FactKZ2}, we note that the single-cycle forms of the orbifold $\sl (2)$ 
Ward identities
\begin{subequations}
\label{SCLWI}
\begin{equation} 
\label{SCLWI0} 
\hat{A}_+ (j;\st,\bz,z,\s) \sum_{\m =1}^{N} \left( \lplm z_\m + \Delta_{\gfraks} (T^{(\m )}) \right) = 0 
\end{equation} 
\begin{equation} 
\label{SCLWIP} 
\hat{A}_+ (j;\st,\bz,z,\s) \sum_{\m =1}^{N} \left[ \left( \lplm z_\m + (1 +\frac{1}{f_j (\s)}) \Delta_{\gfraks} 
(T^{(\m )}) \right) \tau_1^{(\m )} (j,\s) z_{\m}^{\frac{1}{f_j (\s)}} \right] =0
\end{equation} 
\begin{equation} 
\label{SCLWIM} 
\hat{A}_+ (j;\st,\bz,z,\s) \sum_{\m =1}^{N} \left[ \left( \lplm z_\m + (1 -\frac{1}{f_j (\s)}) \Delta_{\gfraks} 
(T^{(\m )}) \right) \tau_{-1}^{(\m )}(j,\s) z_{\m}^{-\frac{1}{f_j (\s)}} \right] =0
\end{equation} 
\end{subequations}
\begin{subequations}
\label{SCRWI}
\begin{equation} 
\label{SCRWI0}
\sum_{\m =1}^N \left( \bz_\m \bar{\pl}_\m + \Delta_{\gfraks} (T^{(\mu)}) \right) \hat{A}_- (j;\st,\bz,z,\s) =0 
\end{equation} 
\begin{equation} 
\sum_{\m =1}^N \left[ \bz_{\m}^\fji \left( \bz_\m \bar{\pl}_\m + (1 +\fji) \Delta_{\gfraks} (T^{(\mu)}) \right)
\tau_{-1}^{(\m )} (j,\s) \right] \hat{A}_- (j;\st,\bz,z,\s) =0 
\end{equation} 
\begin{equation} 
\sum_{\m =1}^N \left[ \bz_{\m}^{-\fji} \left( \bz_\m \bar{\pl}_\m + (1 -\fji) \Delta_{\gfraks} (T^{(\mu)}) 
\right) \tau_1^{(\m )} (j,\s) \right] \hat{A}_- (j;\st,\bz,z,\s) =0 
\end{equation} 
\end{subequations} 
are also satisfied by any solution of the single-cycle twisted KZ system.
 
\subsection{First Form of the Single-Cycle Two-Point Correlator} 
 
The unreduced left-mover two-point correlator was given for the case of the WZW permutation orbifolds in 
Refs.~\cite{deBoer:2001nw,Halpern:2002ab}. Similarly, using the shift condition \eqref{GDiffV} and the global Ward 
identity \eqref{SCGWI}, we have worked out the single-cycle non-chiral two-point correlator 
\begin{subequations} 
\label{TwoPCorr} 
\begin{eqnarray} 
\label{TwoPt} 
\hat{A} (j;1,2)\!\! & \equiv \!\!&\! \hat{A}_- (j;1,2) \cdot \hat{A}_+ (j;1,2) = \langle \hat{g}_j (\st^{(1)} 
(T^{(1)},\s),\bz_1 ,z_1,\s) \hat{g}_j (\st^{(2)} (T^{(2)},\s),\bz_2,z_2,\s) \rangle_\s  \nn \\ 
& = \!\!&\! \C (j;\st,\s) |z_1 |^{-2\Delta_{\gfraks} (T^{(1)}) (1-\fji)} |z_2 |^{-2\Delta_{\gfraks} (T^{(2)}) 
(1-\fji)} \times \nn \\ 
& &\bigspc \times |z_{12} |^{-4\Delta_{\gfraks} (T^{(1)}) /f_j(\s)} \exp \{\frac{2}{2k +Q_{\gfraks}} F(j;1,2) \} 
\bigspc  
\end{eqnarray} 
\begin{eqnarray} 
\label{FDefn} 
F(j;1,2) \equiv \frac{T_a^{(2)} \eta^{ab} T_b^{(1)}}{f_j(\s)} \sum_{\hat{j}=1}^{f_j(\s)-1} \{  
\tau^{(1)}_{\hat{j}} (j,\s) \tau^{(2)}_{-\hat{j}} (j,\s) I_{\jfj} (\frac{\bz_1}{\bz_2},\infty)  \bigspc \nn \\ 
\bigspc + \tau^{(2)}_{\hat{j}}(j,\s) \tau^{(1)}_{-\hat{j}}(j,\s) I_{\jfj} (\frac{z_1}{z_2},\infty) \}  
\end{eqnarray} 
\begin{equation} 
\C (j;\st,\s) = \C_- (j;\st,\s) \cdot \C_+ (j;\st,\s) ,\quad I_{\jfj} (y,\infty) \equiv \int_{\infty}^{y} 
\frac{dx}{x-1} x^{-\jfj}  
\end{equation} 
\end{subequations} 
as the solution to the single-cycle twisted KZ equations. After some algebra, we have verified that this solution satisfies
the orbifold $\sl (2)$ Ward identities \eqref{SCLWI} and \eqref{SCRWI}, as it must.
 
We explain some of the steps used in obtaining \eqref{TwoPCorr}, emphasizing properties of the constant matrix 
$\C (j;\st,\s)$. In the first place, the blocks $\hat{g}_j$ are functions of difference variables, and hence the 
constant matrices $\C_{\pm} ,\C$ are also functions of their respective difference variables 
\begin{subequations} 
\label{CDifVbl} 
\begin{equation} 
O (j;\st,\s)_{\hat{j}_1 +\hat{m}_1;\hat{j}_2}{}^{\hat{l}_1 +\hat{m}_1 ;\hat{l}_2} =O (j;\st,\s)_{\hat{j}_1 
;\hat{j}_2 +\hat{m}_2}{}^{\hat{l}_1 ;\hat{l}_2 +\hat{m}_2} = O (j;\st,\s)_{\hat{j}_1; \hat{j}_2}{}^{\hat{l}_1 
;\hat{l}_2} 
\end{equation} 
\begin{equation} 
O (j;\st,\s) = \C_{\pm} (j;\st,\s) \text{ or } \C (j;\st,\s)  
\end{equation} 
\end{subequations} 
where we have suppressed Lie indices and quantum group indices.  Moreover, the global Ward identity places  
further constraints on the matrix $\C$, namely 
\begin{subequations} 
\begin{equation} 
[T_c^{(2)} \eta^{cd} T_d^{(1)} ,(T_a^{(1)} + T_a^{(2)})] =0 
\end{equation} 
\begin{equation} 
\label{CGWI} 
\C (j;\st,\s) (T_a^{(1)} +T_a^{(2)}) = (T_a^{(1)} +T_a^{(2)}) \C (j;\st,\s) =0 . 
\end{equation} 
\end{subequations} 
Taken together, the conditions \eqref{CDifVbl} and \eqref{CGWI} imply that 
\begin{subequations}
\begin{equation} 
[ \C (j;\st,\s), T_a^{(2)} \eta^{ab} T_b^{(1)}] =0 ,\quad [ \C (j;\st,\s) ,F(j;1,2)] =0 
\end{equation} 
\begin{eqnarray}
\exp \{ \frac{2}{2k +Q_{\gfraks}} \frac{T_a^{(1)} \eta^{ab} T_b^{(2)}}{f_j (\s)} \sum_{\hat{j}=1}^{f_j(\s) -1}
\tau_{\hat{j}}^{(1)} (j,\s) \tau_{-\hat{j}}^{(2)} (j,\s) I_{\jfj} ( \frac{\bz_1}{\bz_2} ,\infty ) \} \times \bigspc \nn \\
\quad \times \C (j;\st,\s) \exp \{ \frac{2}{2k +Q_{\gfraks}} \frac{T_a^{(2)} \eta^{ab} T_b^{(1)}}{f_j (\s)} 
\sum_{\hat{j}=1}^{f_j(\s) -1} \tau_{\hat{j}}^{(2)} (j,\s) \tau_{-\hat{j}}^{(1)} (j,\s) I_{\jfj} ( \frac{z_1}{z_2} ,\infty ) \}
\nn \\
= \C (j;\st,\s) \exp \{ \frac{2}{2k +Q_{\gfraks}} F (j;1,2) \}
\end{eqnarray}
\end{subequations}
which allowed us to pull the matrix $\C$ to the left in the result \eqref{TwoPCorr}.  
 
Furthermore, the complete solution of the global Ward identity \eqref{CGWI} gives a form which is  
equivalent to the standard Haar integration over two Lie group elements, and we find that the  
the matrix $\C (j;\st,\s)$ has the following form 
\begin{subequations} 
\begin{equation} 
\label{CForm} 
\C (j;\st,\s)_{\a_1 \hat{j}_1 ;\a_2 \hat{j}_2}{}^{\be_1 \hat{l}_1 ;\be_2 \hat{l}_2} = \de_{T^{(2)}  
,\bar{T}^{(1)}} \de_{\a_1 \a_2} \de^{\be_1 \be_2} \D (j;\st,\s)_{\hat{j}_1 ;\hat{j}_2}{}^{\hat{l}_1 ;\hat{l}_2} 
\end{equation} 
\begin{equation} 
\D (j;\st,\s)_{\hat{j}_1 +\hat{m}_1;\hat{j}_2}{}^{\hat{l}_1 +\hat{m}_1 ;\hat{l}_2}= \D (j;\st,\s)_{\hat{j}_1 ; 
\hat{j}_2 +\hat{m}_2}{}^{\hat{l}_1 ;\hat{l}_2 +\hat{m}_2} = \D (j;\st,\s)_{\hat{j}_1; \hat{j}_2}{}^{\hat{l}_1 
;\hat{l}_2}  
\end{equation} 
\end{subequations} 
in terms of an as-yet-undetermined constant matrix $\D$. Finally, the single-cycle two-point  
correlator is non-vanishing only when $T \!\equiv \!T^{(2)} \!=\! \bar{T}^{(1)}$, so the untwisted conformal weights 
in \eqref{TwoPt} are equal. This gives the first form of the {\it single-cycle two-point correlator}
\begin{subequations} 
\label{SCTPCorr}
\begin{equation} 
\label{SCTwoPt} 
\hat{A} (j;1,2)= \C (j;\st,\s) |z_1 z_2 |^{-2\Dg (1-\fji)} |z_{12} |^{-4\Dg /f_j(\s)} \exp \{\frac{2}{2k +Q_{\gfraks}} 
F(j;1,2) \} 
\end{equation} 
\begin{equation} 
\label{TPCnfWt}
\C (j;\st,\s) = \de_{T^{(2)} ,\bar{T}^{(1)}} \D(j;\st,\s) ,\quad \Dg \one \equiv \frac{\eta^{ab} T_a T_b}
{2k +Q_{\gfraks}} = \Delta_{\gfraks} (T^{(1)}) \one =\Delta_{\gfraks} (T^{(2)}) \one 
\end{equation} 
\end{subequations} 
where $F(j;1,2)$ is defined in \eqref{TwoPCorr}. We will return in Subsec.~$4.4$ to fix the multiplicative
constant matrix $\D$ in \eqref{TPCnfWt}. 
 
\section{Twisted Affine Primary and Principal Primary States} 
 
\subsection{The Twisted Affine Primary States of Cycle $j$} 
 
The object of this subsection is to obtain the {\it orbifold-modified asymptotic formulae} \cite{Borisov:1997nc} 
for the in- and out-states created by the reduced twisted affine primary fields, and to show that in fact these 
states are {\it twisted affine primary states}, as one might expect. The twisted affine primary states are also 
seen to be primary under the orbifold Virasoro algebra.
 
To obtain some information about excited states in the orbifold, we first consider the following limits of the 
single-cycle two-point correlator in \eqref{SCTPCorr}, 
\begin{equation} 
I_{\jfj} (y,\infty) = \Ord (y^{-\jfj}) \quad \text{for } y>>1 ,\,\,\, \hat{j} =1,\ldots ,f_j (\s)-1  
\end{equation} 
\begin{subequations} 
\begin{eqnarray} 
\lim_{|z_2 | \rightarrow 0} |z_2 |^{2\Dg (1-\fji)} {}_\s \langle 0| \hat{g}_j (\st^{(1)} ,\bz_1 ,z_1 ,\s) 
\hat{g}_j (\st^{(2)} ,\bz_2 ,z_2 ,\s) |0 \rangle_\s \nn \\ 
= \C (j;\st,\s) |z_1 |^{-2\Dg (1+\fji)} 
\end{eqnarray} 
\begin{eqnarray} 
\label{CAsymp} 
\lim_{|z_1 | \rightarrow \infty} \lim_{|z_2 |\rightarrow 0} |z_1 |^{2\Dg (1+\fji)} |z_2 |^{2\Dg (1-\fji)} 
{}_\s \langle 0| \hat{g}_j (\st^{(1)} ,\bz_1,z_1,\s) & \!\!\hat{g}_j (\st^{(2)} ,\bz_2,z_2,\s) |0\rangle_\s 
\nn \\ 
=\C (j;\st,\s) & 
\end{eqnarray} 
\end{subequations} 
where the form of the constant matrix $\C (j;\st,\s)$ is given in \eqref{CForm}. These limits define the `in' 
and `out' states 
\begin{subequations} 
\label{AfPrSt} 
\begin{equation} 
\lim_{|z| \rightarrow 0} |z|^{2\Dg (1-\fji)} \hgjz |0\rangle_\s = |A (j;\st) \rangle_\s  
\end{equation} 
\begin{equation} 
\lim_{|z| \rightarrow \infty} {}_\s \langle 0| |z|^{2\Dg (1+\fji)} \hgjz = {}_\s \langle A (j;\st)|  
\end{equation} 
\end{subequations} 
created by the twisted affine primary field $\hgjz$ on the ground state (scalar twist-field state) of sector $\s$. 
In fact, these states are matrix states 
\begin{equation} 
( |A (j;\st) \rangle_\s )_{\a \hat{j}}{}^{\be \hat{l}} ,\quad (_\s \langle A(j;\st) |)_{\a \hat{j}} {}^{\be 
\hat{l}} 
\end{equation} 
but we will generally suppress their indices. The relations in \eqref{CAsymp} and \eqref{AfPrSt} tell us that 
the constant matrix $\C (j;\st ,\s)$ is the inner product of these states 
\begin{equation} 
\C (j;\st,\s) = {}_\s \langle A (j;\st^{(1)}) | A (j;\st^{(2)}) \rangle_\s  
\end{equation} 
and, according to \eqref{CForm}, both sides of this relation are proportional to $\de_{T^{(2)}, 
\bar{T}^{(1)}}$. 
 
We can similarly define chiral and antichiral states, created by the left- and right-mover twisted affine  
primary fields $\hat{g}_j^{\pm}$ : 
\begin{subequations} 
\label{Acreat} 
\begin{equation} 
\label{LInA} 
|A^+ (j;\st) \rangle_\s \equiv \lim_{z \rightarrow 0} z^{\Dg (1-\fji)} \hgjp |0\rangle_\s 
\end{equation} 
\begin{equation} 
\label{LOutA} 
{}_\s \langle A^+ (j;\st) | \equiv \lim_{z \rightarrow \infty} {}_\s \langle 0| z^{\Dg (1+\fji)} \hgjp 
\end{equation} 
\begin{equation} 
\label{RInA} 
|A^- (j;\st) \rangle_\s \equiv \lim_{\bz \rightarrow 0} \bz^{\Dg (1-\fji)} \hgjm |0\rangle_\s 
\end{equation} 
\begin{equation} 
\label{ROutA} 
{}_\s \langle A^- (j;\st) | \equiv \lim_{\bz \rightarrow \infty} {}_\s \langle 0| \bz^{\Dg (1+\fji)} \hgjm 
\end{equation} 
\begin{equation} 
\label{ATotal} 
|A(j;\st )\rangle_\s = |A^- (j;\st )\rangle_\s \cdot |A^+ (j;\st )\rangle_\s ,\quad {}_\s \langle A(j;\st )|  
= {}_\s \langle A^- (j;\st )| \cdot {}_\s \langle A^+(j;\st )| . 
\end{equation} 
\end{subequations} 
Note that the {\it orbifold-modified asymptotic formulae} \eqref{Acreat} reduce to the conventional 
asymptotic formulae for Virasoro primary fields and states only when the cycle size $f_j (\s) =1$, which 
includes the untwisted sector $\s =0$. These orbifold modifications can be traced back to the $z_\m^{-1}$ 
terms in the single-cycle twisted KZ connections \eqref{FactKZ2}, and these terms in turn reflect the 
translation non-invariance of the scalar twist-field state. Such modified asymptotic formulae were first seen 
in Ref.~\cite{Borisov:1997nc}.

In what follows, we discuss some important properties of the states in \eqref{Acreat}. For brevity, we 
focus on the left-mover `in' state in \eqref{LInA}, but similar properties are easily established for 
all the states in \eqref{Acreat}. 
 
To simplify the following arguments, we will define rescaled affine primary fields 
\begin{equation} 
\label{RescaleG}
\hat{\hat{g}}_j^+ (\st (T,\s),z,\s) \equiv z^{\Dg (1-\fji)} \hat{g}_j^+ (\st (T,\s),z,\s) ,\quad \Ajop = 
\lim_{z \rightarrow 0} \hat{\hat{g}}_j^+ (\st,z,\s) |0 \rangle_\s 
\end{equation} 
and we observe that rescaling does not remove all orbifold modification of the asymptotic formulae; for example,
the formula
\begin{equation}
{}_\s \langle A^+ (j;\st)| = \lim_{z \rightarrow \infty} {}_\s \langle 0| z^{2\Dg /f_j(\s)} 
\hat{\hat{g}}_j^+ (\st,z,\s)
\end{equation}
agrees with the conventional form only when $f_j (\s)=1$. 

Our next observation is that the asymptotic state $\Ajop$ created by the left-mover twisted affine primary field is 
indeed a {\it twisted affine primary state} \footnote{Note that the scalar twist-field state $|0\rangle_\s$ can also 
be considered as a twisted affine primary state with $\st =0$.}: 
\begin{subequations} 
\begin{equation} 
[ \hj_{\hat{j}aj} \modmj ,\hat{\hat{g}}_l^+ (\st,z,\s)] = \de_{jl} z^{\modmj} \hat{\hat{g}}_j^+ (\st,z,\s) 
T_a \tau_{\hat{j}} (j,\s) 
\end{equation} 
\begin{eqnarray} 
\label{BaseAfPr} 
\hj_{\hat{j}aj} (m +\jfj \geq 0) |A^+ (l;\st) \rangle_\s &=& \lim_{z \rightarrow 0} \hj_{\hat{j}aj} 
(m+\jfj \geq 0) \hat{\hat{g}}_l^+ (\st,z,\s) |0 \rangle_\s \nn \\ 
&=& \lim_{z \rightarrow 0} [ \hj_{\hat{j}aj} (m+\jfj \geq 0) ,\hat{\hat{g}}_l^+ (\st,z,\s)]|0 \rangle_\s \nn \\ 
&=& \lim_{z \rightarrow 0} \de_{jl} \,z^{\modmj} \hat{\hat{g}}_j^+ (\st,z,\s) T_a \tau_{\hat{j}} (j,\s) 
|0 \rangle_\s \nn \\ 
&=& \de_{jl} \,\de_{m+ \jfj ,0} \Ajop T_a . 
\end{eqnarray} 
\end{subequations} 
To obtain this result, we used the ground state conditions \eqref{GSCond} and the algebra \eqref{CurAlgg}. 

Moreover, these twisted affine primary states are primary under the orbifold Virasoro algebra
\begin{subequations} 
\begin{eqnarray} 
\label{PPAlg} 
[ \hat{L}_{\hat{j}j} \modmj ,\hat{\hat{g}}_l^+ (\st,z,\s) ] = \bigspc \bigspc \bigspc \bigspc \bigspc \nn \\ 
=\de_{jl} \,\hat{\hat{g}}_j^+ (\st,z,\s) \left( \lpl z +(m+\frac{\hat{j}+1}{f_j(\s)})\Dg \right) 
\tau_{\hat{j}} (j,\s) z^{m+\jfj} 
\end{eqnarray} 
\begin{align} 
\label{BaseCW} 
& \hat{L}_{\hat{j}j} (m+\jfj \geq 0) |A^+ (l;\st) \rangle_\s = \lim_{z \rightarrow 0} \hat{L}_{\hat{j}j} 
(m+\jfj \geq 0) \hat{\hat{g}}_l^+ (\st,z,\s) |0 \rangle_\s = \nn \\ 
&\,\,\,= \lim_{z \rightarrow 0} \left( [ \hat{L}_{\hat{j}j} (m+\jfj \geq 0) ,\hat{\hat{g}}_l^+ (\st,z,\s)] + 
\de_{m+\jfj ,0} \gscfwtj \hat{\hat{g}}_l^+ (\st,z,\s) \right) |0 \rangle_\s \nn \\ 
&\,\,\,= \lim_{z \rightarrow 0} \hat{\hat{g}}_l^+ (\st,z,\s) \left( \de_{jl} (\lpl z + (m+\frac{\hat{j}+1}
{f_j(\s)}) \Dg) \tau_{\hat{j}} (j,\s) z^{m+\jfj} + \de_{m+\jfj ,0} \gscfwtj \right) |0 \rangle_\s \nn \\ 
&\,\,\,= \de_{m+\jfj ,0} (\gscfwtj + \de_{jl} \frac{\Dg}{f_j(\s)} ) |A^+ (l;\st) \rangle_\s \bigspc \bigspc 
\end{align} 
\end{subequations} 
where we have used the ground state relation \eqref{GSEqn} and the commutator \eqref{VirAlgg}.  The partial 
conformal weight $\gscfwtj$ is given in \eqref{GSCnfJ}. 
 
Finally, Eq.~\eqref{BaseCW} tells us that the twisted affine primary state is also a Virasoro primary  
state under the full Virasoro generators 
\begin{equation} 
\label{FullCW} 
L_\s (m \geq 0) \Ajop = \de_{m,0} (\gscfwt + \frac{\Dg}{f_j (\s)}) \Ajop  
\end{equation} 
where the ground state conformal weight $\gscfwt$ is given in \eqref{GSCnf}.

The corresponding relations on the left-mover `out' state 
\begin{subequations} 
\begin{equation} 
{}_\s \langle A^+ (l;\st)| \hj_{\hat{j}aj} (m+\jfj \leq 0) = \de_{jl} \,\de_{m+\jfj ,0} {}_\s \langle
A^+ (j;\st)| T_a 
\end{equation} 
\begin{equation} 
{}_\s \langle A^+ (l;\st)| \hat{L}_{\hat{j}j} (m+\jfj \leq 0)= \de_{m+\jfj ,0} (\gscfwtj +\de_{jl}  
\frac{\Dg}{f_j (\s)}) {}_\s \langle A^+ (l;\st)| 
\end{equation} 
\end{subequations} 
are the adjoint of the relations above. Similarly, the right-mover `in' state in \eqref{RInA} is primary under 
the rectified right-mover currents and the rectified right-mover orbifold Virasoro generators 
\begin{subequations} 
\begin{equation} 
\hjbb_{\hat{j}aj} (m+\jfj \geq 0) |A^- (j;\st) \rangle_\s = -\de_{jl} \,\de_{m+\jfj ,0} T_a |A^- (j;\st) 
\rangle_\s 
\end{equation} 
\begin{equation} 
\hlbb_{\hat{j}j} (m+\jfj \geq 0) |A^- (j;\st) \rangle_\s = \de_{m+\jfj ,0} (\gscfwtj +\de_{jl} \frac{\Dg} 
{f_j(\s)}) |A^- (j;\st) \rangle_\s  
\end{equation} 
\end{subequations} 
where we have used the corresponding right-mover ground state conditions \eqref{RGSCond} and right-mover 
relations (3.19d,e). The adjoint of this result is easily obtained for the right-mover `out' state 
\eqref{ROutA}. 
 
\subsection{The Principal Primary States of Twisted Rep $\st$ and Cycle $j$} 
 
In this subsection, we will use the twisted affine primary state $\Ajop$ to construct the following particular 
set of {\it principal primary states} 
\begin{subequations}
\label{VirPP}
\begin{equation} 
\label{PrPrDfn} 
|\hat{j},j;\st (T,\s) \rangle_\s ,\quad \hat{j} =0,\ldots ,f_j(\s)-1 ,\quad \forall T \text{ for each } \srange 
\end{equation} 
\begin{equation} 
\label{BaseS} 
| 0,j;\st \rangle_\s \equiv \Ajop 
\end{equation} 
\end{subequations}
which are examples of the principal primary states of Ref.~\cite{Borisov:1997nc}. More precisely, the states  
\eqref{PrPrDfn} are the principal primary states of twisted representation $\st$ in block 
$j$ of sector $\s$. A brief description of more general sets of principal primary states is
given in App.~B, including as another set of examples the principal primary states associated
to the twisted currents.

Up to constants of normalization $(\simeq )$, the set of principal primary states \eqref{VirPP} are defined 
as follows
\begin{equation}        
\label{PrPrDef} 
| \hat{j},j;\st \rangle_\s \simeq ( \hat{L}_{-1,j} (-\fji) )^{\hat{j}} |0,j;\st \rangle_\s ,\quad 
\hat{j}=0,\ldots, f_j(\s)-1 
\end{equation} 
where the twisted affine primary state $|0,j;\st \rangle_\s$ serves as the {\it base state} for the set.
Using the properties \eqref{BaseAfPr} and \eqref{BaseCW} of the base state, together with the orbifold  
Virasoro algebra \eqref{Lmodes}, we verify the following additional properties of the principal primary states: 
\begin{subequations} 
\begin{equation} 
\label{PPVirA} 
\hat{L}_{\hat{j}\p j} (m+\frac{\hat{j}\p}{f_j (\s)}) |\hat{j},j;\st \rangle_\s = 0 ,\quad \text{if } 
(m+\frac{\hat{j}\p}{f_j (\s)}) > \jfj 
\end{equation} 
\begin{equation} 
\label{PPVirB} 
\hat{L}_{0l} (m \geq 0) |\hat{j},j;\st \rangle_\s = \de_{m,0} (\hat{\Delta}_{0l} (\s) +\de_{jl} \frac{\Dg 
+\hat{j}}{f_j(\s)}) |\hat{j},j;\st \rangle_\s 
\end{equation} 
\begin{equation} 
L_\s (m \geq 0) |\hat{j},j;\st \rangle_\s = \de_{m,0} (\gscfwt +\frac{\Dg +\hat{j}}{f_j (\s)}) 
|\hat{j},j;\st \rangle_\s . 
\end{equation} 
\end{subequations} 
According to \eqref{PPVirB}, all these principal primary states are primary under the semisimple integral 
Virasoro subalgebra \eqref{ssIVsA}. On the other hand, the base state \eqref{BaseS} with $\hat{j}=0$ is the 
only principal primary state in the set which is both a) twisted affine primary and b) primary under the full 
orbifold Virasoro algebra. The other principal primary states (with $\hat{j} \neq 0$) are not primary under 
either the orbifold affine algebra or the orbifold Virasoro algebra. (The definition \eqref{PrPrDef} can be 
extended to the range $\hat{j} \geq f_j (\s)$, but such states are no longer primary under the integral 
Virasoro subalgebra). 
 
From the definition \eqref{PrPrDef}, we can derive the following {\it first asymptotic formula} for 
the principal primary states
\begin{eqnarray} 
\label{PPAsym} 
|\hat{j},j;\st \rangle_\s &\simeq& ( \hat{L}_{-1,j} (-\fji) )^{\hat{j}} |0,j;\st \rangle_\s \nn \\ 
&=& \lim_{z \rightarrow 0} (\hat{L}_{-1,j} (-\fji) )^{\hat{j}} \hat{\hat{g}}_j^+ (\st,z,\s) |0\rangle_\s \nn \\ 
&=& \lim_{z \rightarrow 0} [( \hat{L}_{-1,j} (-\fji) )^{\hat{j}}, \hat{\hat{g}}_j^+ (\st,z,\s)] 
|0 \rangle_\s \nn \\ 
&=& \lim_{z \rightarrow 0} (z^{1-\fji} \pl )^{\hat{j}} \hat{\hat{g}}_j^+ (\st,z,\s) \tau_{-\hat{j}} 
(j,\s) |0 \rangle_\s 
\end{eqnarray} 
where we have also used \eqref{PPAlg} and the ground state property \eqref{GSEqn}. When $\hat{j}=0$, this result 
reduces to the asymptotic formula \eqref{RescaleG} for the twisted affine primary state.

Combining the left-mover result in Eq.~\eqref{PPAsym} with the analogous right-mover result gives the
following non-chiral form 
\begin{subequations} 
\begin{equation} 
|0,0,j;\st \rangle_\s \equiv |A(j;\st) \rangle_\s =\left( |A^- (j;\st)\rangle_\s \cdot |A^+ (j;\st)
\rangle_\s \right) = \left( |0_- ,j;\st \rangle_\s \cdot |0,j;\st \rangle_\s \right) 
\end{equation} 
\begin{eqnarray} 
|\hat{j}_- ,\hat{j}_+ ,j;\st \rangle_\s \simeq (\hlbb_{-1,j} (-\fji) )^{\hat{j}_-} (\hat{L}_{-1,j} 
(-\fji))^{\hat{j}_+} |0,0,j;\st \rangle_\s \simeq |\hat{j}_- ,j;\st \rangle_\s \cdot |\hat{j}_+ ,j;\st 
\rangle_\s \nn \\ 
=\lim_{|z| \rightarrow 0} (\bz^{1-\fji} \bar{\pl} )^{\hat{j}_-} (z^{1-\fji} \pl )^{\hat{j}_+} |z|^{2\Dg 
(1-\fji)} \tau_{\hat{j}_-} (j,\s) \hgjz \tau_{-\hat{j}_+} (j,\s) |0 \rangle_\s \quad \quad 
\end{eqnarray} 
\end{subequations} 
where we have used Eqs.~\eqref{ATotal}, \eqref{BaseS} and the algebra \eqref{Algg}. 
 
\subsection{The Twisted Affine Primary Fields are Principal Primary Fields} 
 
It is explained in Ref.~\cite{Borisov:1997nc} that every set of principal primary states can be created by a 
corresponding set of fields called the {\it principal primary fields}. A brief description of
general principal primary fields is given in App.~B, where the twisted currents themselves are
discussed as a simple example. As a more intricate example, we will establish here that the {\it matrix 
components} (see \eqref{GDiffV}) of the twisted affine primary fields are themselves the principal primary 
fields for the set of principal primary states \eqref{VirPP}.

To begin this discussion, we note that it is consistent to require as a {\it boundary condition} that  
the matrix component $\hat{\hat{g}}_j^+ (\st,z,\s)_{\hat{j}} {}^{\hat{l}}$ with $\hat{j}=\hat{l}$ mod 
$f_j(\s)$ is the most singular\footnote{One can also require that some other matrix element of $\hat{\hat{g}}$  
is the most singular, but we expect that other choices of this type are equivalent to a relabelling of 
the principal primary fields below.} matrix element of $\hat{\hat{g}}_j^+$, and hence the one that  
creates the twisted affine primary or base state \eqref{BaseS}:
\begin{equation} 
\label{BaseId} 
\lim_{z \rightarrow 0} \hat{\hat{g}}_j^+ (\st,z,\s)_{\hat{j}}{}^{\hat{j}} |0 \rangle_\s =  
\lim_{z \rightarrow 0} \hat{\hat{g}}_j^+ (\st,z,\s)_0 {}^0 |0\rangle_\s = |0,j;\st \rangle_\s . 
\end{equation}   
As we shall see below, this specification corresponds to choosing a particular solution $\hat{g}_j^+$ 
of the twisted vertex operator equation or, equivalently, to fixing the multiplicative constant matrices 
(such as $\C (j;\st,\s)$ in \eqref{TwoPt} and \eqref{SCTwoPt}) in the solutions of the twisted KZ system. 
 
Given the boundary condition \eqref{BaseId}, we will demonstrate below that all the principal primary  
states \eqref{PrPrDef} can be obtained from the matrix elements of $\hat{\hat{g}}_j^+$ by the 
{\it second asymptotic formula} 
\begin{subequations} 
\label{Asymp} 
\begin{equation} 
|\hat{j},j;\st \rangle_\s = \lim_{z \rightarrow 0} z^{-\jfj} \hat{\hat{g}}_j^+ (\st,z,\s)_{0}
{}^{\hat{j}} |0 \rangle_\s ,\,\,\, \hat{j} =0,\ldots, f_j(\s)-1 ,\,\,\, \srange
\end{equation} 
\begin{equation} 
\hat{\hat{g}}_j^+ (\st,z,\s)_0 {}^{\hat{j}}=\hat{\hat{g}}_j^+ (\st,z,\s)_{\hat{l}} {}^{\hat{j}+\hat{l}} 
\end{equation} 
\end{subequations} 
which identifies the matrix components of the twisted affine primary field as {\it principal primary fields}
\cite{Borisov:1997nc}. To compare \eqref{Asymp} to the asymptotic formula given for principal primary fields 
in Eq.~$(3.15)$ of Ref.~\cite{Borisov:1997nc}, it is useful to make the following redefinition:
\begin{subequations}
\label{ChRedef}
\begin{equation} 
\hat{\phi}_{\Dg}^{(\hat{j}j)} (z) \equiv \hgjp_0 {}^{f_j (\s)-1-\hat{j}} ,\quad \hat{j}=0,\ldots ,f_j (\s)-1 
\end{equation} 
\begin{eqnarray}
[\hat{L}_{\hat{j}j} \modmj ,\hgjp_0 {}^{\hat{l}} ]= \bigspc \bigspc \bigspc \bigspc \nn \\
\hgjp_0 {}^{\hat{l}-\hat{j}} (\lpl z + (m+\jfj +1) \Dg )z^{m+\jfj}
\end{eqnarray}
\begin{equation}
[\hat{L}_{\hat{j}j} \modmj ,\hat{\phi}_{\Dg}^{(\hat{l}j)} (z)]=\hat{\phi}_{\Dg}^{(\hat{j}+
\hat{l},j)} (z) (\lpl z +(m+\jfj +1) \Dg )z^{m+\jfj} .
\end{equation}
\end{subequations}
This is essentially the notation
\begin{equation} 
\hat{\phi}_{\Delta}^{(r)} (z) = \hat{\phi}_{\Delta}^{(\hat{j}0)} (z) |_{\hat{j} =r} ,\quad r=0,\ldots ,\lambda -1 
\end{equation} 
given for the principal primary fields of the $\z_{\lambda}$ cyclic permutation orbifolds with $\lambda =$ 
prime in Ref.~\cite{Borisov:1997nc}. In fact, we chose an equality rather than a proportionality in \eqref{Asymp}  
in order to have an exact match with the asymptotic formula of Ref.~\cite{Borisov:1997nc}.  This choice sets 
the relative normalization of the twisted affine primary fields versus the principal primary states, and we will 
continue with this convention below. 
 
We turn finally to the proof of the second asymptotic formula \eqref{Asymp}: 
One may combine \eqref{PPAsym} with the boundary condition \eqref{BaseId} to obtain a matrix form of the 
first asymptotic formula 
\begin{eqnarray} 
\label{FtAsym} 
|\hat{j} ,j;\st \rangle_\s &\simeq& \lim_{z \rightarrow 0} \left( \hat{L}_{-1,j} (-\fji) \right)^{\hat{j}} 
\hat{\hat{g}}_j^+ (\st,z,\s)_0 {}^0 |0\rangle_\s \nn \\ 
&=& \lim_{z \rightarrow 0} (z^{1-\fji} \pl)^{\hat{j}} \,[\hat{\hat{g}}_j^+ (\st,z,\s) \tau_{-\hat{j}} (j,\s) 
]_0 {}^0 |0\rangle_\s \nn \\ 
&=& \lim_{z \rightarrow 0} (z^{1-\fji} \pl)^{\hat{j}} \,\hat{\hat{g}}_j^+ (\st,z,\s)_0 {}^{\hat{j}} |0\rangle_\s . 
\end{eqnarray} 
The solution of \eqref{FtAsym} is 
\begin{equation} 
\hat{\hat{g}}_j^+ (\st,z,\s)_{0}{}^{\hat{j}} |0 \rangle_\s = \Ord (z^{\jfj}) \quad \text{as }  
z \rightarrow 0 ,\quad \hat{j}=0,\ldots ,f_j(\s) -1 
\end{equation} 
and, up to constants of normalization, this is equivalent to the second asymptotic formula. 
 
There is an alternate proof of the second asymptotic formula which uses the twisted vertex operator  
equation for $\hat{\hat{g}}_j^+$ 
\begin{equation} 
\pl \hat{\hat{g}}_j^+ (\st,z,\s) = \frac{2\eta^{ab}}{2k +Q_{\gfraks}} \fji \sum_{\hat{j}=0}^{f_j(\s)-1} 
: \hj_{\hat{j}aj} (z) \hat{\hat{g}}_j^+ (\st,z,\s) :_M T_b \tau_{-\hat{j}} (j,\s)  
\end{equation} 
which follows from \eqref{RescaleG} and \eqref{RedVOE}. In particular, we need the most singular terms of this 
equation when acting on the ground state of sector $\s$: 
\begin{align} 
\label{hhgVOE} 
& \pl \hat{\hat{g}}_j^+ (\st,z,\s)_{\hat{j}}{}^{\hat{l}} |0 \rangle_\s = \frac{2\eta^{ab}}{2k +Q_{\gfraks}} \fji 
\sum_{\hat{m}=0}^{f_j (\s)-1} z^{-\frac{\hat{m}}{f_j (\s)}} \hj_{\hat{m},aj} (-1+\frac{\hat{m}}{f_j (\s)}) 
\hat{\hat{g}}_j^+ (\st,z,\s)_{\hat{j}}{}^{\hat{l}+\hat{m}} T_b |0 \rangle_\s \nn \\ 
& \bigspc \bigspc \bigspc \quad + \text{ less singular terms near } z=0 .  
\end{align} 
Assuming the boundary condition \eqref{BaseId}, a careful analysis of leading powers of $z$ in this equation leads to the 
same conclusion \eqref{Asymp}. As a bonus, Eq.~\eqref{hhgVOE} can also be used to establish that  
\begin{equation} 
\lim_{z \rightarrow 0} z\pl \hat{\hat{g}}_j^+ (\st,z,\s) |0\rangle_\s =0  
\end{equation} 
which was in fact assumed to obtain \eqref{BaseCW} and \eqref{FullCW}. 
 
Similarly, one may establish the second asymptotic formula for the right-mover principal primary states
\begin{subequations} 
\begin{equation} 
\hat{\hat{g}}_j^- (\st (T,\s),\bz,\s) \equiv \bz^{\Dg (1-\fji)} \hat{g}_j^- (\st (T,\s),\bz,\s) 
\end{equation} 
\begin{equation} 
|\hat{j}_- ,j;\st \rangle_\s = \lim_{\bz \rightarrow 0} \bz^{-\frac{\hat{j}_-}{f_j(\s)}} 
\hat{\hat{g}}_j^- (\st,\bz,\s)_{\hat{j}_-} {}^0 |0\rangle_\s ,\quad \hat{j}_- =0,\ldots ,f_j(\s)-1
\end{equation} 
\end{subequations} 
and corresponding results are easily obtained for the left- and right-mover `out' states. 
 
\subsection{Final Form of the Single-Cycle Two-Point Correlator} 
 
In \eqref{SCTPCorr} we presented the form of the single-cycle two-point correlator up to a multiplicative  
constant matrix $\D$. In this subsection, we use some of the information above about the twisted affine 
primary fields and states to determine this constant matrix. 
 
We begin by writing the following proportionalities 
\begin{subequations} 
\label{Diag} 
\begin{equation} 
\label{AfPrDiag} 
\lim_{z \rightarrow 0} z^{\Dg (1-\fji)} \hgjp_{\hat{j}}{}^{\hat{l}} |0\rangle_\s \propto \de_{\hat{j}-\hat{l},0\,
\text{mod } f_j(\s)} 
\end{equation} 
\begin{equation} 
\label{RAfPDiag} 
\lim_{\bz \rightarrow 0} \bz^{\Dg (1-\fji)} \hgjm_{\hat{j}} {}^{\hat{l}} |0\rangle_\s \propto  
\de_{\hat{j}-\hat{l},0\,\text{mod } f_j(\s)} 
\end{equation} 
\begin{equation} 
\lim_{z \rightarrow \infty} z^{\Dg (1+\fji)} {}_\s \langle 0| \hgjp_{\hat{j}}{}^{\hat{l}} \propto 
\de_{\hat{j}-\hat{l},0\,\text{mod } f_j(\s)} 
\end{equation} 
\begin{equation} 
\lim_{bz \rightarrow \infty} \bz^{\Dg (1+\fji)} {}_\s \langle 0| \hgjm_{\hat{j}}{}^{\hat{l}} \propto 
\de_{\hat{j}-\hat{l},0\,\text{mod } f_j(\s)} 
\end{equation} 
\end{subequations} 
which hold because the most singular matrix element of each operator is $\hat{j}=\hat{l}$ mod $f_j(\s)$. 
A more precise form of \eqref{Diag} is the following list of relations: 
\begin{subequations} 
\label{PrecDiag} 
\begin{equation} 
\label{NoBSBase} 
\lim_{z \rightarrow 0} z^{\Dg (1-\fji)} \hgjp_{\hat{j}} {}^{\hat{l}} |0\rangle_\s =(\Ajop )_{\hat{j}} {}^{\hat{l}} 
= \de_{\hat{j}-\hat{l},0\,\text{mod } f_j(\s)} |0,j;\st \rangle_\s 
\end{equation} 
\begin{equation} 
\lim_{\bz \rightarrow 0} \bz^{\Dg (1-\fji)} \hgjm_{\hat{j}} {}^{\hat{l}} |0\rangle_\s =(\Ajom )_{\hat{j}} {}^{\hat{l}} 
= \de_{\hat{j}-\hat{l},0\,\text{mod } f_j(\s)} |0_-,j;\st \rangle_\s 
\end{equation} 
\begin{equation} 
\lim_{z \rightarrow \infty} z^{\Dg (1+\fji)} {}_\s \langle 0| \hgjp_{\hat{j}} {}^{\hat{l}} = (_\s \langle 
A^+ (j;\st) |)_{\hat{j}}{}^{\hat{l}} = \de_{\hat{j}-\hat{l},0\,\text{mod } f_j(\s)} {}_\s \langle 0,j;\st | 
\end{equation} 
\begin{equation} 
\lim_{z \rightarrow \infty} z^{\Dg (1+\fji)} {}_\s \langle 0| \hgjm_{\hat{j}} {}^{\hat{l}} = (_\s \langle 
A^- (j;\st) |)_{\hat{j}}{}^{\hat{l}} = \de_{\hat{j}-\hat{l},0\,\text{mod } f_j(\s)} {}_\s \langle 0_-,j;\st | 
\end{equation} 
\begin{equation} 
|0,0,j;\st \rangle_\s = |0_- ,j;\st \rangle_\s \cdot |0,j;\st \rangle_\s ,\quad 
{}_\s \langle 0,0,j;\st |= {}_\s \langle 0_- ,j;\st | \cdot {}_\s \langle 0,j;\st | . 
\end{equation} 
\end{subequations} 
Note in particular that \eqref{NoBSBase} is the explicit form of Eq.~\eqref{BaseS}, now including all  
$\hat{j},\,\hat{l}$ matrix indices. Together, Eqs.~\eqref{PrecDiag} and \eqref{PrPrDef} tell us that the 
principal primary states $|\hat{j} ,j;\st \rangle_\s$ do not have $\hat{j} ,\,\hat{l}$ matrix indices 
(and similarly for their right-mover analogues).  
 
Abelian analogues of all the relations in \eqref{PrecDiag} can be obtained with the left- and right-mover 
twisted vertex operators of Ref.~\cite{Halpern:2002ab}. In particular, the left-mover relations 
\begin{subequations} 
\begin{equation} 
\lim_{z \rightarrow 0} z^{\Delta (T) (1-\fji)} \hgjp |0 \rangle_\s = \hat{\Gamma}_j (T\tau_0 (j,\s), \hj_{0j} (0),\s) 
| \hj_{0aj} (0) = T_a \rangle_\s 
\end{equation} 
\begin{equation} 
\hat{\Gamma}_j (T\tau_0 (j,\s),\hj_{0j} (0),\s)_{\hat{j}}{}^{\hat{l}} \propto \de_{\hat{j}-\hat{l} ,0\, \text{mod } f_j(\s)} ,\quad 
| \hj_{0aj} (0) =T_a \rangle_\s \equiv e^{i \hat{q}^{aj} (\s) T_a} |0\rangle_\s 
\end{equation} 
\end{subequations} 
are obtained from \eqref{GJGamma}. This tells us that the natural assumption \eqref{GammaJ} is the abelian 
analogue of our boundary condition \eqref{BaseId}. 
 
We now apply the asymptotic relations \eqref{PrecDiag} to the single-cycle two-point correlator 
\eqref{SCTPCorr}. In particular, these relations and the limit \eqref{CAsymp} give us further 
information about the multiplicative constant matrix $\C (j;\st,\s)$:
\begin{align} 
\label{CFormOne} 
&\lim_{|z_1 | \rightarrow \infty} \lim_{|z_2 |\rightarrow 0} |z_1 |^{2\Dg (1+\fji)} |z_2 |^{2\Dg (1-\fji)}  
{}_\s \langle 0| \hat{g}_j (\st^{(1)} ,\bz_1,z_1,\s)_{\hat{j}_1} {}^{\hat{l}_1} \hat{g}_j (\st^{(2)}  
,\bz_2,z_2,\s)_{\hat{j}_2} {}^{\hat{l}_2} |0\rangle_\s \nn \\ 
&\quad =\C (j;\st,\s)_{\hat{j}_1 ;\hat{j}_2} {}^{\hat{l}_1 ;\hat{l}_2} = \de_{\hat{j}_1-\hat{l}_1,0\,\text{mod }  
f_j(\s)} \de_{\hat{j}_2-\hat{l}_2,0\,\text{mod } f_j(\s)} \,{}_\s \langle 0,0,j;\st^{(1)}  
|0,0,j;\st^{(2)} \rangle_\s .
\end{align} 
Combining this relation with the form of $\C (j;\st,\s)$ in \eqref{CForm}, we find that: 
\begin{subequations} 
\begin{equation} 
\label{OrthoN} 
(_\s \langle 0,0,j;\st |)_{\a_1}{}^{\be_1} (|0,0,j;\st \rangle_\s )_{\a_2}{}^{\be_2} = c\, \de_{T^{(2)} ,\bar{T}^{(1)}}
\de_{\a_1 \a_2} \de^{\be_1 \be_2}  
\end{equation} 
\begin{equation} 
\D (j;\st,\s)_{\hat{j}_1 ;\hat{j}_2}{}^{\hat{l}_1 ;\hat{l}_2} = c\, \de_{\hat{j}_1-\hat{l}_1,0\,\text{mod }  
f_j(\s)} \de_{\hat{j}_2-\hat{l}_2,0\,\text{mod } f_j(\s)} 
\end{equation} 
\begin{equation} 
\C (j;\st,\s)_{\a_1 \hat{j}_1 ;\a_2 \hat{j}_2}{}^{\be_1 \hat{l}_1 ;\be_2 \hat{l}_2} = c\, \de_{T^{(2)} ,\bar{T}^{(1)}} 
\de_{\a_1 ,\a_2} \de^{\be_1 ,\be_2} \de_{\hat{j}_1-\hat{l}_1,0\,\text{mod } f_j(\s)}  
\de_{\hat{j}_2-\hat{l}_2,0\,\text{mod } f_j(\s)} . 
\end{equation} 
\end{subequations} 
We will also choose the overall normalization constant $c=1$, which gives the final, completely-determined 
form of the {\it single-cycle two-point correlator}: 
\begin{subequations}
\label{FinTwoPt} 
\begin{align} 
&\hat{A} (j;1,2) \equiv \langle \hat{g}_j (\st^{(1)} \bz_1,z_1,\s) \hat{g}_j (\st^{(2)} ,\bz_2,z_2,\s) \rangle_\s \nn \\ 
& \quad \quad = \de_{T^{(2)}, \bar{T}^{(1)}} |z_1 z_2 |^{-2\Dg (1-\fji)} |z_{12} |^{-4\Dg /f_j (\s)} \exp \{\frac{2}
{2k +Q_{\gfraks}} F(j;1,2) \}  
\end{align} 
\begin{equation}
\forall j ,\,\, \forall T^{(1)}, T^{(2)} \text{ for each } \srange .
\end{equation}
\end{subequations}
Here $\Dg$ is the untwisted conformal weight in \eqref{TPCnfWt} and $F(j;1,2)$ is defined in \eqref{TwoPCorr}. 
Our relative normalization \eqref{Asymp} and our orthonormality condition (\eqref{OrthoN} with $c\!=\!1$) 
\begin{equation} 
\label{AfPNorm} 
{}_\s \langle 0,0,j;\st^{(1)} |0,0,j;\st^{(2)} \rangle_\s = \de_{T^{(2)} ,\bar{T}^{(1)}} 
\end{equation} 
can be used to compute the constants in all the proportionalities $(\simeq)$ above. 
 
We remark that the single-cycle two-point correlator \eqref{FinTwoPt} is symmetric under $1 \leftrightarrow 2$ 
exchange 
\begin{equation} 
\hat{A} (j;1,2) = \hat{A} (j;2,1) 
\end{equation} 
as expected. To establish this one needs to verify the symmetry relation 
\begin{equation} 
F(j;1,2) = F(j;2,1) 
\end{equation} 
which follows by steps that are essentially identical to those given in Eq.~$(9.32)$ of Ref.~\cite{deBoer:2001nw} and 
App.~C of Ref.~\cite{Halpern:2002ab}.
 
Finally, we record the completely-determined form of the original {\it unreduced} two-point correlator 
\begin{align} 
& \langle \hat{g} (\st^{(1)} ,\bz_1,z_1,\s)_{\a_1 \hat{j}_1 j_1}{}^{\be_1 \hat{l}_1 l_1} \hat{g} (\st^{(2)}, 
\bz_2,z_2,\s)_{\a_2 \hat{j}_2 j_2}{}^{\be_2 \hat{l}_2 l_2} \rangle_\s = ( \de_{j_1}^{l_1} \de_{j_2}^{l_2}) 
\de_{j_1 j_2} (\de_{T^{(2)} ,\bar{T}^{(1)}} \de_{\a_1 \a_2} \de^{\be_1 \be_2}) \times \nn \\ 
&\quad \times |z_1 z_2 |^{-2\Dg (1-\fji)} |z_{12} |^{-4\Dg /f_j(\s)} \exp \{ \frac{2}{2k+Q_{\gfraks}} F(j;1,2) 
\}_{\hat{j}_1 ;\hat{j}_2} {}^{\hat{l}_1 ;\hat{l}_2} 
\end{align} 
which follows immediately from \eqref{FacTwoPt} and \eqref{FinTwoPt}.

\section{General WZW Orbifolds} 
  
In this section, we consider the following topics

$\bullet$ asymptotic formulae

$\bullet$ conformal weights

$\bullet$ reducibility

\noindent for the twisted affine primary fields and states of general WZW orbifolds.
 
We begin by reminding the reader that twisted KZ equations are known \cite{deBoer:2001nw,Halpern:2002ab} 
for the correlators in the scalar twist-field state\footnote{For the inner-automorphic WZW orbifolds,
a different set of twisted KZ equations \cite{deBoer:2001nw} was given for the correlators in the untwisted affine vacuum state.}
\begin{equation} 
\hj_{\nrm} (m+\frac{n(r)}{\r(\s)} \geq 0) |0\rangle_\s = {}_\s \langle 0| \hj_{\nrm} (m+\frac{n(r)}{\r(\s)}  
\leq 0) =0 
\end{equation} 
in each sector of any WZW orbifold, where $\hj_\nrm (\mnrrs)$ are the modes of the twisted current algebra 
\cite{deBoer:1999na, Halpern:2000vj, deBoer:2001nw} 
of that sector. Such scalar twist-field states exist for all sectors of all current-algebraic orbifolds. The reason is easy to 
understand in the equivalent form
\begin{equation}
\hj_\nrm (m +\nrrs >0) |0\rangle_\s =0 ,\quad \hj_{0\m} (0)|0\rangle_\s =0 \,.
\end{equation}
The first condition holds for any primary state of any (infinite-dimensional) Lie algebra and the second condition restricts our
attention to the ``s-wave'' or trivial representation of the untwisted residual symmetry algebra of the sector. 
 
In further detail, the general twisted KZ system which describes the correlators of the twisted left-mover 
affine primary fields $\hat{g}_+ (\st,z,\s)$ in the scalar twist-field states reads as follows \cite{deBoer:2001nw,Halpern:2002ab}: 
\begin{subequations} 
\label{tlmkzeq} 
\begin{equation} 
\hat A_+ (\T,z,\s) \equiv {}_\s\langle 0|  \hgp(\T^{(1)},z_1,\s) \hgp(\T^{(2)},z_2,\s) 
\cdots \hgp(\T^{(N)},z_N,\s) |0 \rangle_\s 
\end{equation} 
\begin{equation} 
\part_\kappa \hat A_+(\T,z,\s) = \hat A_+ (\T,z,\s) \hat W_\kappa (\T,z,\s) ,\quad \kappa = 1 \ldots N \sp \s = 0, \ldots ,N_c -1 
\end{equation} 
\begin{equation} 
\label{kzct2} 
\hat W_{\kappa}(\T,z,\s) = 2 \lr^{n(r)\m;-n(r),\n} (\s) 
 \left[ \sum_{\r \neq \k}\left( \frac{z_\r}{z_\k} 
\right)^{ \srac{\bar n(r)}{\r(\s)}} \frac{1}{z_{\k \r }} 
\T_{n(r)\m}^{(\r)} \T_{-n(r),\n}^{(\k)} 
- \srac{\bar n(r)}{\r(\s)} \frac{1}{z_{\k}} 
\T_{n(r)\m}^{(\k)} \T_{-n(r),\n}^{(\k)} \right] 
\end{equation} 
\begin{equation} 
\label{gwig} 
\hat A_+ (\T,z,\s) \left(\sum_{\r =1}^N \T_{0 \m}^{(\r)}\right) = 0 \sp \forall \;\m . 
\end{equation} 
\end{subequations} 
Here $\r(\s)$ is the order of $h_\s \in H$, $n(r)$ is determined from the appropriate $H$-eigenvalue 
problem, and $\bar{n}(r) \in \{ 0,\ldots,\r(\s)-1 \}$ is the pullback of $n(r)$ to the fundamental range. 
General formulae for the twisted inverse inertia tensor ${\cL}_{\sgb (\s)} (\s)$ and the twisted representation 
matrices $\st_{\nrm} (T,\s)$ are given in Refs.~\cite{deBoer:2001nw,Halpern:2002ab}.  For the special case of
the WZW permutation orbifolds the general twisted KZ system \eqref{tlmkzeq} reduces 
\begin{equation}
\nrm \,\,\,\rightarrow \,\,\,n(r)aj \,\,\,\rightarrow \,\,\,\hat{j}aj
\end{equation}
to the twisted KZ system \eqref{twistKZ}, and the explicit form of the system \eqref{tlmkzeq} is also given for the 
(outer-automorphic) charge conjugation orbifold on $\su(n \geq 3)$ in Ref.~\cite{Halpern:2002ab}. 

Using the general twisted KZ system \eqref{tlmkzeq}, and in particular the last term of \eqref{kzct2}, we find the 
following asymptotic formulae for the twisted affine primary fields
\begin{subequations} 
\label{GenAfPr} 
\begin{equation} 
\hat{\hat{g}}_+ (\st,z,\s) \equiv \hat{g}_+ (\st,z,\s) z^{\gamma (\st,\s)} 
\end{equation} 
\begin{equation} 
\gamma (\st,\s) \equiv 2 {\cL}_{\sgb (\s)}^{\nrm ;\mnrn} (\s) \frac{\bar{n}(r)}{\r(\s)} \st_{\nrm} (T,\s) \st_{\mnrn} (T,\s) 
\end{equation} 
\begin{equation} 
|A^+ (\st) \rangle_\s \equiv \lim_{z \rightarrow 0} \hat{\hat{g}}_+ (\st,z,\s) |0 \rangle_\s 
\end{equation} 
\end{subequations}
where $\gamma (\st;\s)$ is called the {\it matrix exponent} of the twisted affine primary field $\hggp$. We also
find as expected that the states $|A^+ (\st) \rangle_\s$ are twisted affine primary states:
\begin{equation} 
\hj_{\nrm} (m+ \frac{n(r)}{\r(\s)} \geq 0) |A^+ (\st) \rangle_\s = \de_{m+\frac{n(r)}{\r(\s)} ,0} 
|A^+ (\st) \rangle_\s \st_{\nrm} (T,\s) .
\end{equation} 
This relation follows from the commutation relations \cite{deBoer:2001nw,Halpern:2002ab} of the general twisted 
current modes $\hj_{\nrm} (\mnrrs)$ with the twisted affine primary field.  

Moreover, the twisted affine primary state is Virasoro primary under the full Virasoro generators $L_\s (m)$ of sector 
$\s$, and we can evaluate the {\it total conformal weight matrix} $\hat{\Delta} (\st,\s)$ of the twisted affine primary 
state as follows: 
\begin{subequations} 
\begin{equation} 
L_\s (m \geq 0) |A^+ (\st) \rangle_\s = \de_{m,0} |A^+ (\st) \rangle_\s \hat{\Delta} (\st,\s) 
\end{equation} 
\begin{equation} 
\hat{\Delta} (\st,\s) \equiv \gscfwt \one + \D_{\sgb (\s)} (\st) - \gamma (\st,\s) 
\end{equation} 
\begin{equation} 
\gscfwt \equiv {\cL}_{\sgb (\s)}^{\nrm ;\mnrn} (\s) \frac{\bar{n}(r)}{2\r(\s)} (1- \frac{\bar{n}(r)}{\r(\s)}) 
\sG_{\nrm ;\mnrn} (\s) 
\end{equation} 
\begin{equation} 
\D_{\sgb (\s)} (\st) \equiv {\cL}_{\sgb (\s)}^{\nrm ;\mnrn} (\s) \st_{\nrm} (T,\s) \st_{\mnrn} (T,\s) .
\end{equation} 
\end{subequations} 
Here $\gscfwt$ and $\D_{\sgb (\s)} (\st)$ are respectively the conformal weight of the scalar twist-field
state and the twisted conformal weight matrix, which appears in the $\hat{T}_\s \hat{g}_+$ OPE \cite{deBoer:2001nw}:
\begin{equation}
\hat{T}_\s (z) \hat{g}_+ (\st,w,\s) = \hat{g}_+ (\st,w,\s) \left( \frac{\D_{\sgb (\s)} (\st)}{(z-w)^2}
+ \frac{\lplw}{z-w} \right) +\Ord (z-w)^0 .
\end{equation}
For the WZW permutation orbifolds, the explicit forms of $\gscfwt$ and $\D_{\sgb (\s)} (\st)$ are given in 
Eqs.~\eqref{GSCnf} and \eqref{TwCfWt}. 
 
With these tools in hand, we turn next to the question of the reducibility of general twisted affine primary states
and fields. 
 
In the case of the WZW permutation orbifolds we have shown above that the twisted affine primary fields $\hat{g}_+$ are  
generically reducible into blocks $\hat{g}_j^+$, and hence the unreduced fields create generically reducible states. 
As a simpler analogue, we note first for the abelian permutation orbifolds that
\begin{subequations} 
\begin{equation} 
\gamma (\st (T,\s),\s) = \Delta (T) \sum_j (1-\fji) t_{0j} (\s) 
\end{equation} 
\begin{equation} 
|A^+ (\st) \rangle_\s \equiv \lim_{z \rightarrow 0} z^{\Delta (T) \sum_j (1-\fji) t_{0j} (\s)}  
\hat{g}_+ (\st,z,\s) |0 \rangle_\s = \hat{\Gamma} (\st, \hj_{0} (0),\s) \{ \otimes_j |\hj_{0aj} (0)=T_a \rangle_\s \} 
\end{equation} 
\end{subequations} 
where $\hat{g}_+ (\st,z,\s)$ is the twisted left-mover vertex operator \cite{Halpern:2002ab} given in \eqref{TwVerOp}. 
The results for the unreduced twisted affine primary fields of the WZW permutation orbifolds are quite similar
\begin{subequations} 
\begin{equation} 
\gamma (\st (T,\s),\s) = \Dg \sum_j (1-\fji) t_{0j} (\s) 
\end{equation} 
\begin{equation} 
|A^+ (\st) \rangle_\s \equiv \lim_{z \rightarrow 0} z^{\Dg \sum_j (1-\fji) t_{0j} (\s)} \hat{g}_+ (\st,z,\s) |0\rangle_\s = \otimes_j \Ajop 
\end{equation} 
\end{subequations} 
where $\Ajop$ is the base state \eqref{BaseS} of twisted representation $\st$ in block $j$ of sector $\s$.  In this 
case we have computed the total conformal weight matrix 
\begin{subequations} 
\begin{equation} 
\gscfwt = \frac{\cg}{24} (K - \sum_j \fji ) ,\quad \D_{\sgb (\s)} (\st (T,\s)) = \Dg \one  
\end{equation} 
\begin{equation} 
\hat{\Delta} (\st (T,\s),\s) = \gscfwt \one + \Dg \sum_j \frac{t_{0j} (\s)}{f_j (\s)} 
\end{equation} 
\begin{equation} 
\hat{\Delta} (\st(T,\s),\s)_{\a \hat{j}j}{}^{\be \hat{l}l} = \de_j^l \,\de_{\a}^{\be} \,\de_{\hat{j}-\hat{l}  
,0\, \text{mod } f_j(\s)} (\gscfwt +\frac{\Dg}{f_j(\s)}) 
\end{equation} 
\end{subequations} 
which shows reducibility with respect to $j$, as expected. 

As another set of examples, consider the single twisted sector $\s =1$ of the charge conjugation orbifold 
\cite{Halpern:2002ab} on $\su (n \geq 3)$.  Ref.~\cite{Halpern:2002ab} explains that charge conjugation has a 
simple diagonal action in the standard Cartesian basis, with the irregularly-embedded $\so(n)$ subalgebra 
\begin{equation} 
\so (n)_{2\tau x} \subset \su (n)_x ,\quad x=\frac{2k}{\psi_{\su(n)}^2} ,\quad \tau = \left\{ 
\begin{array}{ll} 
2 & {\rm for}\,\; n =3 \\ 
1 & {\rm for}\,\; n \geq 4  \  
\end{array} \right.
\end{equation} 
as an invariant subalgebra. Moreover, we read from this reference that 
\begin{subequations} 
\begin{equation} 
\hat{\Delta}_0 = \frac{(n-1)(n+2)x}{32 (x+n)} 
\end{equation} 
\begin{equation} 
\gamma (\st) = \frac{1}{2k +Q_{\gfraks}} \sum_I \st_{1I} (T) \st_{1I} (T) 
\end{equation} 
\begin{equation} 
( \D_{\sgb (\s)} (\st) - \gamma (\st) ) = \frac{1}{2k +Q_{\gfraks}} \sum_A \st_{0A} (T) \st_{0A} (T) 
\end{equation} 
\begin{equation} 
A \in \so(n) ,\quad I \in \frac{\su (n)}{\so(n)} 
\end{equation} 
\end{subequations} 
for any twisted representation $\st (T)$.   
 
Further details depend on whether the untwisted representation $T$ is complex or real. In the case of the
(complex) fundamental representation $T=T^{(n)}$ of $\su(n)$, the matrix exponent and the total conformal weight  
matrix are 
\begin{subequations} 
\begin{equation} 
\gamma (\st (T^{(n)})) = \frac{(n-1)(n+2)}{4 n(x+n)} \left( \begin{array}{cc} 
1 & 0 \\ 0 & 1 \end{array} \right) \one_n  
\end{equation}  
\begin{equation} 
\label{TCfWtn} 
\hat{\Delta} (\st (T^{(n)})) = \frac{n-1}{4 (x+n)} (\frac{(n+2)x}{8} +1) \left( \begin{array}{cc} 
1 & 0 \\ 0 & 1 \end{array} \right) \one_n . 
\end{equation} 
\end{subequations} 
These matrices show a reducibility in the 2x2 space, which has in fact already been discussed \cite{Halpern:2002ab}
for all complex representations.  Next, we have computed the matrix elements of the matrices $\gamma$ and 
$\hat{\Delta}$ in the case of the (real) adjoint representation $T^{\text{adj}}$ of $\su(n)$:  
\begin{subequations} 
\begin{equation} 
\gamma (\st (T^{\text{adj}}))_{AB} = \de_{AB} \frac{n+2}{2(x+n)} ,\quad \gamma (\st (T^{\text{adj}}))_{IJ} =\de_{IJ} \frac{n}{2(x+n)} 
\end{equation} 
\begin{equation} 
\gamma (\st (T^{\text{adj}}))_{AI} = \gamma (\st (T^{\text{adj}}))_{IA} =0 
\end{equation} 
\begin{equation} 
\hat{\Delta} (\st (T^{\text{adj}}))_{AB} = \frac{\de_{AB}}{x+n} \left( \frac{(n-1)(n+2) x}{32} +\frac{n-2}{2} 
\right) 
\end{equation} 
\begin{equation} 
\hat{\Delta} (\st (T^{\text{adj}}))_{IJ} = \frac{\de_{IJ}}{x+n} \left( \frac{(n-1)(n+2) x}{32} +\frac{n}{2} 
\right)  
\end{equation} 
\begin{equation} 
\hat{\Delta} (\st (T^{\text{adj}}))_{AI} = \hat{\Delta} (\st (T^{\text{adj}}))_{IA} =0 . 
\end{equation} 
\end{subequations} 
These matrices show reducibility of the adjoint of $\su (n)$ into representations of $\so (n)$ 
\begin{equation} 
(n^2 -1) = (\frac{n(n-1)}{2} ) \oplus (\frac{n(n+1)}{2} -1) 
\end{equation} 
which are identified in this case as an adjoint of $\so(n)$ plus a second rank symmetric traceless 
$\so(n)$ tensor. No such $\su(n) \rightarrow \so(n)$ splitting is observed in the total conformal weight 
matrix \eqref{TCfWtn} associated to the fundamental representation because, in this situation, both the $n$  
and the $\bar{n}$ of $\su(n)$ are interpreted as $n$'s of $\so(n)$.  
 
Using Eq.~\eqref{tlmkzeq} and the data of Refs.~\cite{Halpern:2000vj,deBoer:2001nw}, one finds the results for the 
inner-automorphic WZW orbifolds
\begin{subequations}
\begin{align}
& \hat{W}_{\kappa} (\st ,z,\s)\! = \!\frac{2}{2k +Q_{\gfraks}} \left[ \sum_{\r \neq \kappa} \frac{1}{z_{\kappa \r}} 
\!\left( \sum_A T_A^{(\r)} T_A^{(\kappa)}\! + \!\sum_\a \left( \srac{z_\r}{z_\kappa} \right)^{\frac{\bar{n}_\a}{\r(\s)}} 
T_\a^{(\r)} T^{(\kappa)}_{-\a} \right) \!-\! \frac{1}{z_\kappa} \sum_\a \srac{\bar{n}_\a}{\r(\s)} T_\a^{(\kappa)} 
T^{(\kappa)}_{-\a} \right]
\end{align}
\begin{equation}
\hat{A}_+ (\st ,z,\s) \left( \sum_{\kappa =1}^N T_A^{(\kappa)} \right) =0 ,\quad \frac{n_\a}{\r(\s)} = -\s \a \cdot d ,\quad
\frac{\bar{n}_\a}{\r(\s)} = - \s \a \cdot d - \lfloor -\s \a \cdot d \rfloor
\end{equation}
\end{subequations}
\begin{subequations}
\begin{equation}
\gscfwt = \frac{x/4}{x +\hg} \sum_\a \frac{\bar{n}_\a}{\r(\s)} ( 1- \frac{\bar{n}_\a}{\r(\s)}) ,\quad \D_{\sgb (\s)} (\st (T,\s)) 
= \Dg \one
\end{equation}
\begin{equation}
\gamma (\st (T,\s),\s) = \frac{2}{2k +Q_{\gfraks}} \sum_\a \frac{\bar{n}_\a}{\r(\s)} T_\a T_{-\a}
\end{equation}
\end{subequations}
where $d$ is the shift vector \cite{Halpern:1988fr}, $(T_A ,T_\a )$ is irrep $T$ of $g$ in the Cartan-Weyl basis, and $\lfloor x
\rfloor$ is the floor of $x$. These twisted affine primary fields are also seen to be reducible due to symmetry breaking, as expected.

On the basis of these examples, one expects that the twisted affine primary fields of general WZW orbifolds 
are generically reducible. We remind the reader that there are special cases in which the twisted affine primary fields are  
irreducible, including all the twisted affine primary fields of the single-cycle sectors of the WZW  
permutation orbifolds.  
 
\vskip 1cm 
\noindent {\bf Acknowledgements} 
 
We wish to thank J.~Evslin, N.~Obers and C.~Schweigert for helpful discussions. 
The work of MBH was supported in part by the Director, Office of Energy Research, 
Office of High Energy and Nuclear Physics, Division of High Energy Physics of 
the U.S. Department of Energy under Contract DE-AC03-76SF00098 and in part by 
the National Science Foundation under grant PHY00-98840. 
  
\appendix
  
\section{About the Orbifold $\sl (2)$ Ward Identities} 
 
As stated in the text, the Ward identities \eqref{NewSLWI} associated to the centrally-extended 
$\sl (2)$ algebra are in fact implied by the twisted KZ system \eqref{twistKZ}. The proof of this 
statement is non-trivial, and the reader may find helpful the following hints. 
 
We begin with the $\hat{L}_{0j} (0)$ Ward identity in \eqref{NewWI0}, for which we list the intermediate 
steps:
\begin{subequations} 
\label{WrdHlp0} 
\begin{equation} 
2\eta^{ab} \sum_{{\m, \n \atop \m \neq \n}} \sum_{\hat{j} =1}^{f_j (\s)-1} 
\left( \frac{z_\n}{z_\m} \right)^{\frac{\hat{j}}{f_j(\s)}} \frac{z_\m}{z_{\m \n}} T_a^{(\n)} T_b^{(\m)} 
t_{\hat{j}j}^{(\n)} (\s) t_{-\hat{j},j}^{(\m)}(\s) t_{0j}^{(\m)} (\s) = 0 
\end{equation} 
\begin{equation} 
2\eta^{ab} \sum_{{\m, \n \atop \m \neq \n}} \frac{z_\m}{z_{\m \n}} T_a^{(\n)} T_b^{(\m)} t_{0j}^{(\n)} (\s) t_{0j}^{(\m)}(\s)  
t_{0j}^{(\m)} (\s) = \eta^{ab} \sum_{{\m, \n \atop \m \neq \n}} T_a^{(\n)} T_b^{(\m)} t_{0j}^{(\n)}  
(\s) t_{0j}^{(\m)}(\s) 
\end{equation} 
\begin{align} 
\label{R0Step3} 
& \hat{A}_+ (\s) \sum_{\m} \left( \lplm z_\m + \dg (T^{(\m)}) \right) t_{0j}^{(\m)}(\s) = \nn \\ 
& \bigspc \bigspc =\hat{A}_+ (\s) \sum_{\m} (z_\m \hat{W}_\m (\s) + \dg (T^{(\m)}) )t_{0j}^{(\m)}(\s)  \nn \\ 
&\bigspc \bigspc \propto \eta^{ab} \hat{A}_+ (\s) \, (\sum_\n T_a^{(\n)} t_{0j}^{(\n)} (\s)) \, (\sum_\m  
T_b^{\m} t_{0j}^{(\n)} (\s)) \\ 
\label{Result0} 
&\bigspc \bigspc =0 . 
\end{align} 
\end{subequations} 
The first relation in \eqref{WrdHlp0} holds for each $(\m \n) + (\n \m)$ pair separately, by the variable change  
$\hat{j} \rightarrow f_j(\s)-\hat{j}$ in the second term.  Eqs.~(A.1a,b) and the form of the twisted 
KZ connection in \eqref{twistKZ} are used to obtain \eqref{R0Step3}, and then \eqref{Result0} follows 
from the global Ward identity \eqref{GWI1}. The reader will note that the steps given for this case are quite similar 
to those followed in the proof of the $L_\s (0)$ Ward identity in Ref.~\cite{deBoer:2001nw}.
 
Similarly, we can use the form of the twisted KZ connection to simplify the $\hat{L}_{1j} (\fji)$ Ward identity  
\eqref{NewWIP}, as follows: 
\begin{subequations} 
\label{WdHlpP} 
\begin{equation} 
2\eta^{ab} \sum_{{\m, \n \atop \m \neq \n}} \sum_{\hat{j} =2}^{f_j (\s)-1} 
\left( \frac{z_\n}{z_\m} \right)^{\frac{\hat{j}}{f_j(\s)}} \frac{{z_{\m}}^{(1 + \frac{1}{f_j(\s)})}}  
{z_{\m \n}} T_a^{(\n)} T_b^{(\m)} t_{\hat{j}j}^{(\n)} (\s) t_{-\hat{j},j}^{(\m)}(\s) t_{1j}^{(\m)} (\s) = 0 
\end{equation} 
\begin{align} 
&\eta^{ab} \sum_{{\m, \n \atop \m \neq \n}} [ \frac{{z_\m}^{(1 + \fji)}}{z_{\m \n}} T_a^{(\n)}  
T_b^{(\m)} t_{0j}^{(\n)} (\s) t_{1j}^{(\m)} (\s) +\left( \frac{z_\n}{z_\m} \right)^{\fji} 
\frac{{z_\m}^{(1 + \fji)}}{z_{\m \n}} T_a^{(\n)} T_b^{(\m)} t_{1j}^{(\n)} (\s) t_{0j}^{(\m)} (\s) ] \nn \\ 
&\bigspc \bigspc \quad =\eta^{ab} \sum_{{\m, \n \atop \m \neq \n}} {z_\m}^{\fji} T_a^{(\n)} T_b^{(\m)} t_{0j}^{(\n)} (\s) t_{1j}^{(\m)} (\s)  
\end{align} 
\begin{align} 
\label{RPStep3} 
&\hat{A}_+ (\s) \sum_{\m} (\lplm z_\m + (1+\frac{1}{f_j(\s)}) \dg (T^{(\m)}) )t_{1j}^{(\m)}(\s) z_{\m}^{\frac{1}{f_j(\s)}} \nn \\ 
&\quad = \hat{A}_+ (\s) \sum_{\m} (z_\m \hat{W}_\m (\s)+ (1+\frac{1}{f_j(\s)}) \dg (T^{(\m)}) )t_{1j}^{(\m)}(\s) 
z_{\m}^{\frac{1}{f_j(\s)}} \nn \\ 
&\quad= \hat{A}_+ (\s) \left( \sum_\n T_a^{(\n)} t_{0j}^{(\n)} (\s) \right) \frac{2\eta^{ab}}{2k + Q_{\gfraks}} \frac{1}{f_j(\s)}  
\left( \sum_\m z_{\m}^{\frac{1}{f_j(\s)}} T_b^{(\m)} t_{1j}^{(\m)}(\s) \right)  \\ 
\label{ResultP} 
&\quad = 0 .  
\end{align} 
\end{subequations} 
Again, the first equation in \eqref{WdHlpP} holds for each $(\m \n) + (\n \m)$ pair separately, by taking  
$\hat{j} \rightarrow (f_j(\s)+1 - \hat{j})$ in the second term. Eq.~\eqref{RPStep3} follows from  
(A.2a,b), and the result \eqref{ResultP} then follows from the global Ward identity. 
 
Finally for the $\hat{L}_{-1,j} (-\fji)$ Ward identity in \eqref{NewWIM}, follow the steps: 
\begin{subequations} 
\label{WrdHpM} 
\begin{equation} 
\label{RMStep1} 
2\eta^{ab}\sum_{{\m, \n \atop \m \neq \n}} \sum_{\hat{j} =0}^{f_j (\s)-1} 
\left( \frac{z_\n}{z_\m} \right)^{\frac{\hat{j}}{f_j(\s)}} \frac{{z_{\m}}^{(1 - \frac{1}{f_j(\s)})}}  
{z_{\m \n}} T_a^{(\n)} T_b^{(\m)} t_{\hat{j}j}^{(\n)} (\s) t_{-\hat{j},j}^{(\m)}(\s) t_{-1,j}^{(\m)} (\s) = 0 
\end{equation} 
\begin{equation} 
\label{RMStep2} 
\sum_\m {z_{\m}}^{(1 - \frac{1}{f_j(\s)})} \hat{W}_\m (\s) t_{-1,j}^{(\m)} (\s) = -(1-\frac{1}{f_j (\s)}) 
\sum_\m z_{\m}^{-\frac{1}{f_j(\s)}} \dg (T^{(\m)}) t_{-1,j}^{(\m)} (\s) 
\end{equation} 
\begin{eqnarray} 
\label{ResultM} 
\hat{A}_+ (\s) \sum_{\m} (\lplm z_\m + (1-\frac{1}{f_j(\s)}) \dg (T^{(\m)}) )t_{-1,j}^{(\m)}(\s) z_{\m}^{-\frac{1}{f_j(\s)}}= 
\bigspc \bigspc \quad \quad \quad \nn \\ 
= \hat{A}_+ (\s) \sum_{\m} (z_\m \hat{W}_\m (\s)+ (1-\frac{1}{f_j(\s)}) \dg (T^{(\m)}) )t_{-1,j}^{(\m)}(\s) z_{\m}^{-\frac{1}
{f_j(\s)}}= 0 . \quad 
\end{eqnarray} 
\end{subequations} 
Here the first equation in \eqref{WrdHpM} holds for each $(\m \n) + (\n \m)$ pair separately, by taking
$\hat{j} \rightarrow (f_j(\s)-1 -\hat{j})$ in the second term. Eq.~\eqref{RMStep2} is obtained from 
\eqref{RMStep1} and the form of the twisted KZ connection, and the result \eqref{ResultM} is implied immediately. 
 
\section{More General Sets of Principal Primary States and Fields}

We know from the {\it orbifold induction procedure} of Ref.~\cite{Borisov:1997nc} that each Virasoro primary state (with 
conformal weight $\Delta$ in the untwisted symmetric theory) gives us a set of {\it principal primary
states} $|\Delta ,\bar{\hat{j}},j \rangle_\s$ in each sector of the corresponding permutation orbifold. Following 
Ref.~\cite{Borisov:1997nc}, we may choose the base state of cycle $j$ as the state with $\bar{\hat{j}} =f_j(\s)-1$ 
(although this is a relabelling relative to that used in Subsec.~$4.2$). The base state is always primary under the 
orbifold Virasoro algebra and each of the principal primary states is primary under the semisimple integral Virasoro 
subalgebra. Each set of principal primary states is created asymptotically by its corresponding set of principal 
primary fields
$\hat{\phi}_{\Delta}^{(\hat{j}j)} (z)$,
\begin{subequations}
\begin{equation}
[\hat{L}_{\hat{j}j} \modmj ,\hat{\phi}_{\Delta}^{(\hat{l}l)} (z)]=  \de_{jl} z^{m+\jfj} (z\pl + \Delta (m+\jfj +1))
\hat{\phi}_{\Delta}^{(\hat{j}+\hat{l} ,j)} (z)
\end{equation}
\begin{equation}
|\Delta ,\bar{\hat{j}},j \rangle_\s = \lim_{z \rightarrow 0} z^{\Delta (1-\fji) -1 +\frac{\bar{\hat{j}}+1}
{f_j(\s)}} \hat{\phi}_{\Delta}^{(\hat{j}j)} (z) |0\rangle_\s ,\quad \hat{\phi}_{\Delta}^{(\hat{j} \pm f_j(\s),j)}
(z) = \hat{\phi}_{\Delta}^{(\hat{j}j)} (z)
\end{equation}
\end{subequations}
whose correlators obey orbifold $\sl (2)$ Ward identities as discussed in Ref.~\cite{Borisov:1997nc} and Subsec.~$2.6$.

In the text (see Eq.~\eqref{ChRedef}) we discussed the principal primary states 
\begin{subequations}
\begin{equation}
| \bar{\hat{j}},j;\st (T,\s) \rangle_\s = | \Dg ,f_j(\s)-1-\bar{\hat{j}} ,j\rangle_\s
\end{equation}
\begin{equation}
\hat{g}_j^+ (\st(T,\s),z,\s)_0{}^{\hat{j}} = \hat{\phi}_{\Dg}^{(f_j(\s)-1-\hat{j} ,j)} (z)
\end{equation}
\end{subequations}
associated to each block of each twisted affine primary field. As a simpler example, we mention 
here (see also Ref.~\cite{Borisov:1997nc}) the set of principal primary states associated to the twisted currents
\begin{subequations}
\begin{equation}
|\Delta =1,\bar{\hat{j}},j \rangle_\s = \lim_{z \rightarrow 0} z^{(1-\fji )-1 +\frac{\bar{\hat{j}}+1}
{f_j (\s)} } \hj_{\hat{j}aj} (z) |0\rangle_\s =\hj_{\hat{j}aj} (-1 +\frac{\bar{\hat{j}}}{f_j (\s)}) |0\rangle_\s
\end{equation}
\begin{equation}
L_\s (0) |\Delta =1,\bar{\hat{j}},j \rangle_\s = \left( \gscfwt +\frac{f_j(\s) -\bar{\hat{j}}}{f_j(\s)}
 \right) |\Delta =1,\bar{\hat{j}},j \rangle_\s
\end{equation}
\end{subequations}
where the ground state conformal weight $\gscfwt$ is given in Eq.~\eqref{GSCnf}. More generally, one finds that
\begin{equation}
\hat{\Delta}_{\bar{\hat{j}} j} (\s) = \left( \gscfwt +\frac{f_j(\s) -\bar{\hat{j}}-1 +\Delta}{f_j(\s)} \right) 
\end{equation}
is the conformal weight of the state $|\Delta ,\bar{\hat{j}},j \rangle_\s$.

\vskip .5cm 
\addcontentsline{toc}{section}{References} 
 
\renewcommand{\baselinestretch}{.4}\rm 
{\footnotesize 
 
\providecommand{\href}[2]{#2}\begingroup\raggedright\endgroup

\end{document}